\documentclass{aa}  

\usepackage{graphicx}
\usepackage{txfonts}
\usepackage{hyperref}
\usepackage{esvect}
\usepackage{lipsum}
\usepackage{fourier}
\usepackage{enumitem}
\usepackage{amsmath}
\usepackage{nccmath}
\begin{document} 
   \title{From electrons to Janskys: Full stokes polarized radiative transfer in 3D relativistic particle-in-cell jet simulations}

   \author{N. R. MacDonald
          \inst{1}
          \and
          K.-I. Nishikawa\inst{2}
          }

   \institute{Max-Planck-Institut f\"{u}r Radioastronomie, Auf dem H\"{u}gel 69, D- 53121 Bonn, Germany\\
              \email{nmacdona@mpifr-bonn.mpg.de}
         \and
             Department of Physics, Chemistry and Mathematics, Alabama A\&M University, Huntsville, AL 35811, USA\\
             \email{kenichi.nishikawa@aamu.edu}
             }

   \date{Received Month, Day, Year; accepted Month, Day, Year}

 
  \abstract
   {Despite decades of dedicated observation and study, the underlying plasma composition of relativistic extragalactic jets remains largely unknown.}
   {Relativistic magnetohydrodynamic (RMHD) models are able to reproduce many of the observed macroscopic features of these outflows (e.g., recollimation shocks, jet sheaths \& spines, bow shocks, and enshrouding jet cocoons). The nonthermal synchrotron emission detected by very long baseline interferometric (VLBI) arrays, however, is a by-product of the kinetic-scale physics occurring within the jet, physics that is not modeled directly in most RMHD codes. This paper attempts to discern the radiative differences between distinct plasma compositions within relativistic jets using small-scale 3D relativistic particle-in-cell (PIC) simulations.}
   {We made use of a polarized radiative transfer scheme to generate full Stokes imaging of two PIC jet simulations, one in which the jet is composed of an electron-proton ($e^{-}$-$p^{+}$) plasma (i.e., a normal plasma jet), and the other in which the jet is composed of an electron-positron ($e^{-}$-$e^{+}$)  plasma (i.e., a pair plasma jet). We examined the differences in the morphology and intensity of the linear polarization (LP) and circular polarization (CP) emanating from these two jet simulations.}
   {Our PIC simulations, when scaled into physical units, are $\sim$ 150 cubic kilometers in size. We find that the fractional level of CP (measured relative to integrated total intensity) emanating from the $e^{-}$-$p^{+}$ plasma jet is orders of magnitude larger than the level emanating from an $e^{-}$-$e^{+}$ plasma jet of a similar speed and magnetic field strength. In addition, we find that the morphology of both the linearly and circularly polarized synchrotron emission is distinct between the two jet compositions. These results highlight the following: (i) the potential of high-resolution full-Stokes polarimetric imaging to discern between normal plasma and pair plasma jet emission in larger scale systems and (ii) the challenges faced by kinetic simulations in modeling this emission self-consistently. We also demonstrate the importance of slow-light interpolation and we highlight the effect that a finite light-crossing time has on the resultant polarization when ray-tracing through relativistic plasma. Placing a firm constraint on the plasma content of relativistic extragalactic jets will help to advance our understanding of jet feedback.}
   {}

  \titlerunning{Polarized Radiative Transfer Through 3-D PIC Jet Simulations}
  \authorrunning{MacDonald \& Nishikawa}
   \maketitle
%

\section{Introduction}
\label{Introduction}

Relativistic extragalactic jets are among the most persistent energetic objects in the universe. They are composed of collimated beams of magnetized relativistic plasma that can extend up to thousands (and in some cases millions) of parsecs from their host galaxies. The current theoretical paradigm postulates that the ultimate physical mechanism powering these relativistic outflows is the energy released from matter accreting onto spinning supermassive black holes \citep{blandford77, blandford82}. The plasma content of these jets (and their central engines) remains an active area of research (see, e.g., \citealt{croston18, fan18, thum18, myserlis18a, myserlis18b, enblin19, anantua20, sikora20, emami21, moscibrodzka21, ricarte21}).

With the advent of global (and recently space-based) millimeter-wave very long baseline interferometry (VLBI), we are able to probe the polarized emission emanating from the innermost regions of a number of jets. In particular, the linearly and circularly polarized synchrotron emission from these jets carry imprints of both the strength and orientation of the collimating magnetic fields as well as the plasma content of each jet. Studying the nature of this synchrotron emission can be used to infer the physical conditions both within the jet and in the surrounding environment into which the jet propagates.  
   
In parallel to this observational advance, modern computational resources have allowed for increasingly sophisticated numerical plasma simulations. In particular, 3D particle-in-cell (PIC) simulations (e.g., \citealt{nishikawa14, nishikawa16a, alves18, guo20, nishikawa20a}) have enabled, for the first time, a self-consistent treatment of the kinetic effects occurring within relativistic plasma outflows, such as the following: plasma instabilities, jet shear and entrainment, and magnetic reconnection (see \citealt{birdsall95} and \citealt{nishikawa20b} for a summary of PIC methods). These PIC simulations, however, are numerically intensive, and this kinetic precision comes at the cost of small (relative to relativistic magnetohydrodynamic - RMHD) simulation sizes. There exists, therefore, a synergy between PIC and RMHD jet modeling that should be exploited in order to gain a more holistic understanding of both the micro and macro physics of relativistic outflows. While RMHD simulations can effectively model the large-scale fluid motions of the jet (e.g., \citealt{mizuno15, tchekhovskoy16, fuentes18, fromm19, mukherjee20, mukherjee21}), PIC can be used to model the microphysics and radiative processes occurring within the jet plasma (e.g., \citealt{sironi15b, zhang18, petropoulou19, zhang20, hosking20, davelaar20, sironi21}). These kinetic-scale processes form a direct link to VLBI observations of the polarized synchrotron emission.

In this paper, we compare the radiative differences between two PIC jet simulations (computed using the TRISTAN-MPI code\footnote{\href{https://ascl.net/1908.008}{https://ascl.net/1908.008}}; see \citealt{niemiec08} \& \citealt{buneman93}), one in which the jet is composed of an electron-proton ($e^{-}$-$p^{+}$) plasma (i.e., a normal plasma jet), and the other in which the jet is composed of an electron-positron ($e^{-}$-$e^{+}$) plasma (i.e., a pair plasma jet). We make use of a polarized radiative transfer scheme (see \citealt{macdonald18}) that has been embedded into a ray-tracing code for post-process imaging of each numerical jet simulation. We use this ray-tracing code to create synthetic full Stokes (I, Q, U, \& V) images of each numerical PIC jet simulation. This paper is organized as follows: In \S\ref{Scaling} we summarize the scaling relations used in our PIC jet simulations. In \S\ref{Polarized Radiative Transfer} \& \ref{Slow-Light Interpolation} we outline the radiative transfer theory adopted in our study. In \S\ref{Results} we present the results of our ray-tracing calculations through the $e^{-}$-$p^{+}$ plasma and $e^{-}$-$e^{+}$ plasma jet simulations. Finally, in \S\ref{Summary and Conclusions} we present our summary and conclusions. We adopt the following cosmological parameters: $H_{\rm o} = 71 ~ \rm km ~ \rm s^{-1} ~ \rm Mpc^{-1}$, $\Omega_{\rm m} = 0.27$, and $\Omega_{\Lambda} = 0.73$.

\begin{figure*}
\centering
\includegraphics[width=0.49\textwidth]{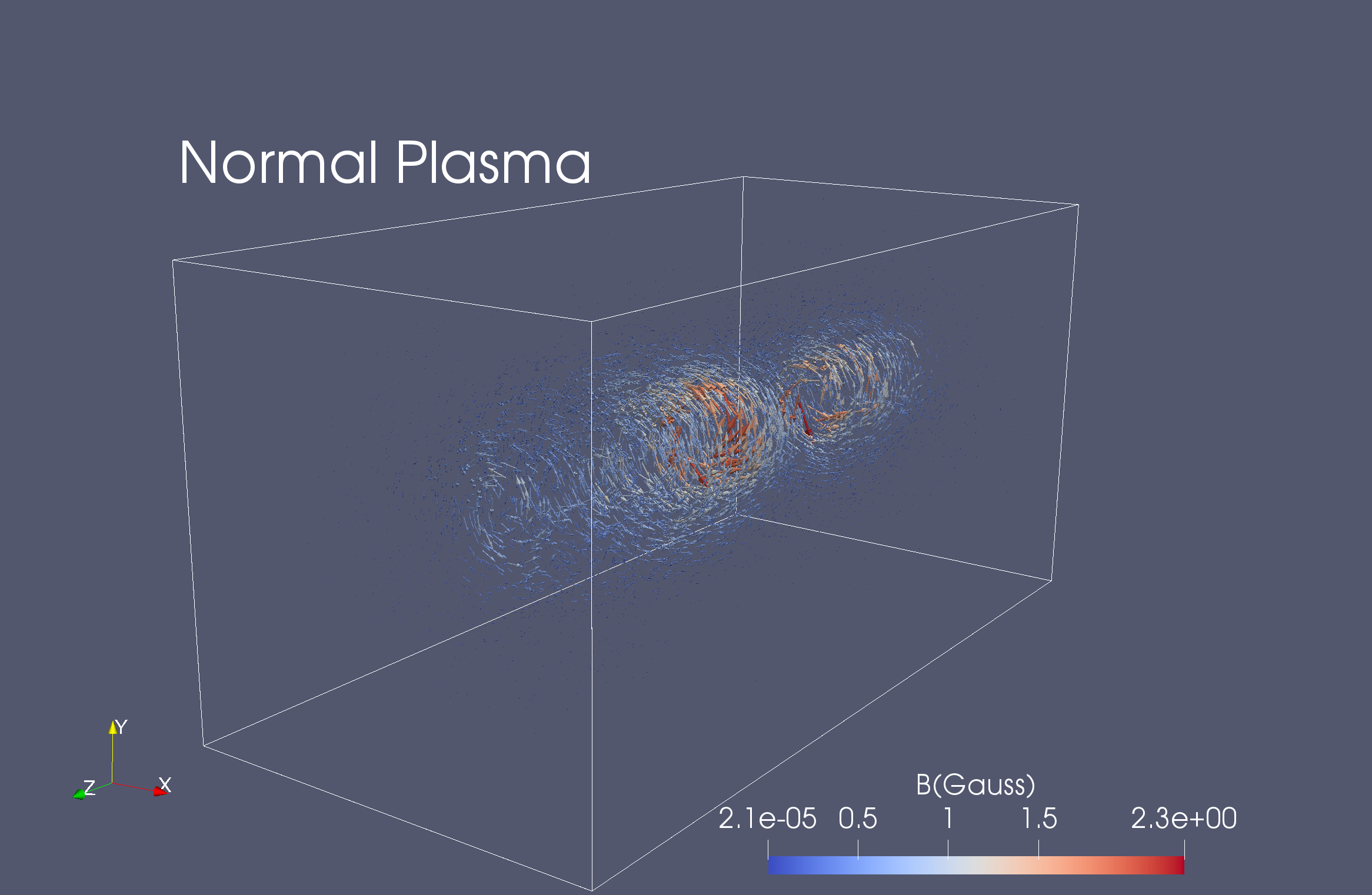}
\includegraphics[width=0.49\textwidth]{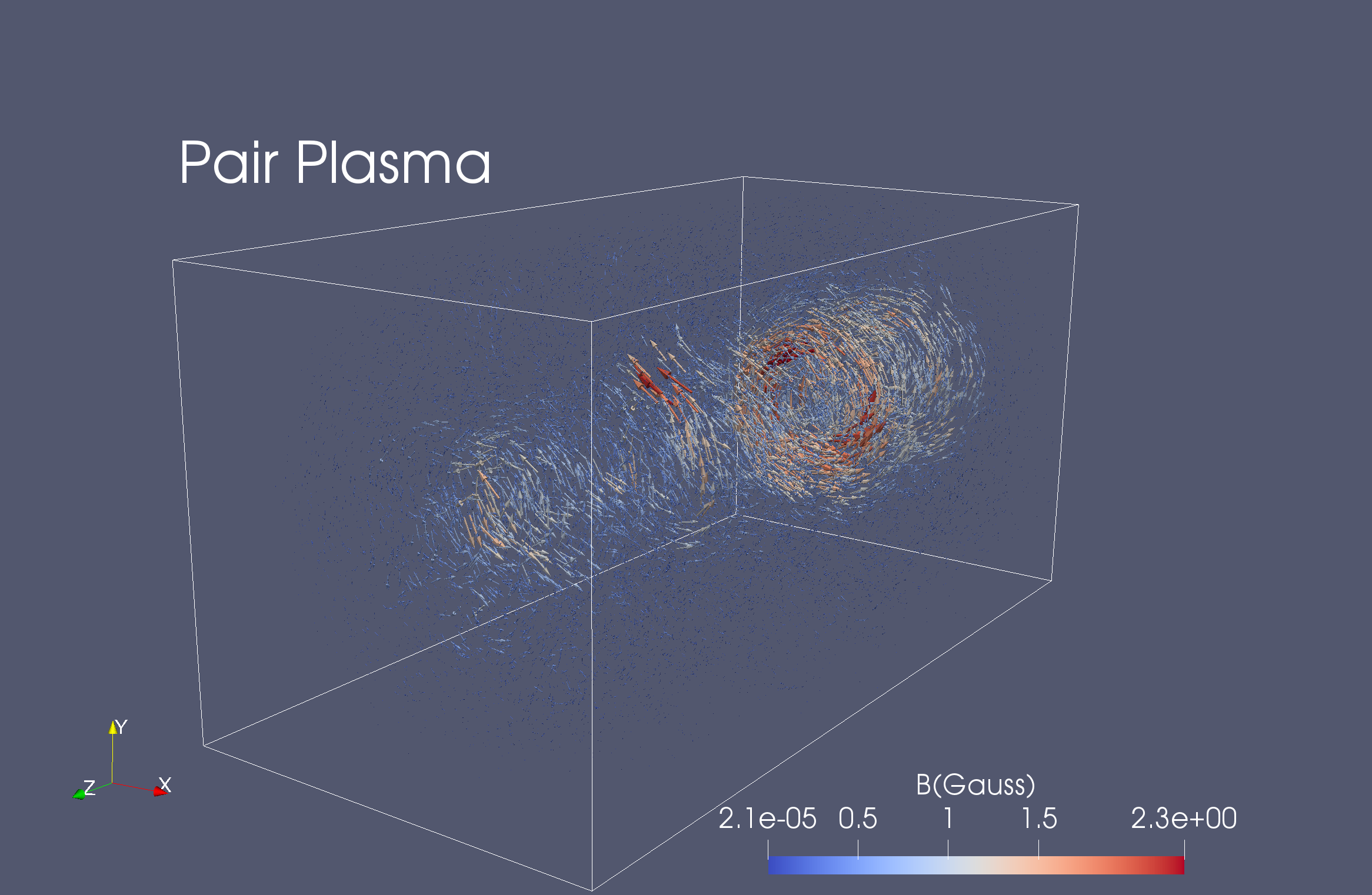}
\caption{\label{fig1}\textbf{Left panel:} 3D visualization of the magnetic field within the normal plasma ($e^{-}$-$p^{+}$) PIC jet simulation. Each vector highlights the magnetic field strength within an individual computational cell. \textbf{Right panel:} 3D visualization of the magnetic field within the pair plasma ($e^{-}$-$e^{+}$) PIC jet simulation. The same convention (normal plasma on the left and pair plasma on the right) is used in the other figures of this paper.}
\end{figure*}

\section{Scaling}
\label{Scaling}

PIC calculations are carried out in dimensionless grid units which must be scaled (via scaling factors) into physical units (i.e., cgs) as a post-process step for use in our ray-tracing calculations. The PIC simulations presented in this paper are small ($120 \times 120 \times 240$ cells) segments of the $e^{-}$-$p^{+}$ and $e^{-}$-$e^{+}$ PIC jet simulations published in \cite{nishikawa17}. 

The simulation cell size ($\Delta l$), the time step ($\Delta t$), and the speed of light ($c$) are computed in the following manner: 
\begin{align}
\Delta l &= \Delta l_{\rm{scale}} ~ \times ~ \Delta l_{\rm{grid}} \label{eqn1} \\
\Delta t &= \Delta t_{\rm{scale}} ~ \times ~ \Delta t_{\rm{grid}} \label{eqn2} \\
c &= ~ ~ v_{\rm{scale}} ~ \times ~ ~ ~ c_{\rm{grid}} \label{eqn3} ~,
\end{align}

\noindent where by definition the grid values: $\Delta l_{\rm{grid}} = \Delta t_{\rm{grid}} = c_{\rm{grid}} = 1$. The quantities: $\Delta l_{\rm{scale}}$, $\Delta t_{\rm{scale}}$, and $v_{\rm{scale}}$ represent scaling factors that set the physical dimensions of each jet simulation. In both jet simulations presented in this paper, we set $v_{\rm{scale}} = 3 \times 10^{10} ~ \rm{cm} ~ \rm{s}^{-1}$. 

The plasma skin depth ($\equiv c/ \omega_{pe}$) parameterizes the length scale across which low-frequency electromagnetic radiation can propagate within a collisionless plasma. The quantity $\omega_{pe}$ denotes the relativistic electron plasma frequency and is given by:
\begin{equation}\label{eqn4}
\omega_{pe} =  \sqrt{ \frac{ 4 \pi ~ n_{e} ~ e^{2} }{ \gamma_{e} m_{e} } } ~,
\end{equation}
where $n_{e}$, $e$, and $m_{e}$ denote the number density [$\rm{cm}^{-3}$], charge [$\rm{cm}^{3/2} ~ \rm{g}^{1/2} ~ \rm{s}^{-1}$], and mass [$\rm{g}$] of the electrons within the plasma and $\gamma_{e}$ is the electron Lorentz factor. The plasma frequency is the fundamental frequency of oscillation of the electrons and together with the plasma skin depth sets a fundamental length scale within our PIC jet simulations.

In order to model the kinetic scales within our jet simulations, the plasma skin depth of the jet plasma is resolved across ten computational grid cells by the TRISTAN code:
\begin{equation}\label{eqn5}
10 ~ \Delta l = \frac{ c }{ \omega_{pe} } ~.
\end{equation} 
Equation \ref{eqn5}, therefore, defines the characteristic length scale of the numerical cells within our PIC jet simulations. We define a \textit{fiducial} jet electron number density as follows: 
\begin{equation}\label{eqn6}
n_{e} \simeq 10^{1} ~ \rm{ cm }^{ -3 } ~,
\end{equation} 
to which we scale our PIC values. By combining Equations \ref{eqn1}, \ref{eqn3}, \ref{eqn4}, \ref{eqn5}, and \ref{eqn6} we obtain the following PIC cell size scale factor:
\begin{equation}\label{eqn7}
\Delta l_{\rm{scale}} = \frac{ v_{\rm{scale}} }{10} ~ \sqrt{ \frac{ \gamma_{e} m_{e} }{ 4 \pi ~ n_{e} ~ e^{2} } } \simeq 6.5 \times 10^{4} ~ \rm{ cm } ~,
\end{equation}
having set $m_{e} = 9.1094 \times 10^{-28} \rm{g}$,  $e = 4.8066 \times 10^{-10} \rm{cm}^{3/2} \rm{g}^{1/2} \rm{s}^{-1}$, and $\gamma_{e} \simeq 15$ (the limiting value at the jet base in our models). This scaling implies that the jet axis of both of our PIC simulations (i.e., 240 cells in length) corresponds to a physical size of $\sim 150 ~ \rm{ km }$. This physical size demonstrates the vastly different spatial scales probed by PIC simulations in comparison to RMHD and represents an inherent challenge PIC codes face when attempting to model the plasma of parsec scale astrophysical jets.

As in Equations \ref{eqn1}, \ref{eqn2}, and \ref{eqn3}, we scale the magnetic field within our PIC simulations in the following manner:
\begin{equation}\label{eqn8}
B = B_{\rm{scale}} ~ \times ~ B_{\rm{grid}} ~,
\end{equation}
where the dimensionless grid value at the base of our jet models is $B_{\rm{grid}} \simeq 2.7$ (and varies throughout each simulation box), and $B_{\rm{scale}}$ is a magnetic field scaling factor $[\rm{Gauss}]$. We define a \textit{fiducial} electron plasma beta ($\beta_{e}$) at the jet base in each of our PIC simulations to which we scale the magnetic field strength:
\begin{equation}\label{eqn9}
\beta_{e} \equiv \frac{ 8 \pi ~ n_{e} ~ k T_{e} }{ B^{2} } ~ \simeq \frac{ 8 \pi ~ n_{e} }{ B^{2} } ~ \frac{ \gamma_{e} m_{e} c^{2} }{ 3 } ~ \simeq 10^{-3} ~.
\end{equation}
The value of $\beta_{e}$ has been chosen to be indicative of a highly-magnetized synchrotron emitting relativistic plasma. By combining Equations \ref{eqn3}, \ref{eqn8}, and \ref{eqn9} we arrive at the following magnetic field scale factor:
\begin{equation}\label{eqn10}
B_{\rm{scale}} = \frac{ 1 }{ B_{\rm{grid}} } ~ \sqrt{ \frac{ 8 \pi }{ 3 } \frac{ n_{e} \gamma_{e} m_{e} v_{\rm{scale}}^{2} }{ \beta_{e} } } ~ \simeq 0.4 ~ \rm{Gauss} ~, 
\end{equation}
again having set $n_{e} \simeq 10^{1} ~ \rm{ cm }^{ -3 }$ and $\gamma_{e} \simeq 15$. The above scale factor results in magnetic field strengths of $B \simeq 10^{0} ~ \rm{Gauss}$ at the base of each jet (see Figure \ref{figB1}), which in combination with our fiducial electron number density of $n_{e} \simeq 10^{1} ~ \rm{cm}^{-3}$ produces numerically tractable levels of synchrotron emission at radio/mm wavelengths within our PIC simulations. This choice of physical scaling also helps to ensure that the electron Larmor radii do not vastly exceed the scaled PIC cell sizes (see Appendix \ref{appA}).

The electron number density [$\rm{cm}^{-3}$] within any PIC computational cell is by definition:
\begin{equation}\label{eqn11}
n_{e} = \frac{ n_{e ~ \rm{grid}} }{ \Delta l^{3} } ~.
\end{equation}
The dimensionless electron grid value at the base of our jet models is $n_{e ~ \rm{grid}} \simeq 64$ (and varies throughout each simulation box). Combining Equations \ref{eqn1}, \ref{eqn7}, and \ref{eqn11} results in a scaled electron number density of $n_{e} \simeq 2.3 \times 10^{-13} ~ \rm{cm}^{-3}$ at the jet base in each of our models. This value is many orders of magnitude below the levels inferred from observational and theoretical modeling of relativistic jets (i.e., our fiducial value of $n_{e} \simeq 10^{1} ~ \rm{cm}^{-3}$). This difference in particle number again highlights the challenge faced by PIC codes when attempting to model the plasma of parsec scale astrophysical jets. 

To help mitigate the numerical limitation in our ability to simulate astrophysically plausible numbers of electrons in our PIC jet calculations, we introduce a PIC "super particle" parameter ($f_{p}$) which we apply to each computational cell within our jet models: $n_{e ~ \rm{grid}}^{\prime} = f_{p} \times n_{e ~ \rm{grid}}$, where $n_{e ~ \rm{grid}}^{\prime}$ is a proxy for the number of `real' electrons represented by our `simulation' electrons. As in Equations \ref{eqn1}, \ref{eqn2}, \ref{eqn3}, and \ref{eqn8}, we formulate the following electron number density scaling relation which we apply throughout our simulations:
\begin{equation}\label{eqn12}
n_{e} = n_{e ~ \rm{scale}} ~ \times ~ n_{e ~ \rm{grid}} ~,
\end{equation}
where:
\begin{equation}\label{eqn13}
n_{e ~ \rm{scale}} = \frac{ f_{p} }{ \Delta l^{3} } ~. 
\end{equation}
Since $\Delta l \simeq 6.5 \times 10^{4} ~ \rm{cm}$ (Equations \ref{eqn1} and \ref{eqn7}) and $n_{e ~ \rm{grid}} \simeq 64$, in order to scale the dimensionless grid electron number densities at the jet base to our fiducial jet value of $n_{e} \simeq 10^{1} ~ \rm{cm}^{-3}$, we need to tune our PIC `super particle' parameter to $f_{p} \simeq 4.6 \times 10^{13}$. This immense value highlights a numerical limit PIC codes face in simulating plasmas on astrophysical scales and also sets a numerical benchmark for future simulations with larger particle populations.

With our dimensionless PIC grid values scaled into physical units (i.e., cgs), we are now in a position to apply polarized radiative transfer via ray-tracing, in order to infer both the level and the morphology of the polarized synchrotron emission emanating from the normal plasma jet and the pair plasma jet (both of which are illustrated in Figure \ref{fig1}).  

\begin{figure*}
\centering
\scalebox{0.67}{\includegraphics[trim={0 1.25cm 0 5.5cm}, clip, width=2.0\columnwidth,clip]{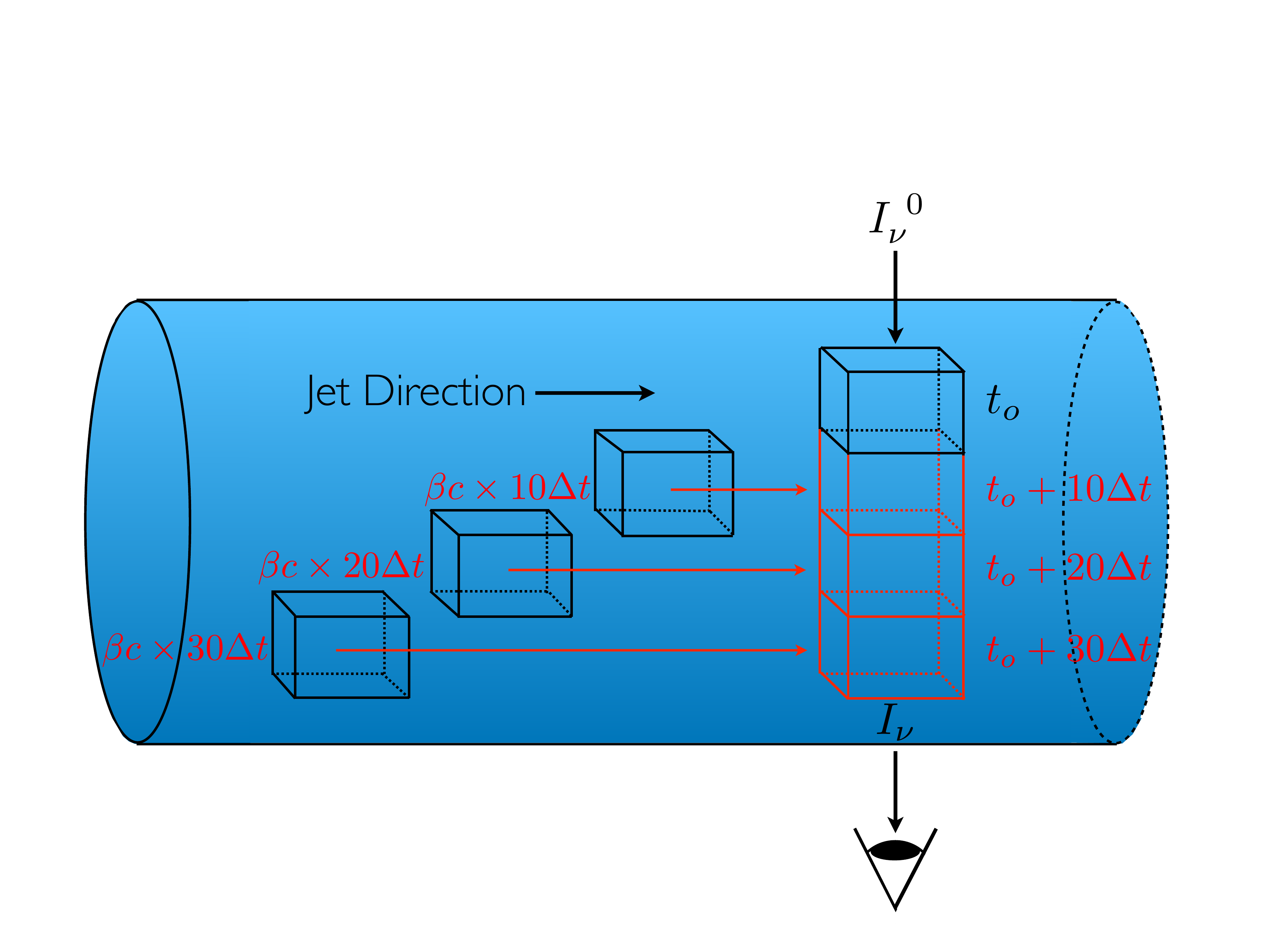}}
\caption{\label{fig2} Schematic representation of our PIC slow-light interpolation scheme.  A ray ($I_{\nu}^{~ 0}$) enters the jet and encounters successive upstream plasma cells as the ray propagates across the jet (similar to stepping through a fast moving stream). These successive encounters contribute to the emission stored in each ray and result in the final total intensity value ($I_{\nu}$) that exits the jet and produces one pixel in our synthetic radio maps.}
\end{figure*}  
  
\section{Polarized radiative transfer}
\label{Polarized Radiative Transfer} 

\citealt{jones77a, jones77b, jones88} present solutions to the full Stokes equations of polarized radiative transfer for synchrotron emission emanating from a homogeneous and an inhomogeneous magnetized plasma containing isotropic distributions of electrons (see Appendix \ref{appG} for further discussion). We use these solutions to compute the levels of LP and CP emanating from our PIC jet simulations at radio frequencies. In particular, we solve the matrix presented in Equation \ref{eqn14} along individual rays passing through our PIC simulations. 

\begin{equation}\label{eqn14}\tiny{
\left( \begin{array}{cccc}
\left( \dfrac{d}{dl} + \kappa_{I} \right) & \kappa_{Q} & \kappa_{U} & \kappa_{V} \\
\kappa_{Q} & \left( \dfrac{d}{dl} + \kappa_{I} \right) & \kappa^{*}_{~ V} & -\kappa^{*}_{~ U} \\
\kappa_{U} & -\kappa^{*}_{~ V} & \left( \dfrac{d}{dl} + \kappa_{I} \right) & \kappa^{*}_{~ Q} \\
\kappa_{V} &  \kappa^{*}_{~ U} & -\kappa^{*}_{~ Q} & \left( \dfrac{d}{dl} + \kappa_{I} \right) \end{array} \right) 
\left( \begin{array}{c}
I_{\nu} \\[11pt]
Q_{\nu} \\[11pt]
U_{\nu} \\[11pt]
V_{\nu} \end{array} \right) = 
\left( \begin{array}{c}
\eta_{\nu}^{~ I} \\[10pt]
\eta_{\nu}^{~ Q} \\[10pt]
\eta_{\nu}^{~ U} \\[10pt]
\eta_{\nu}^{~ V} \end{array} \right) }~.
\end{equation}

\noindent Here, $I_{\nu}$, $Q_{\nu}$, $U_{\nu}$, and $V_{\nu}$ denote the frequency-dependent Stokes parameters, while $(\kappa_{I}, \kappa_{Q}, \kappa_{U}, \kappa_{V})$ and $(\eta_{\nu}^{~ I}, \eta_{\nu}^{~ Q}, \eta_{\nu}^{~ U}, \eta_{\nu}^{~ V})$ represent, respectively, the synchrotron absorption and emission coefficients for each Stokes parameter. The terms $(\kappa^{*}_{~ Q}, \kappa^{*}_{~ U})$ and $(\kappa^{*}_{~ V})$ account for the effects of Faraday conversion and rotation (see Appendix \ref{appB}), respectively, within our PIC plasma. The term $l$ denotes the path length of each ray through the computational cells of our PIC calculations. \cite{jones77a} present an analytic solution to this matrix which we apply along each ray. The radiative transfer is carried out in the `co-moving' frame of the plasma with: (i) a relativistic abberation correction being applied, cell-by-cell, to obtain the angle between each cell's local magnetic field vector and the observer's inclination to the jet axis , and (ii) a rotation correction being applied, cell-by-cell, to transform the linear polarization ellipse from the local co-moving frame onto the plane of the sky (see Appendix \ref{appF} for further discussion). Once the Stokes parameters have been generated for each sightline (i.e., for each pixel/ray in our synthetic maps) a Doppler factor is applied (which incorporates the velocity/angle dependence of the larger scale jet) to obtain the flux levels in the observer's frame. This Doppler boosting/de-boosting is apparent upon comparison of the relative flux levels between images generated when viewing each simulation edge-on and at right angles to the jet axis. We have embedded this polarized radiative transfer scheme into the ray-tracing code RADMC-3D\footnote{\href{http://ascl.net/1202.015}{http://ascl.net/1202.015}}. For the images presented in this paper, RADMC-3D casts $640,000$ individual rays through our cartesian PIC grids forming $800\times800$ square-pixel images of the polarized emission produced by our jet models. The resultant emission maps from these ray-tracing calculations for the normal plasma and pair plasma jet simulations are presented in Figures \ref{fig3}, \ref{fig4}, and \ref{fig6}.
   
\section{Slow-Light Interpolation}
\label{Slow-Light Interpolation}

In order to resolve (both spatially and temporally) the kinetic-scale processes occurring within a relativistic jet plasma, the time step ($\Delta t$) of the TRISTAN code is set such that:
\begin{equation}\label{eqn15}
10 ~ \Delta t = \frac{ \Delta l }{ c } ~.
\end{equation}
Therefore, it takes ten numerical time steps for a light ray to traverse the length of an individual plasma cell within each jet calculation. Equation \ref{eqn15} has profound implications for our ray-tracing calculations which typically assume/invoke the \textit{fast-light} approximation. Fast-light ray-tracing makes the assumption that the light-crossing time of the jet is far smaller than the dynamical time step of the jet simulation being imaged (i.e., that we are taking a snapshot of static jet plasma). This assumption \textit{is not} in general valid for PIC simulations. The fact that the plasma is evolving in time as each ray propagates through the computational grid must be taken into account within our ray-tracing calculations. In particular, we have constructed a slow-light interpolation scheme that accounts for this effect and is illustrated in Figure \ref{fig2}. In essence, our interpolation scheme builds up a \textit{hybrid} computational grid composed of the stratified components of successive time steps within each jet simulation. This scheme ensures that the plane parallel rays of our ray-tracing calculations encounter the correct upstream plasma values as each ray propagates through the jet. Within our PIC jet simulations:

\begin{enumerate}

\item[$\bullet$] The time scale for light crossing is $\sim 500$ simulation time steps.

\item[$\bullet$] The time scale for jet plasma evolution is $\sim 10$ simulation time steps.

\item[$\bullet$] The time scale for particle acceleration is $< 10$ simulation time steps.

\end{enumerate}

\noindent Since the time scale for light crossing $\gg$ than the time scale for jet plasma evolution, slow light interpolation has a noticeable effect on the resultant synchrotron emission (see Figures \ref{fig3} and \ref{fig6}). Our radiative transfer scheme, however, is unable to incorporate the smallest particle acceleration time scales within our emission calculations, since we implicitly assume that the plasma properties of a given cell remain constant during the 10 simulation time steps it takes a light ray to traverse each cell. We also emphasize that when scaled into physical units, a simulation time step within our jet models corresponds to $\sim 2.2^{-7} ~ \rm{s}$. This infinitesimal value is again a reminder of the drastically smaller spatial and temporal scales probed by PIC models in comparison to RMHD jet calculations.  

\section{Results}
\label{Results}

To gain a better understanding of the effects of slow-light interpolation within our ray-traced images, we first construct fast-light images for the purposes of comparison between the two ray-tracing methods. Both the normal plasma jet and the pair plasma jet simulations were run for a sufficient time span to allow both jets to propagate across the full extent of both computational grids (shown in Figure \ref{fig1}). At each time step within our PIC simulations we apply the scaling relations presented in \S\ref{Scaling} to our PIC grid values in order to create ray-tracing output files. These output files consist of three dimensional arrays containing: 
\begin{itemize}
      \item[$\bullet$] magnetic field strength (and orientation): $\hat{B}[i,j,k]$
      \item[$\bullet$] electron number densities: $n_{e}[i, j, k]$
      \item[$\bullet$] minimum electron Lorentz factors: $\gamma_{\rm{min}}[i, j, k]$
\end{itemize}
where in general $E = \gamma_{e} m_{e} c^{2}$, and the indices $[i, j, k]$ denote the $[x, y, z]$ directions within our cartesian computational grids, respectively. In contrast to magnetohydrodynamic calculations where prescriptions for the electron values must be applied in order to infer these quantities from the thermal fluid variables (e.g., \citealt{porth11, fromm16}), in our PIC calculations, we are able to compute these values directly from our jet simulations.

The output files from the TRISTAN particle-in-cell code then become input files for the RADMC-3D ray-tracing code. Given the fact that the entire computational grid in each PIC calculation (when scaled into physical units) only spans roughly $\sim$ 150 km (as discussed in \S\ref{Scaling}) we have arbitrarily placed each jet at a distance of 1 Astronomical Unit (AU) from the `observer' (i.e., the distance to the Sun). This distance scale, in
combination with our simulation box sizes, generates radio maps with an angular extent of $\sim$ 200 milli-arcseconds (mas). Clearly, 150 cubic kilometers of relativistic jet does not exist at such a close distance to the Earth, but, in the spirit of theoretical study, we proceed with these calculations to infer what: (i) the morphology and (ii) the fractional level of polarized synchrotron emission would be if we could image each of these plasma jets with an idealized interferometric array at such close proximity.

\subsection{Fast-light Images}
\label{Fast-light Images}

\begin{figure*}
\centering
\includegraphics[width=0.49\linewidth]{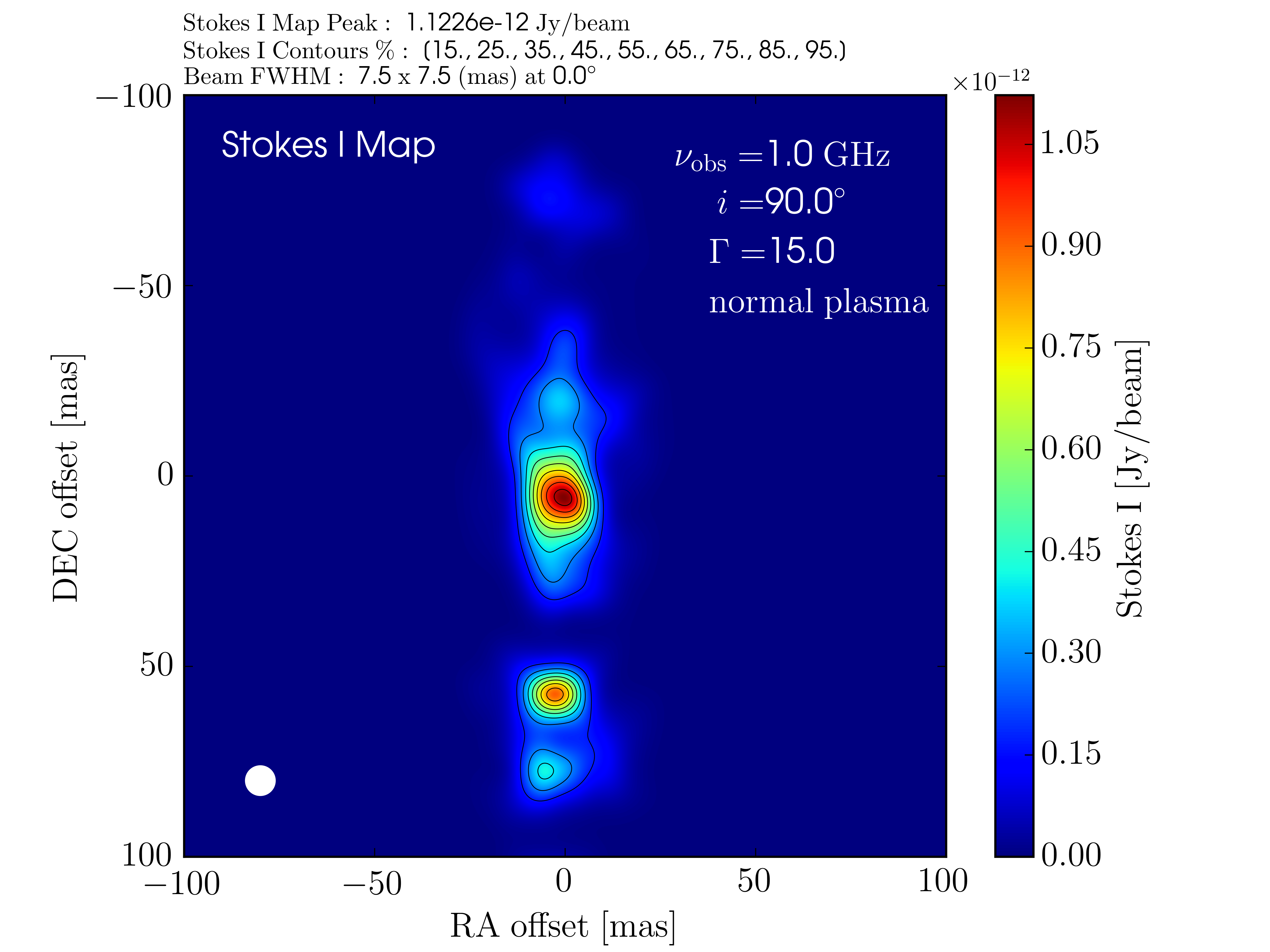}
\includegraphics[width=0.49\linewidth]{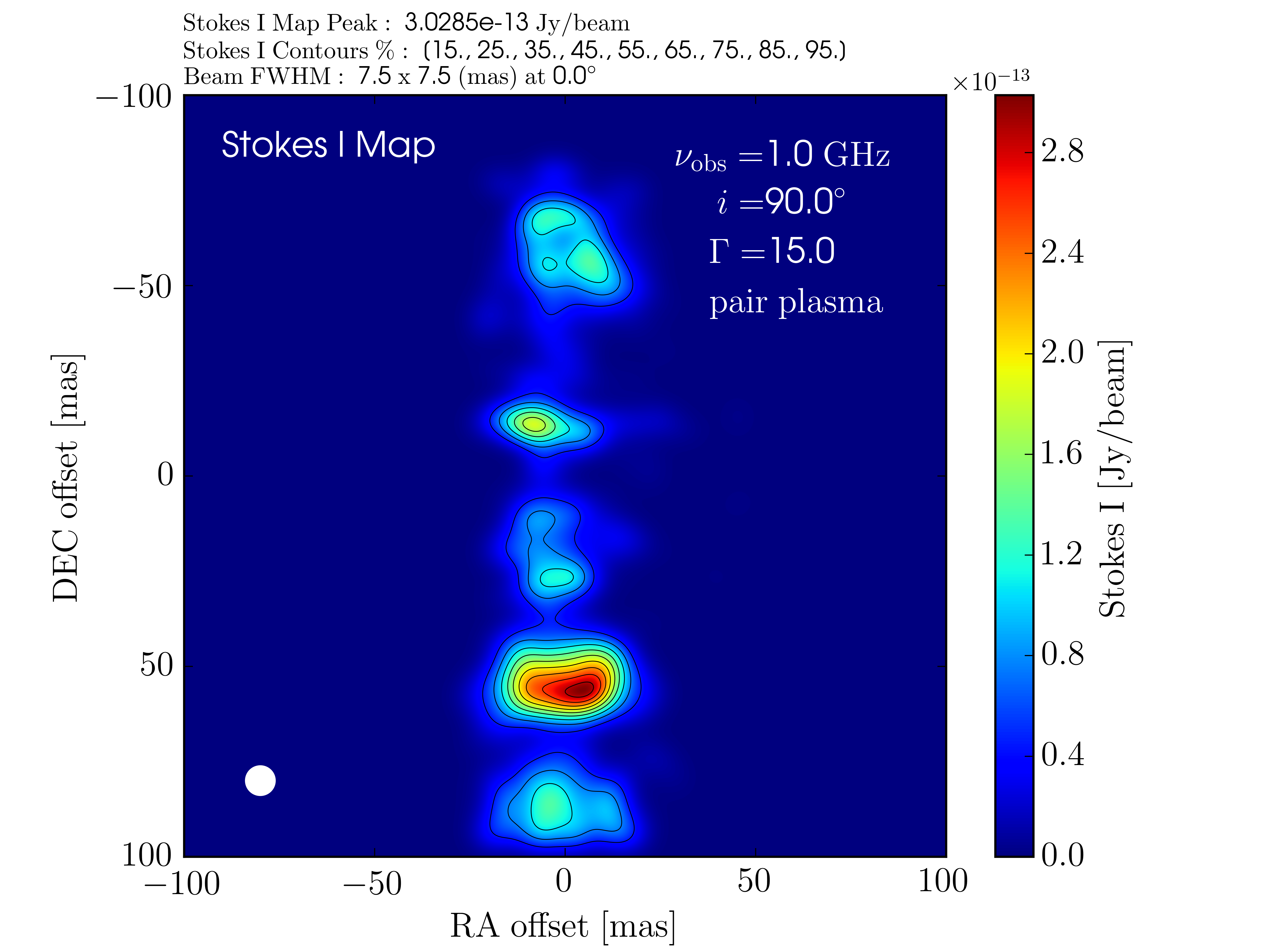}
\includegraphics[width=0.49\linewidth]{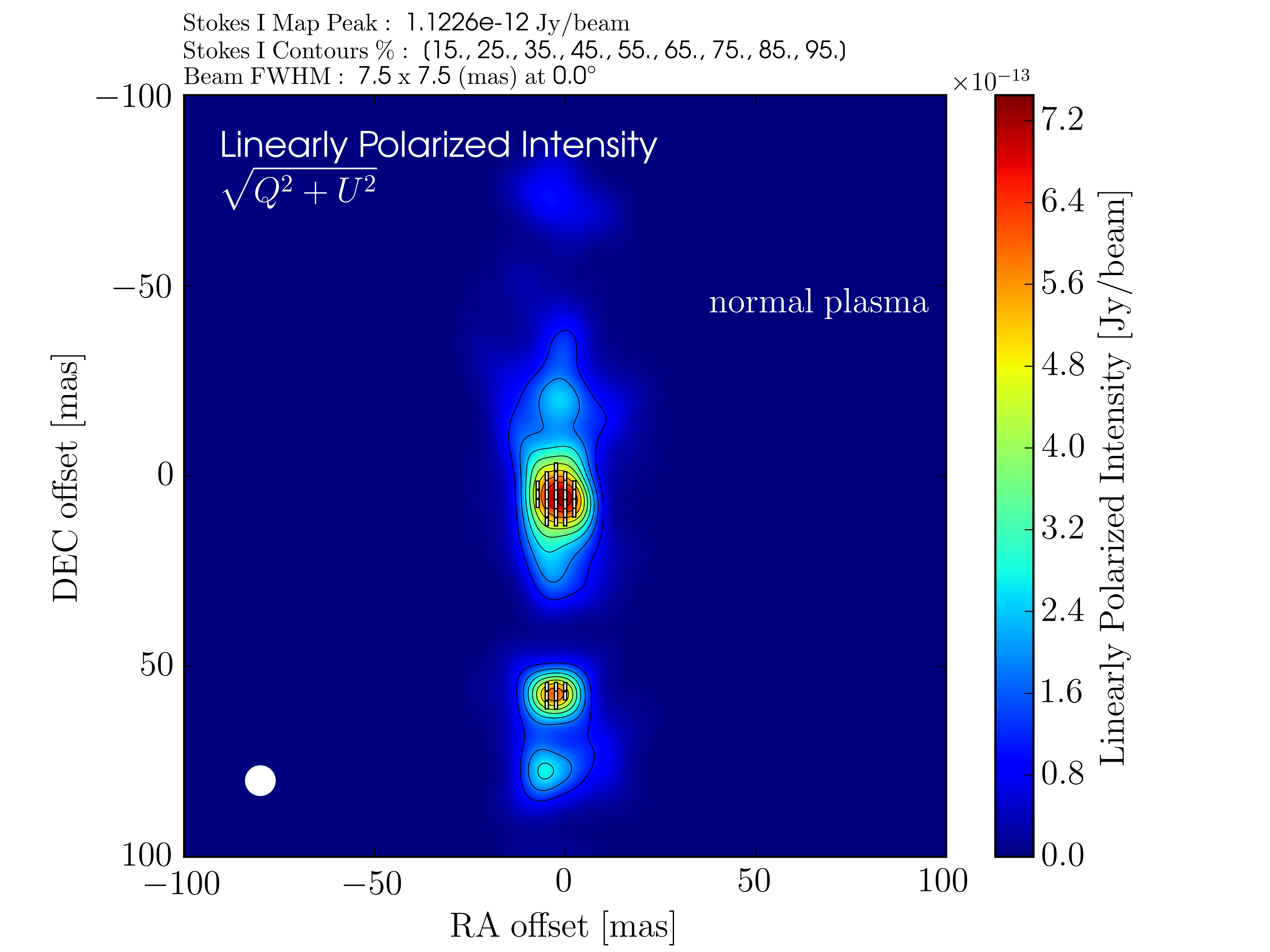}
\includegraphics[width=0.49\linewidth]{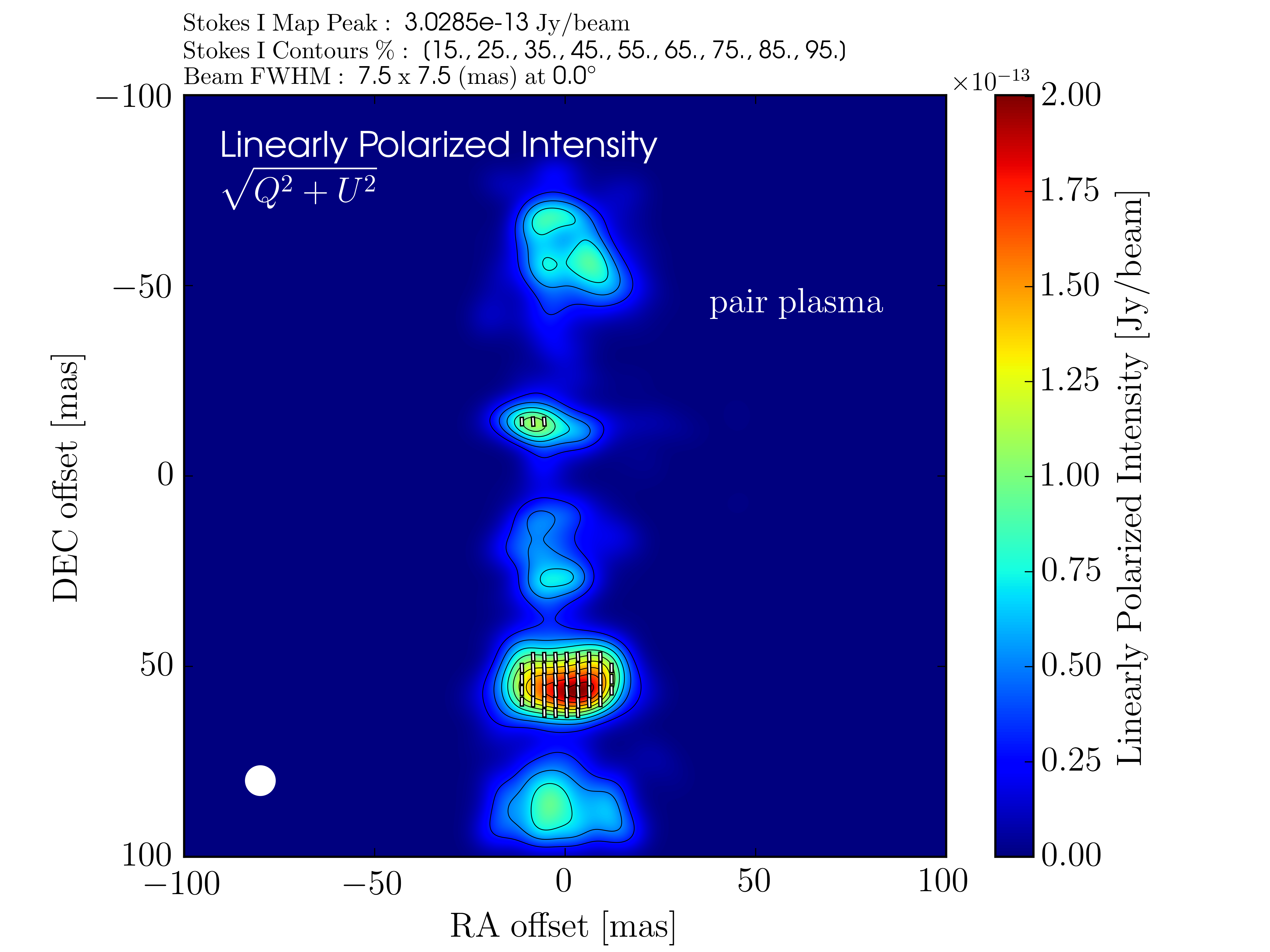}
\includegraphics[width=0.49\linewidth]{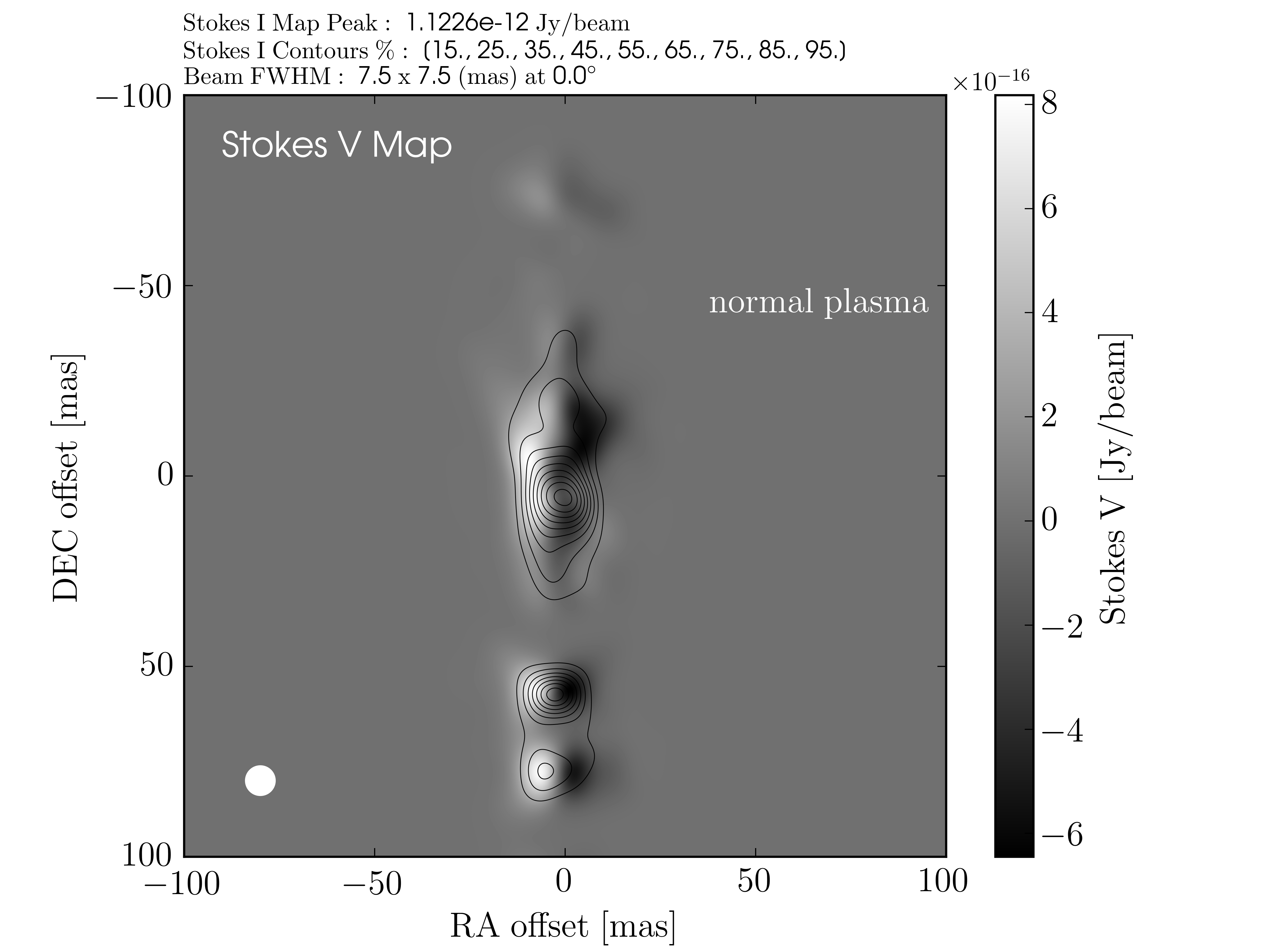}
\includegraphics[width=0.49\linewidth]{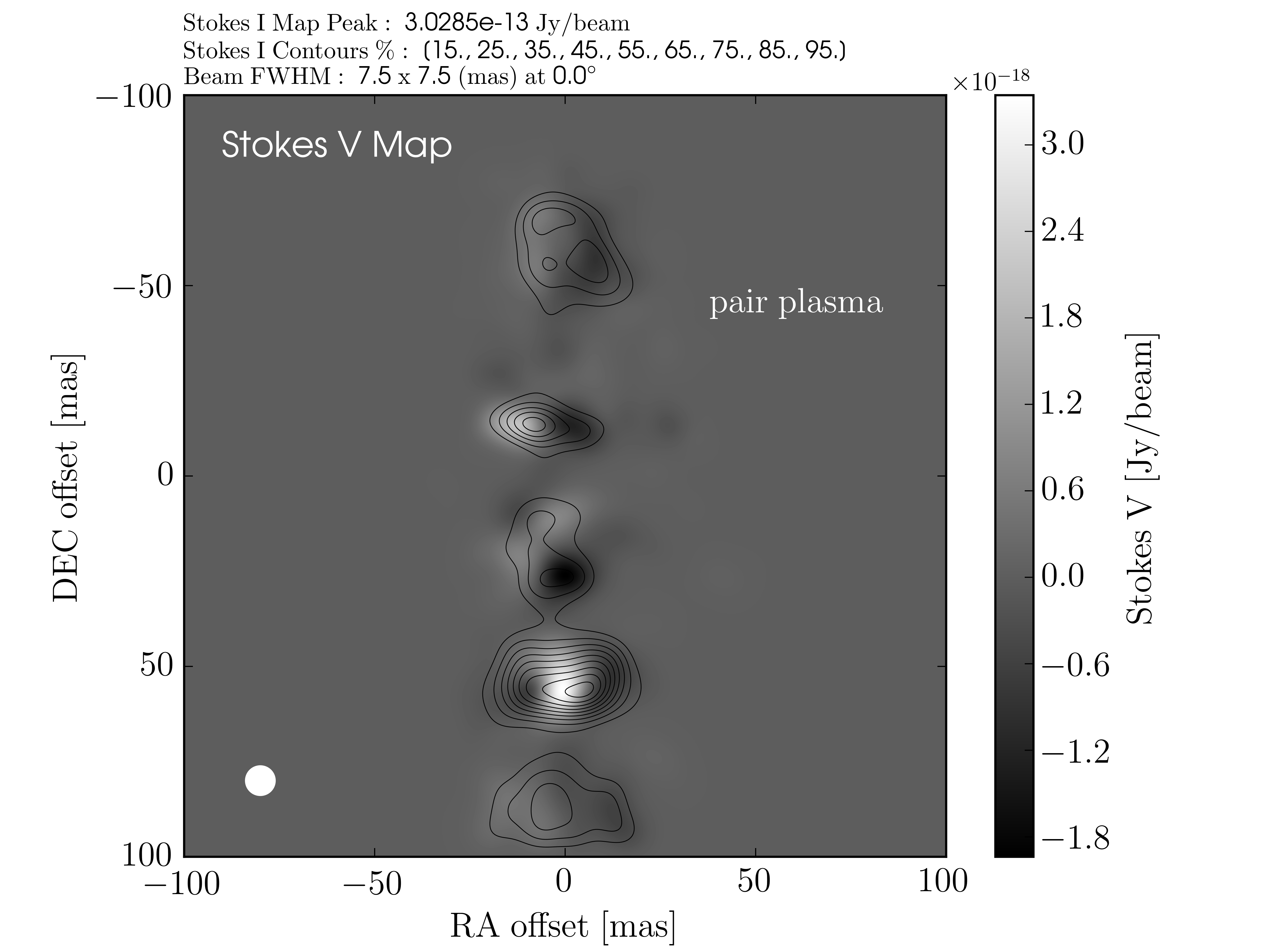}
\caption{\label{fig3}\textbf{Upper left panel:} Fast-light Stokes $I$ image (at an observing frequency of $\nu_{\rm{obs}} = 1 ~ \rm{GHz}$) of the normal plasma ($e^{-}$-$p^{+}$) jet. The internal structure of the normal plasma jet's magnetic field is illustrated in the left panel of Figure \ref{fig1}. \textbf{Upper right panel:} Corresponding fast-light Stokes $I$ image of the pair plasma ($e^{-}$-$e^{+}$) jet. The internal structure of the pair plasma jet's magnetic field is similarly illustrated in the right panel of Figure \ref{fig1}. \textbf{Middle left panel:} Rendering of the LP intensity of the normal plasma jet. White line segments indicate the electric vector position angles (EVPAs) as projected onto the plane of the sky. The effects of relativistic aberration (see \citealt{lyutikov05}) on the orientation of these EVPAs have been included in these calculations. \textbf{Middle right panel:} Corresponding LP intensity image of the pair plasma jet. \textbf{Lower left panel:} A Stokes V image of the normal plasma jet highlighting the different regions within the jet that produce positive and negative CP. \textbf{Lower right panel:} Corresponding Stokes V image of the pair plasma jet. All six images have been convolved with a circular Gaussian beam of FWHM $7.5 ~\rm{mas}$ (shown in the lower left of each panel). The projected PIC cell size is $\sim 0.9 ~ \rm{mas}$ at a source distance of $1$ AU.}
\end{figure*}

\begin{figure*}
\centering
\includegraphics[width=0.49\linewidth]{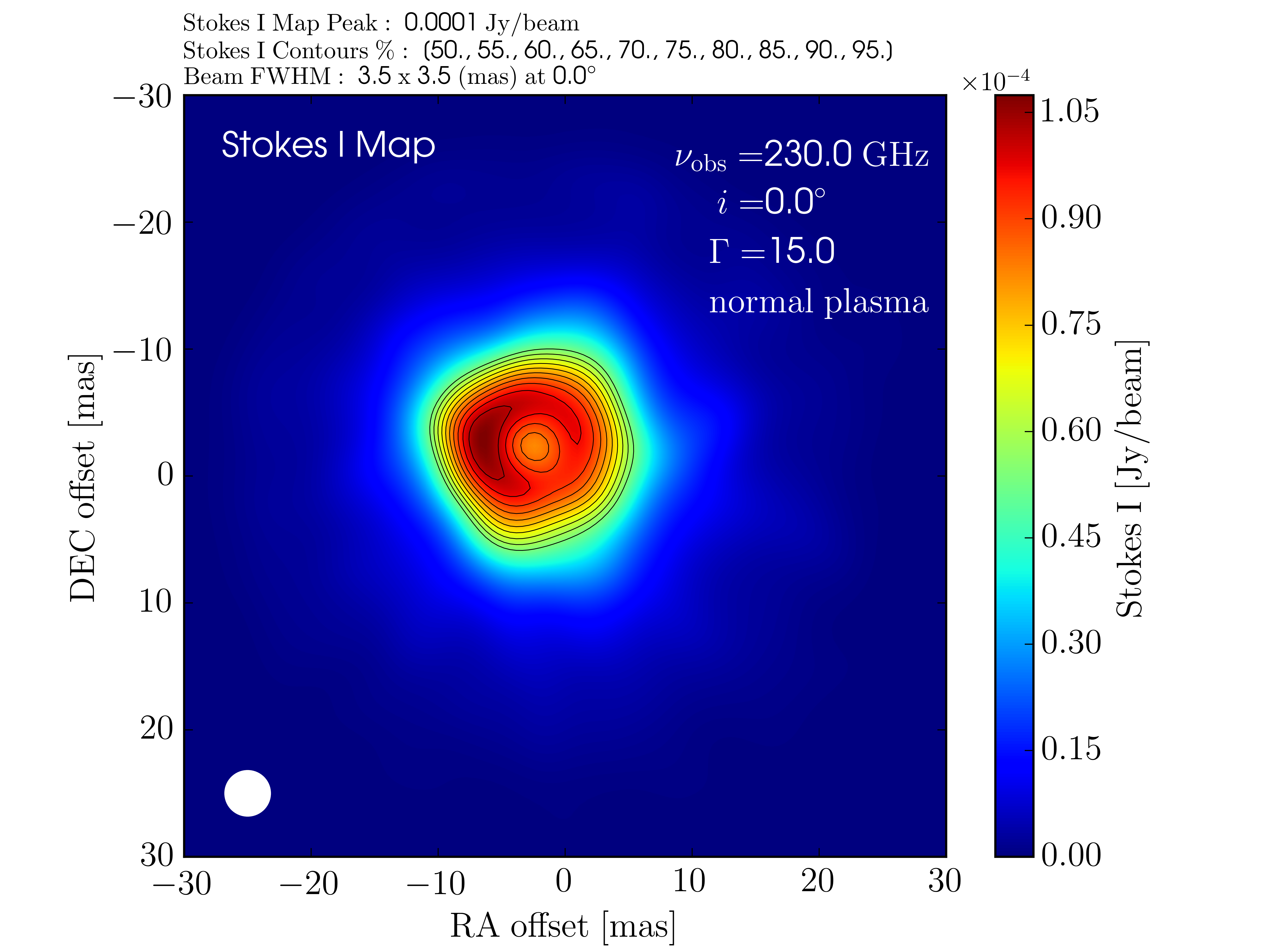}
\includegraphics[width=0.49\linewidth]{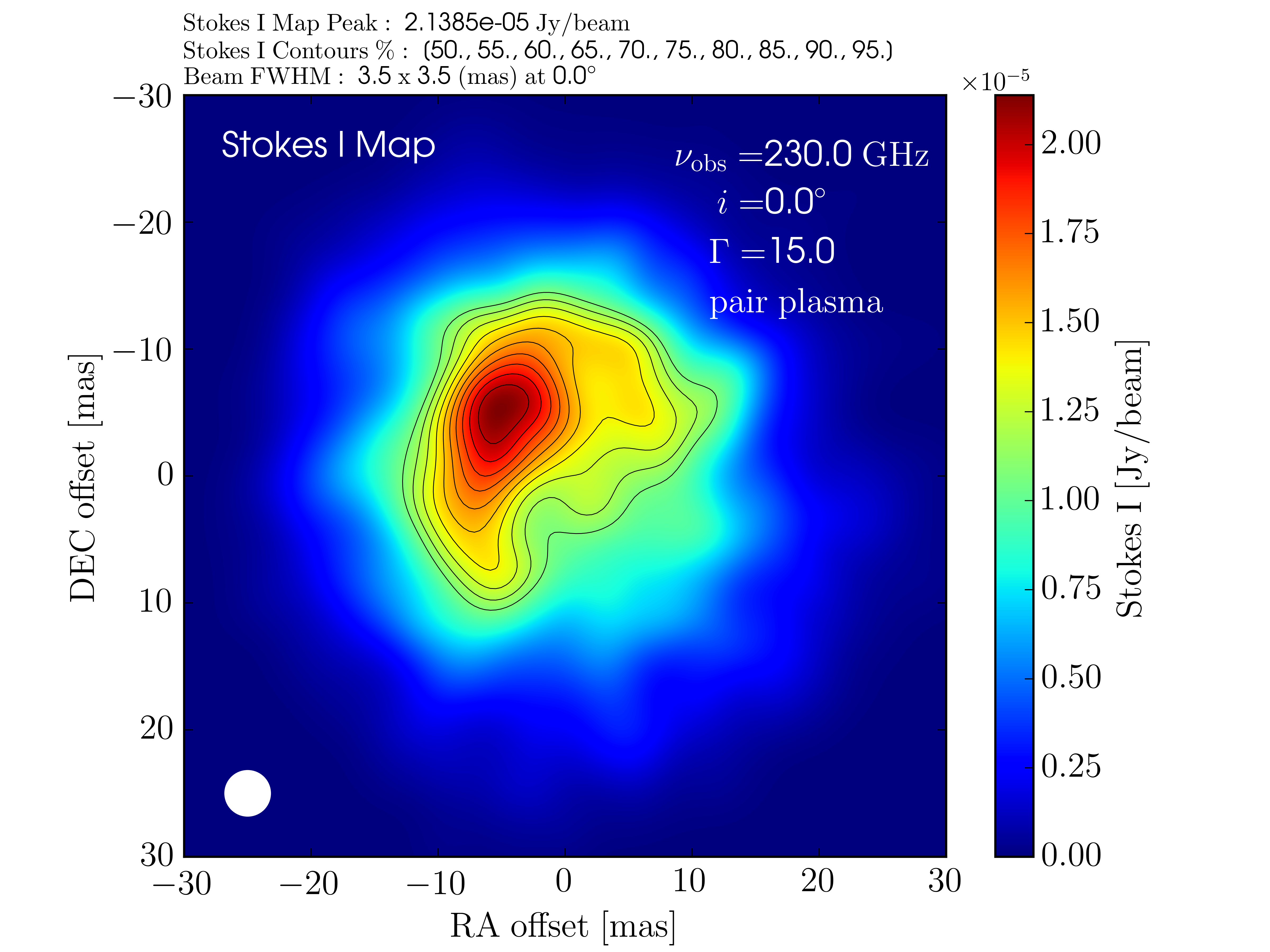}
\includegraphics[width=0.49\linewidth]{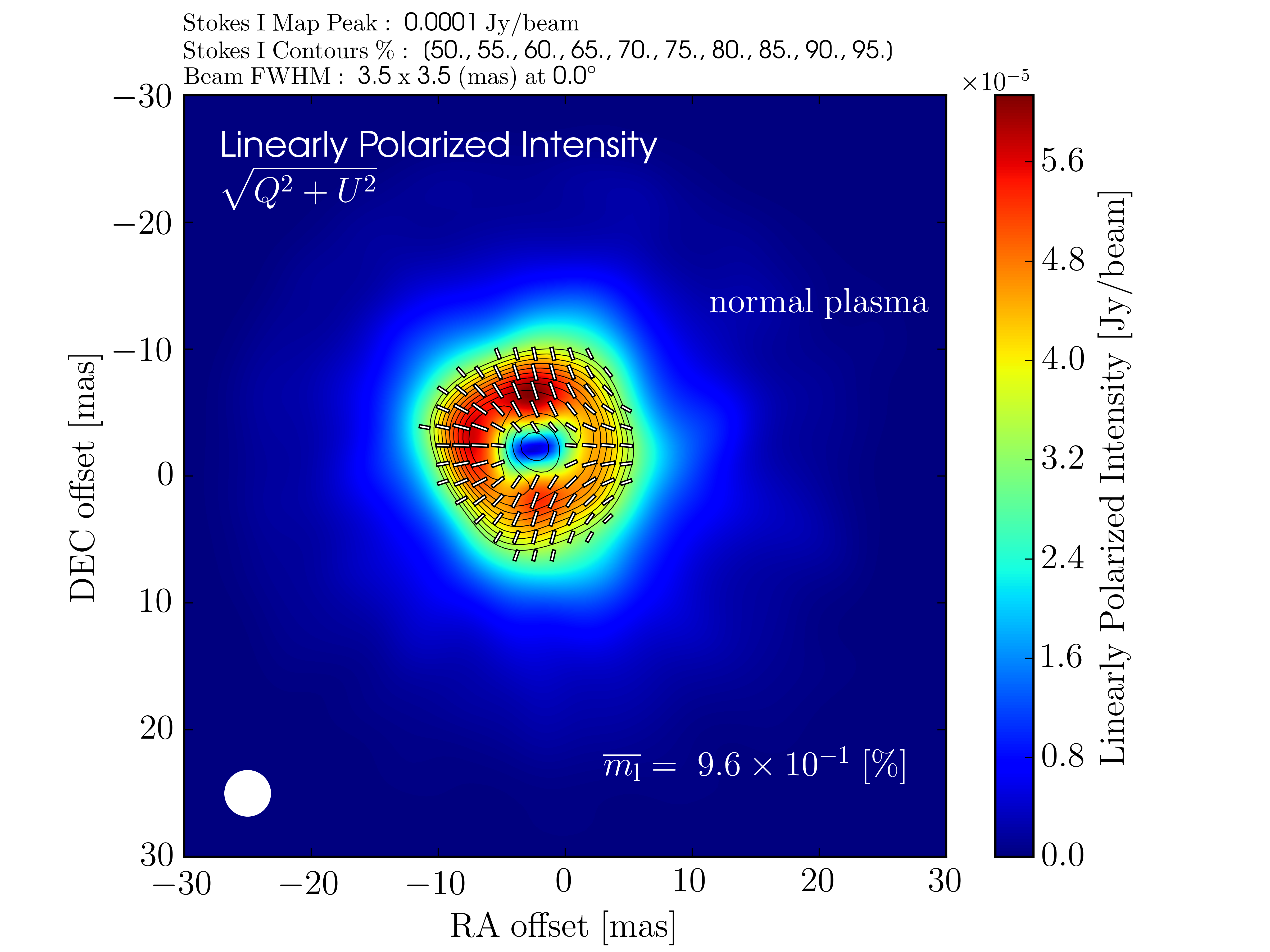}
\includegraphics[width=0.49\linewidth]{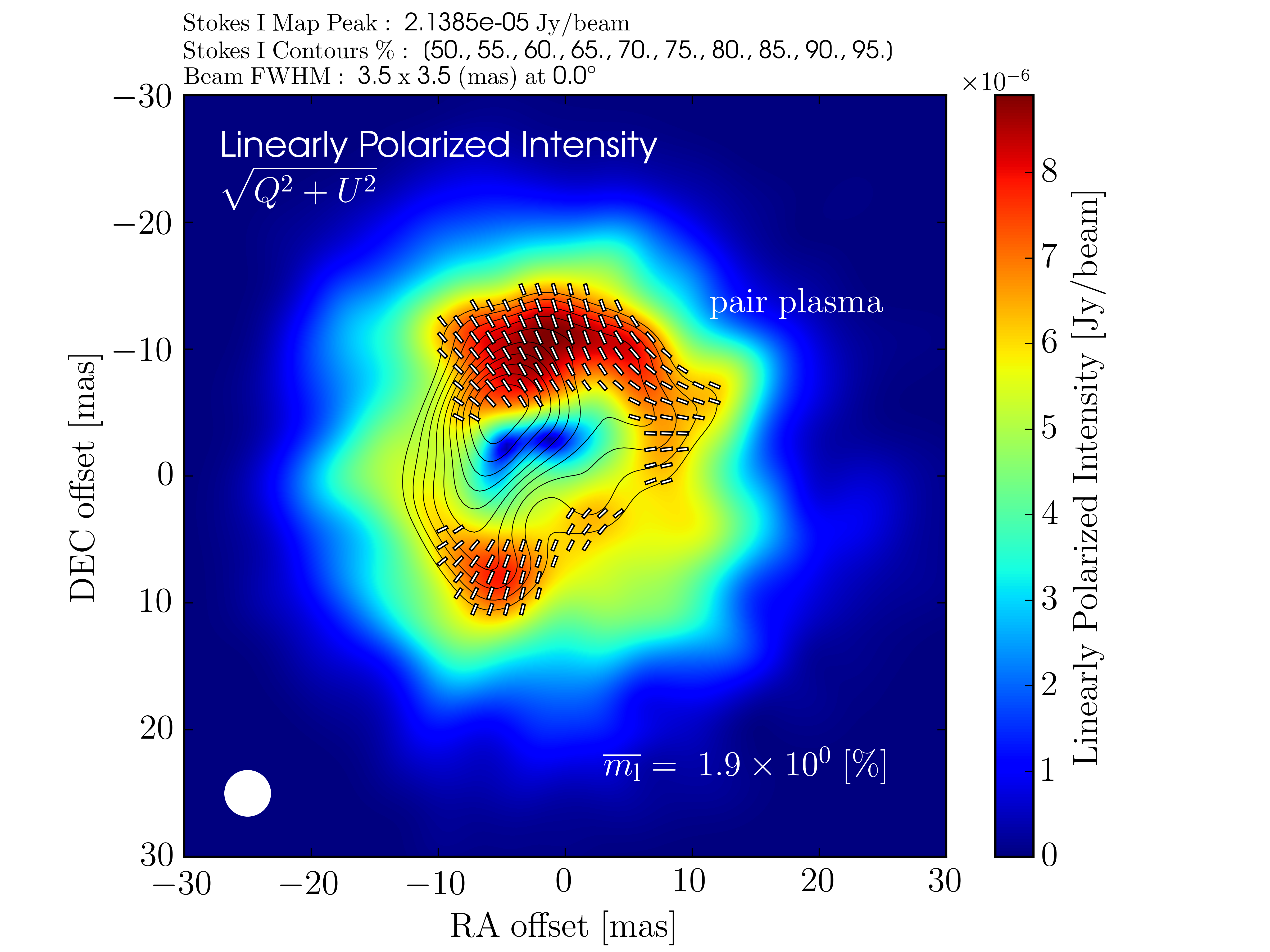}
\includegraphics[width=0.49\linewidth]{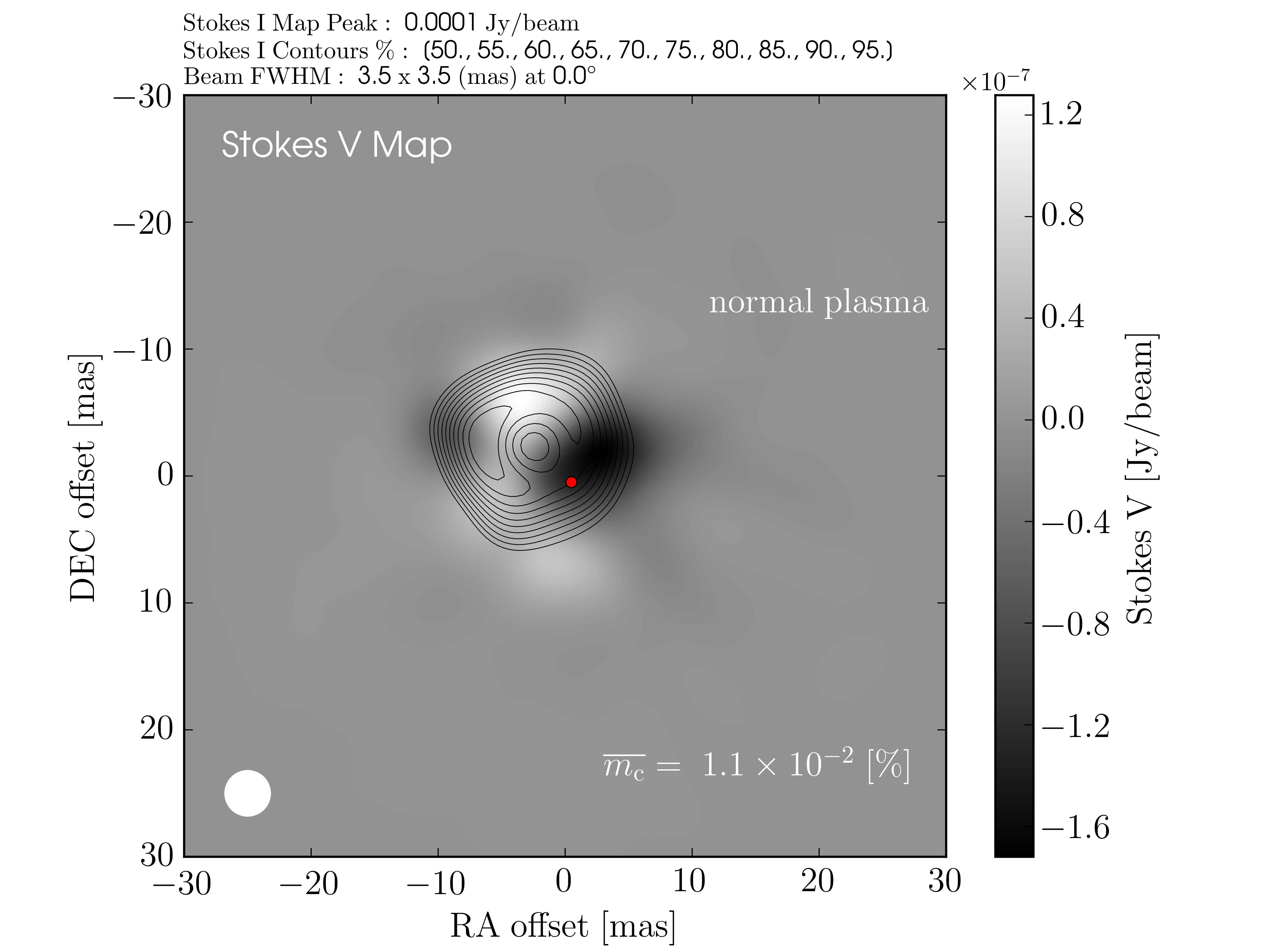}
\includegraphics[width=0.49\linewidth]{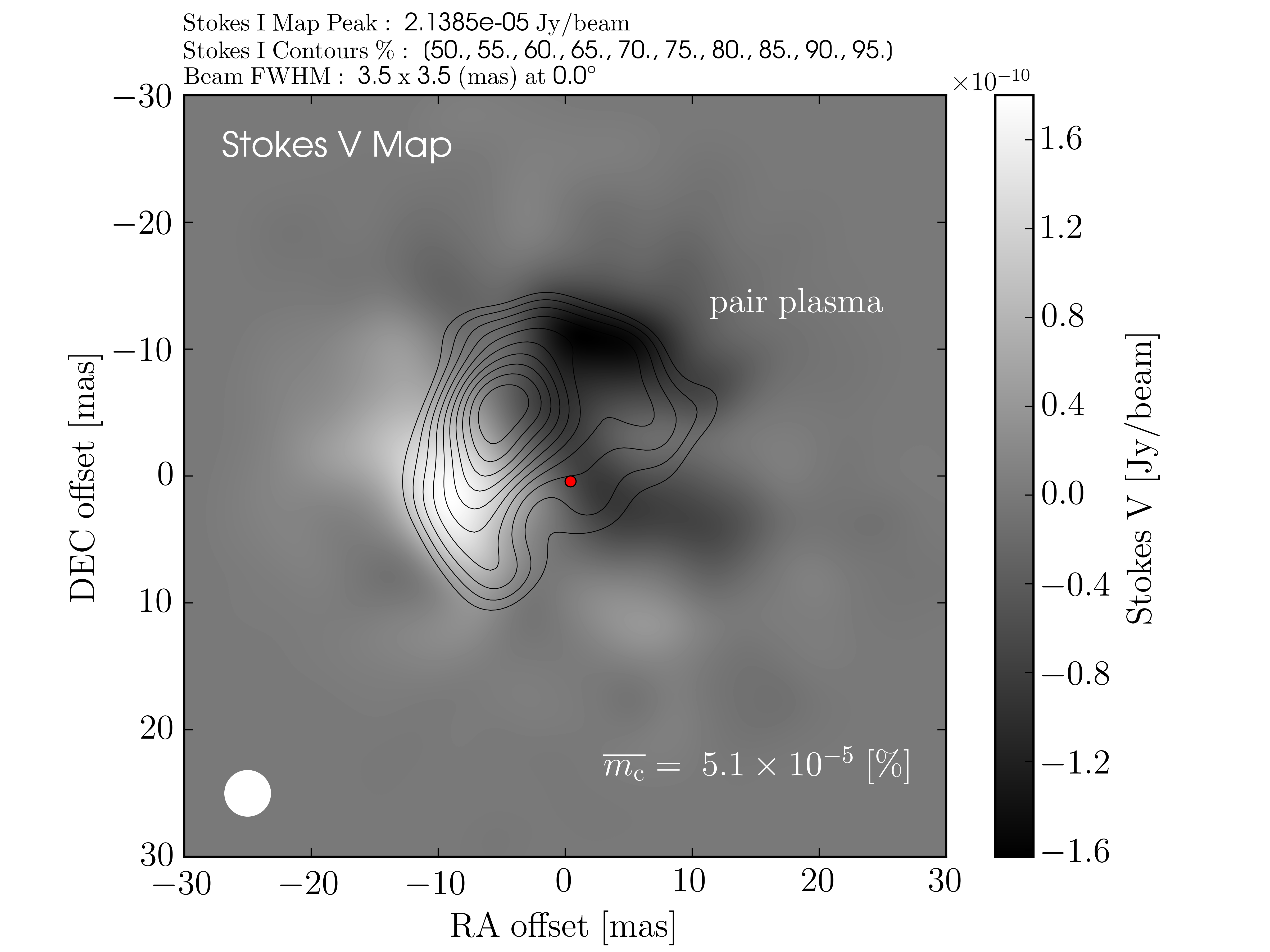}
\caption{\label{fig4}\textbf{Upper left panel:} Fast-light Stokes $I$ image (at an observing frequency of $\nu_{\rm{obs}} = 230 ~ \rm{GHz}$ and $\theta_{\rm{obs}} = 0^{\circ}$) of the normal plasma ($e^{-}$-$p^{+}$) jet. \textbf{Upper right panel:} Corresponding Stokes $I$ image of the pair plasma ($e^{-}$-$e^{+}$) jet. \textbf{Middle left panel:} Rendering of the LP intensity of the normal plasma jet. White line segments indicate the electric vector position angles (EVPAs) as projected onto the plane of the sky. \textbf{Middle right panel:} Corresponding LP intensity image of the pair plasma jet. Integrated levels of fractional linear polarization ($\overline{m_{l}} \equiv ( \overline{Q}^{2} + \overline{U}^{2} )^{1/2}/\overline{I}$) are listed to the lower right in each panel. \textbf{Lower left panel:} Stokes V image of the normal plasma jet. \textbf{Lower right panel:} Stokes V image of the pair plasma jet. The red dots in the lower panels highlight individual sightlines through each simulation which are illustrated in Figure \ref{fig5}. Integrated levels of fractional circular polarization ($\overline{m_{c}} \equiv -\overline{V}/\overline{I}$) are listed to the lower right in each panel. All images have been convolved with a lower resolution circular Gaussian beam of FWHM $3.5 ~\rm{mas}$ for comparison to Figure \ref{fig3}.}
\end{figure*}

The results of our fast-light ray-tracing calculations for both simulations are presented in Figures \ref{fig3} and \ref{fig4}. We explore the nature of the polarized synchrotron emission when viewing each jet at right angles to the jet axis (i.e., $\theta_{\rm{obs}} = 90^{\circ}$, shown in Figure \ref{fig3}) and when viewing each jet edge-on to the jet axis (i.e., $\theta_{\rm{obs}} = 0^{\circ}$, shown in Figure \ref{fig4}). For both sets of images we have performed our ray-tracing calculations through the last time step of each simulation (illustrated in Figure \ref{fig1}). 

We point out that in the case of a `pure' pair plasma, Stokes V would be exactly zero, since the equal numbers of electrons and positrons would cancel out their respective contributions to the circularly polarized synchrotron emission. Similar to the methods presented in \citealt{wardle03, homan09}, and more recently \cite{anantua20}, we parametrize within each plasma cell the proton-to-position ratio ($r_{+} \equiv n_{p^{+}}/n_{e^{+}}$). This ratio is incorporated into the radiative transfer coefficients, including the CP specific terms: $\eta_{\nu}^{~ V}$, $\kappa_{V}$, and $\kappa^{*}_{~ V}$ (see Appendix \ref{appB}; see also \citealt{macdonald18}). We have initially fixed this ratio to $r_{+}=100$ (i.e., 100 protons for every positron) in the normal plasma jet and $r_{+}=0.01$ (i.e., 100 positrons for every proton) in the pair plasma jet. We plan on running a larger set of PIC plasma simulations in the future in which we explore a wider range of jet plasma compositions than the two cases investigated here.

\begin{figure*}
\centering
\includegraphics[width=0.49\linewidth]{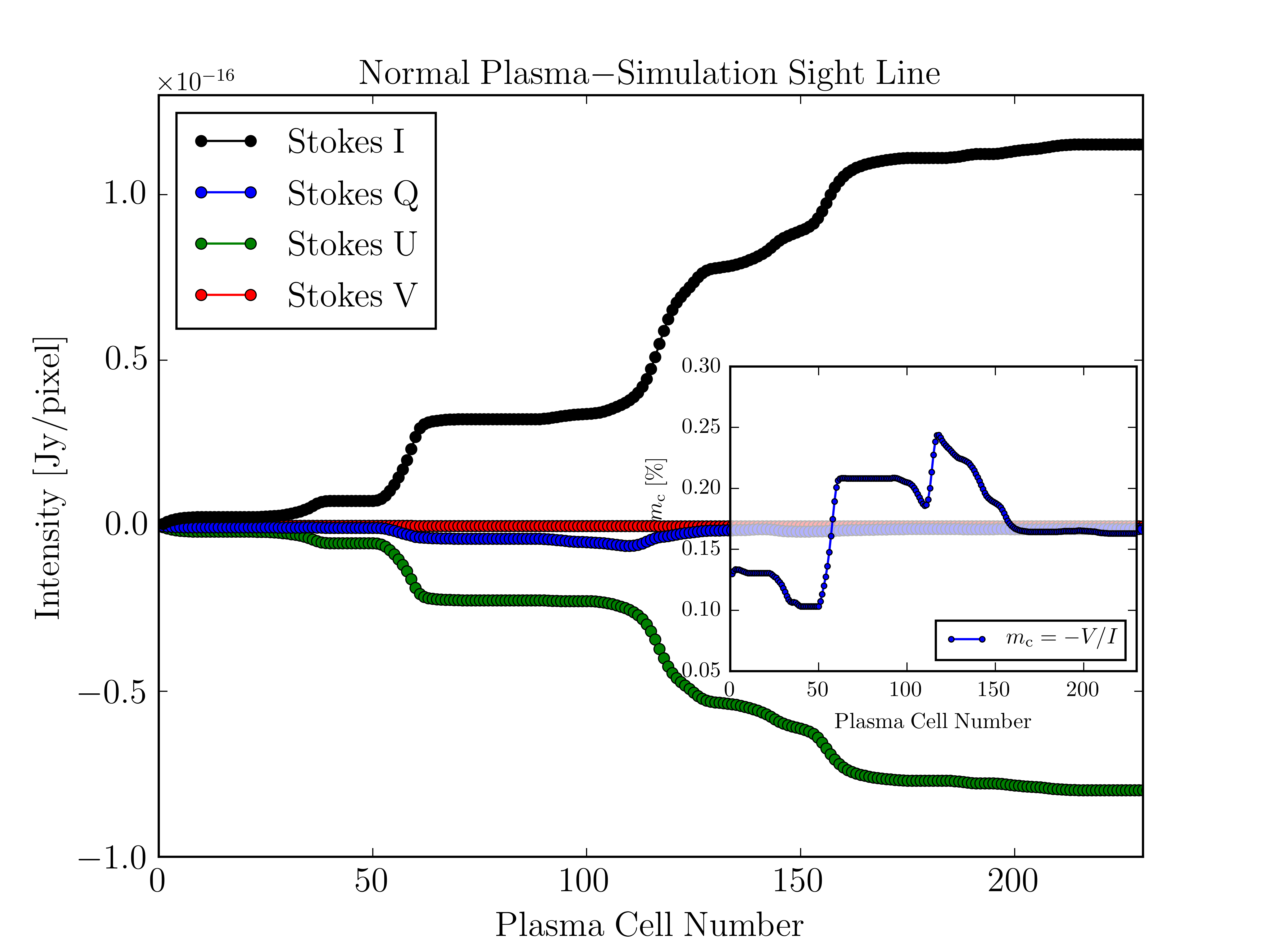}
\includegraphics[width=0.49\linewidth]{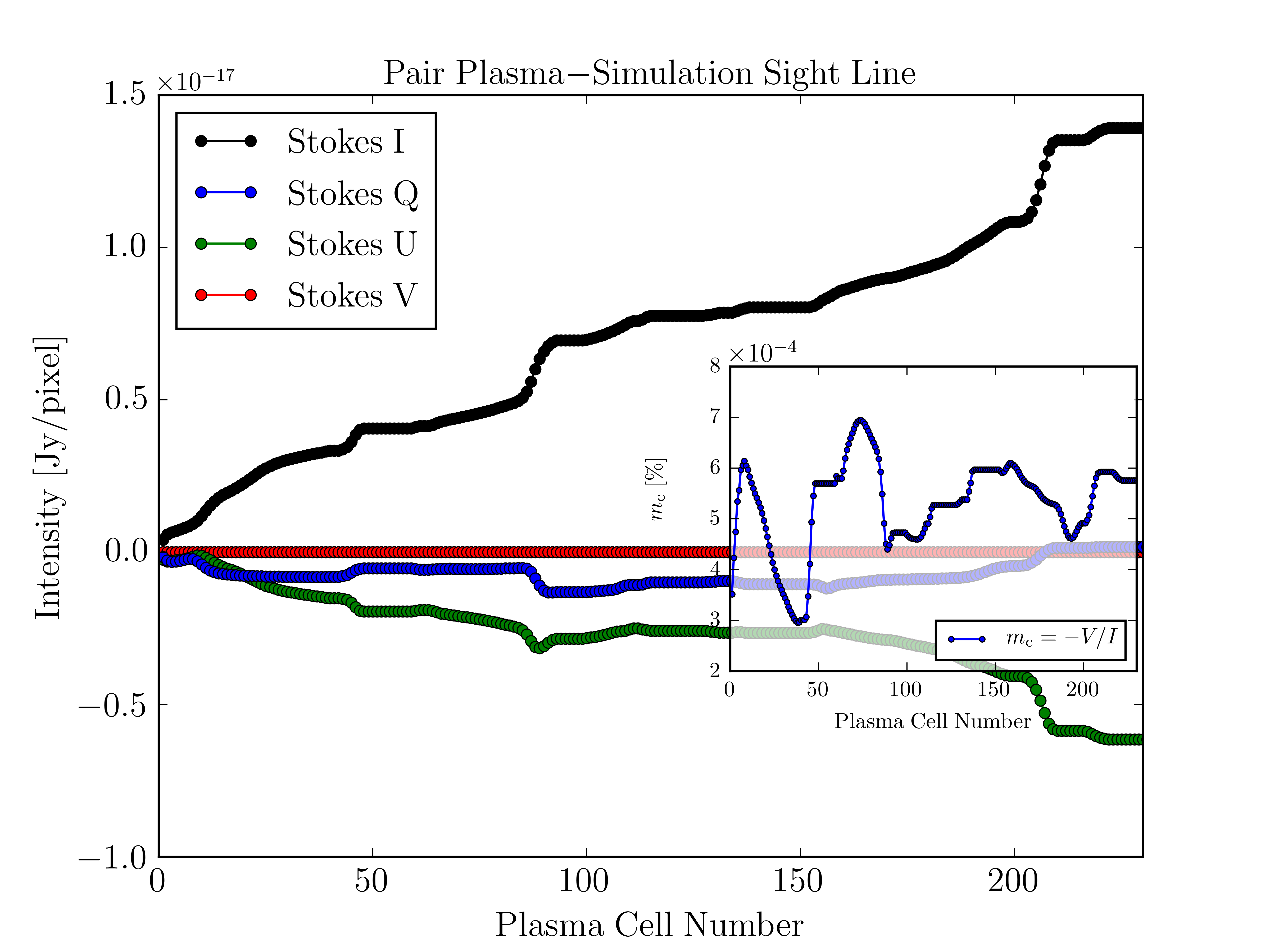}
\includegraphics[width=0.49\linewidth]{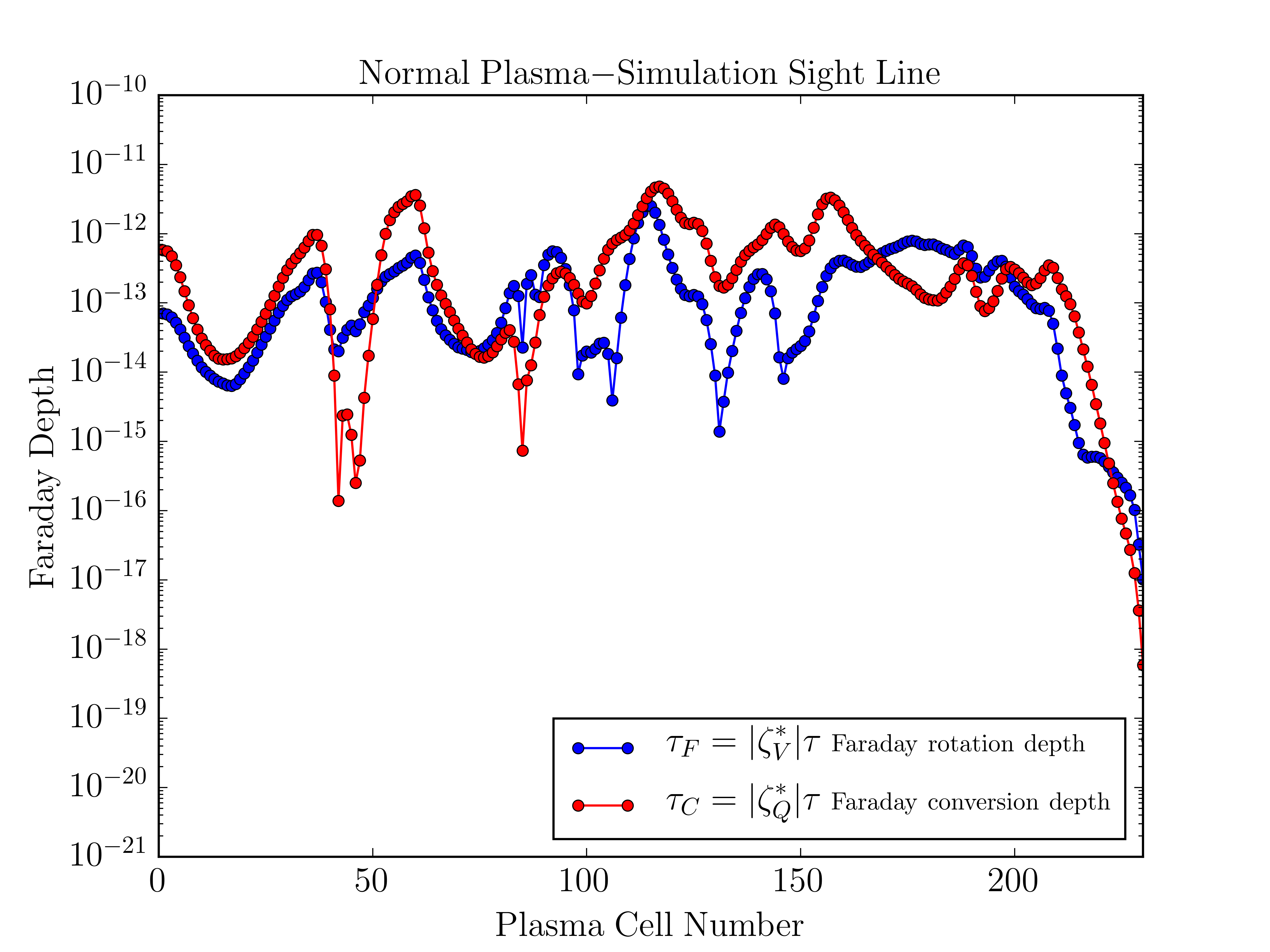}
\includegraphics[width=0.49\linewidth]{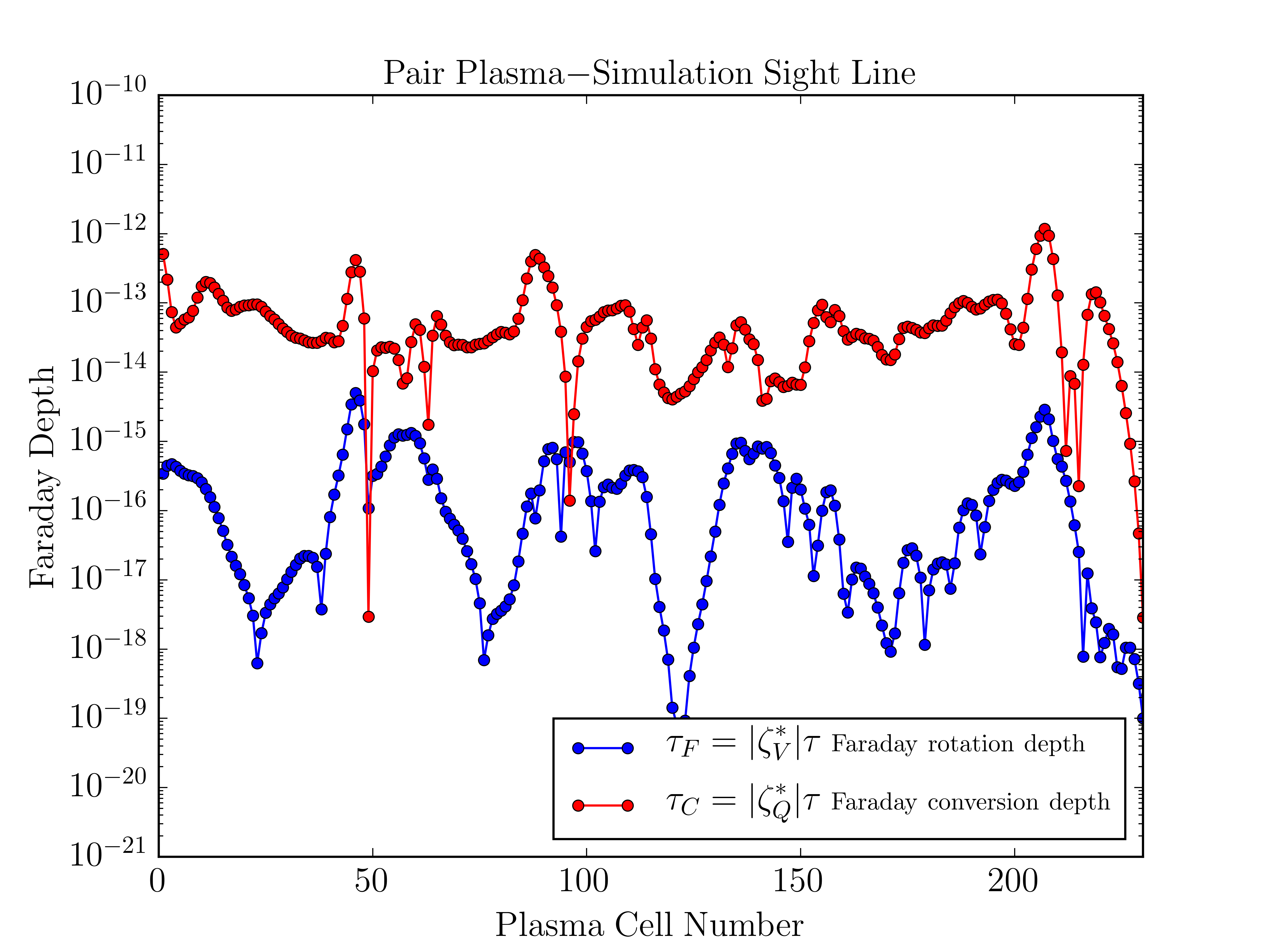}
\caption{\label{fig5}\textbf{Upper left panel:} Variations of the Stokes parameters for the normal plasma ($e^{-}$-$p^{+}$) jet along the sightline indicated by the red dot in the lower left panel of Figure \ref{fig4}. The radiative transfer progresses from left to right: starting on the far side of the jet relative to the observer (cell 0) and then advancing through the jet plasma toward the near side of the jet (cell 240). The inset shows the variation in the fractional circular polarization along the same sightline. \textbf{Upper right panel:} Corresponding ray properties along the pair plasma ($e^{-}$-$e^{+}$) jet sightline indicated by the red dot in the lower right panel of Figure \ref{fig4}. \textbf{Lower left panel:} Variations of the Faraday rotation ($\tau_{F}$) and conversion ($\tau_{C}$) depths along the normal plasma jet sightline. \textbf{Lower right panel:} Corresponding Faraday depths along the pair plasma jet sightline.}
\end{figure*}

\subsubsection{Radio Jet Orientation}
\label{Radio Jet Orientation}

In Figure \ref{fig3}, we present ray-traced images of each jet simulation at $\theta_{\rm{obs}} = 90^{\circ}$ for an observing frequency of $\nu_{\rm{obs}} = 1 ~ \rm{GHz}$ (see Appendix \ref{appC} for further discussion). In the upper panels we present Stokes I maps of both jets (normal plasma on the left and pair plasma on the right), in the middle panels we present the corresponding maps of linearly polarized intensity ($P \equiv \sqrt{ Q^{2} + U^{2} }$) for each jet with electric vector position angles ( EVPAs; $\chi \equiv \frac{1}{2} \rm{arctan} \left[ \frac{ U }{ Q } \right]$ ) overlaid in white. In the lower panels of Figure \ref{fig3} we present the corresponding Stokes V images for both simulations (again normal plasma on the left and pair plasma on the right). To mimic the effects of interferometric resolution, all images have been convolved with a circular Gaussian beam of FWHM $7.5 ~\rm{mas}$. Upon inspection of Figure \ref{fig3}, one can see that the pair plasma jet exhibits a slightly more \textit{filamentary} emission morphology in contrast to the normal plasma jet. This morphological difference in emission is a reflection of the distinct jet dynamics occurring within the two plasmas: the $e^{-}$-$e^{+}$ plasma jet is prone to larger plasma instabilities (such as the Kelvin-Helmholtz and Weibel instabilities) and is, as a result, less stable than an $e^{-}$-$p^{+}$ plasma jet of similar speed and magnetic field strength (see, e.g., \citealt{nishikawa16b, nishikawa17, nishikawa19}).

\subsubsection{Blazar Jet Orientation}
\label{Blazar Jet Orientation}

In Figure \ref{fig4}, we present ray-traced images of each jet simulation at $\theta_{\rm{obs}} = 0^{\circ}$ for an observing frequency of $\nu_{\rm{obs}} = 230 ~ \rm{GHz}$ (see Appendix \ref{appC} for further discussion). In the upper panels we present Stokes I images of both jets (normal plasma on the left and pair plasma on the right), in the middle panels we present LP images, and in the lower panels we present the corresponding CP images for both simulations. We have convolved these higher frequency images with a beam size similar to the lower frequency 1 GHz images for ease of comparison with Figure \ref{fig3}. The effect of Doppler beaming is apparent upon comparison of the Stokes I flux levels in Figures \ref{fig3} and \ref{fig4}. The fractional circular polarization ($m_{c} \equiv -V/I$) is minimal in each simulation ($\ll 1 \%$) but is many orders of magnitude larger in the $e^{-}$-$p^{+}$ plasma jet in comparison to the $e^{-}$-$e^{+}$ plasma jet (integrated values are listed to the lower right in each Stokes V image). We present spectropolarimetry of the integrated levels of fractional polarization for each jet in Appendix \ref{appD}. 

In addition to mimicking the resolution of an interferometric array (via beam convolution), our ray-tracing algorithm can also mimic the sensitivity of an interferometric array by including a synthetic Gaussian noise floor within our ray-traced images. This is discussed further in relation to the \textit{detectability} of our PIC jets in Appendix \ref{appE}.

\subsubsection{Individual Ray-Properties}
\label{Individual Ray-Properties}

In Figure \ref{fig5}, we illustrate the variations of the Stokes parameters (upper panels) and the Faraday rotation ($\tau_{F}$) and conversion ($\tau_{C}$) depths (lower panels) of both jet plasmas along the sightlines indicated by the red dots in the lower panels of Figure \ref{fig4}. The Faraday depths (which are written out explicitly in Appendix \ref{appB} and are themselves functions of $B$, $n_{e}$, and $\gamma_{\rm{min}}$) parameterize the ability of the plasma cells along each sightline to attenuate both the linearly and circularly polarized synchrotron emission within the jet. It is evident upon comparison of the lower panels of Figure \ref{fig5} that, for these particular sightlines, the Faraday rotation depth is larger in the $e^{-}$-$p^{+}$ plasma jet. Both jets, however, are optically (and Faraday) thin and the Stokes V images presented in Figure \ref{fig4} are intrinsic in origin (see the lower panels of Figure \ref{figD1} in Appendix \ref{appD}). 

\subsection{Slow-light Images}
\label{Slow-light Images}

\begin{figure*}
\centering
\includegraphics[width=0.49\linewidth]{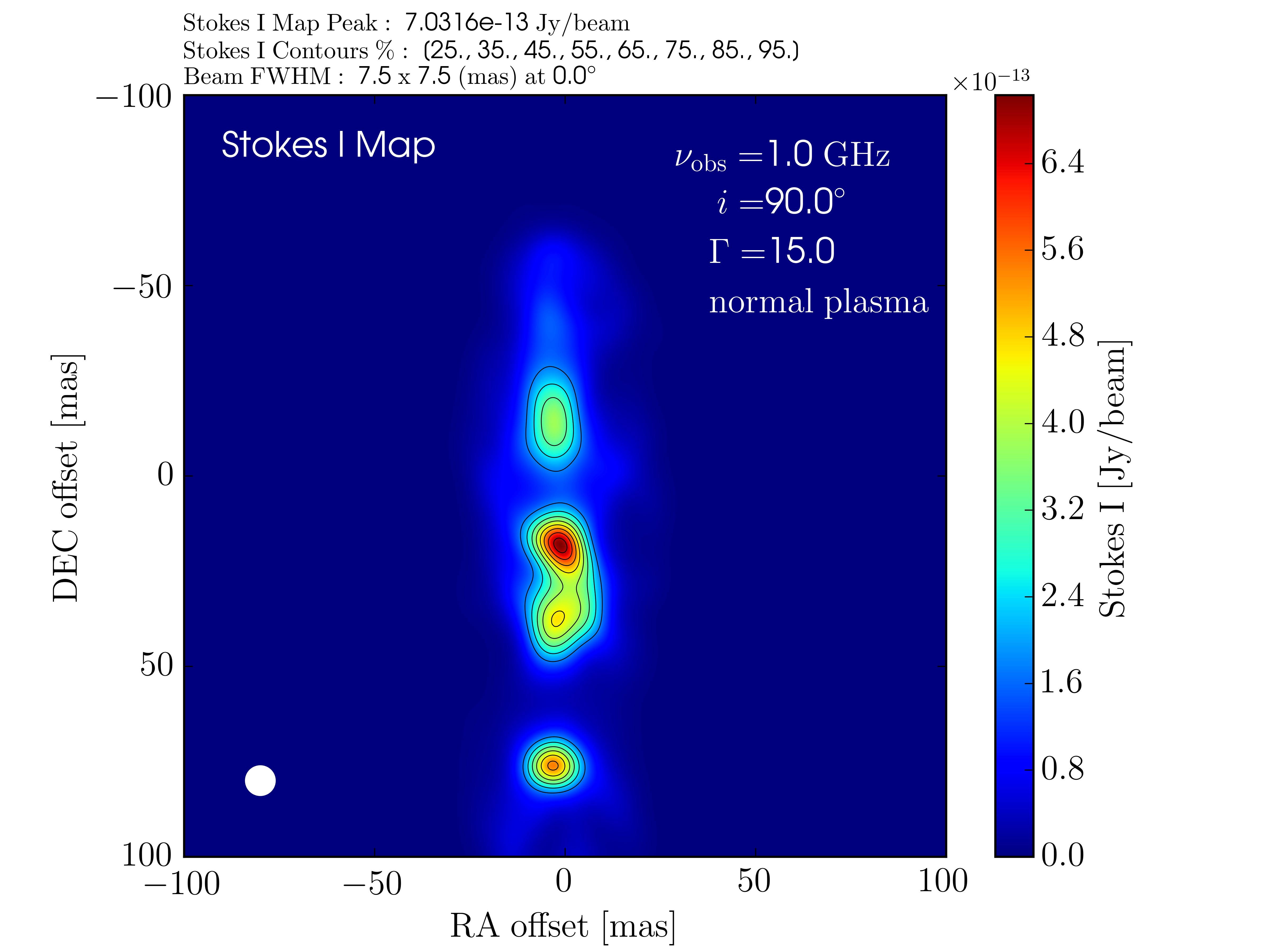}
\includegraphics[width=0.49\linewidth]{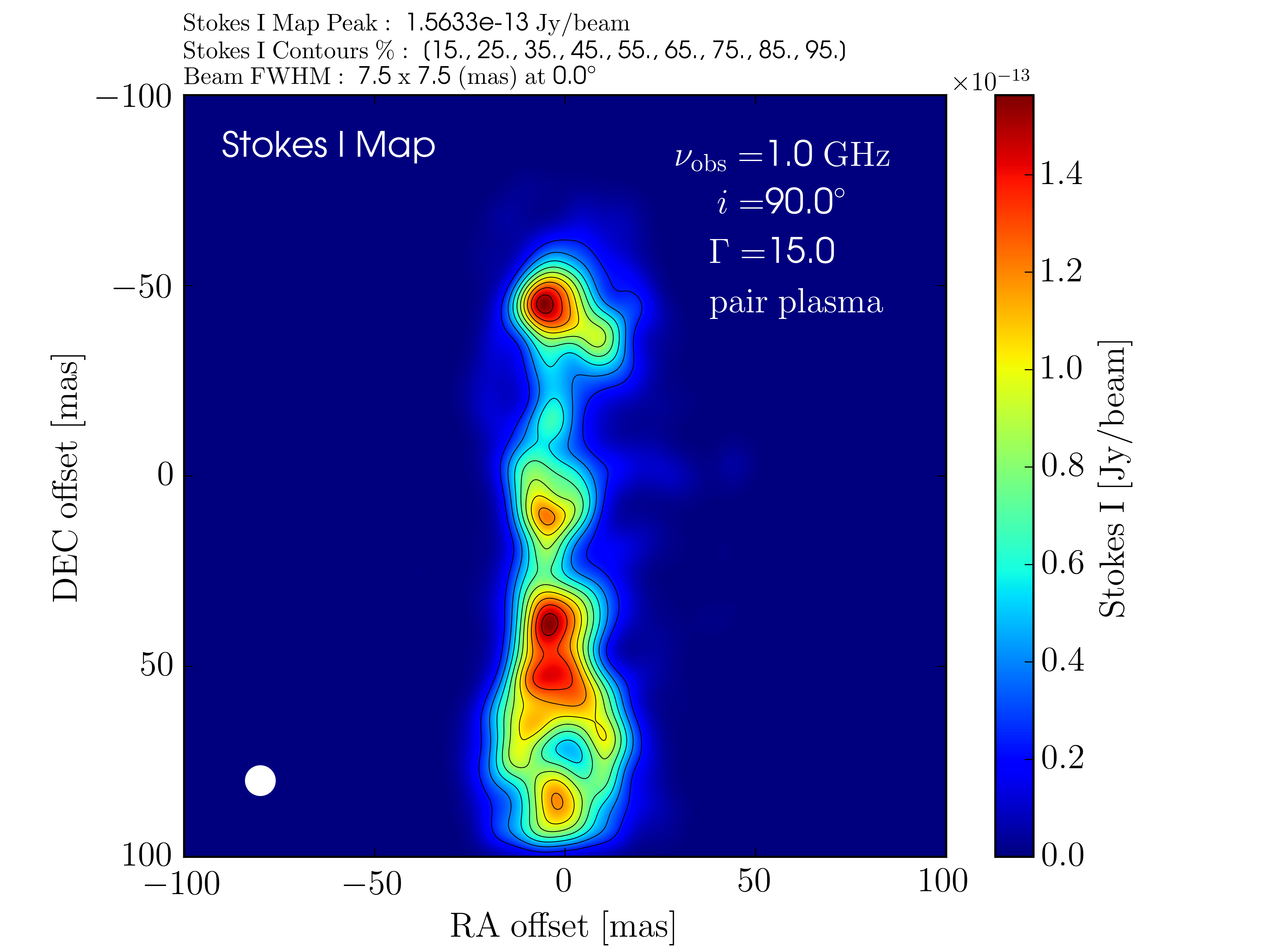}
\includegraphics[width=0.49\linewidth]{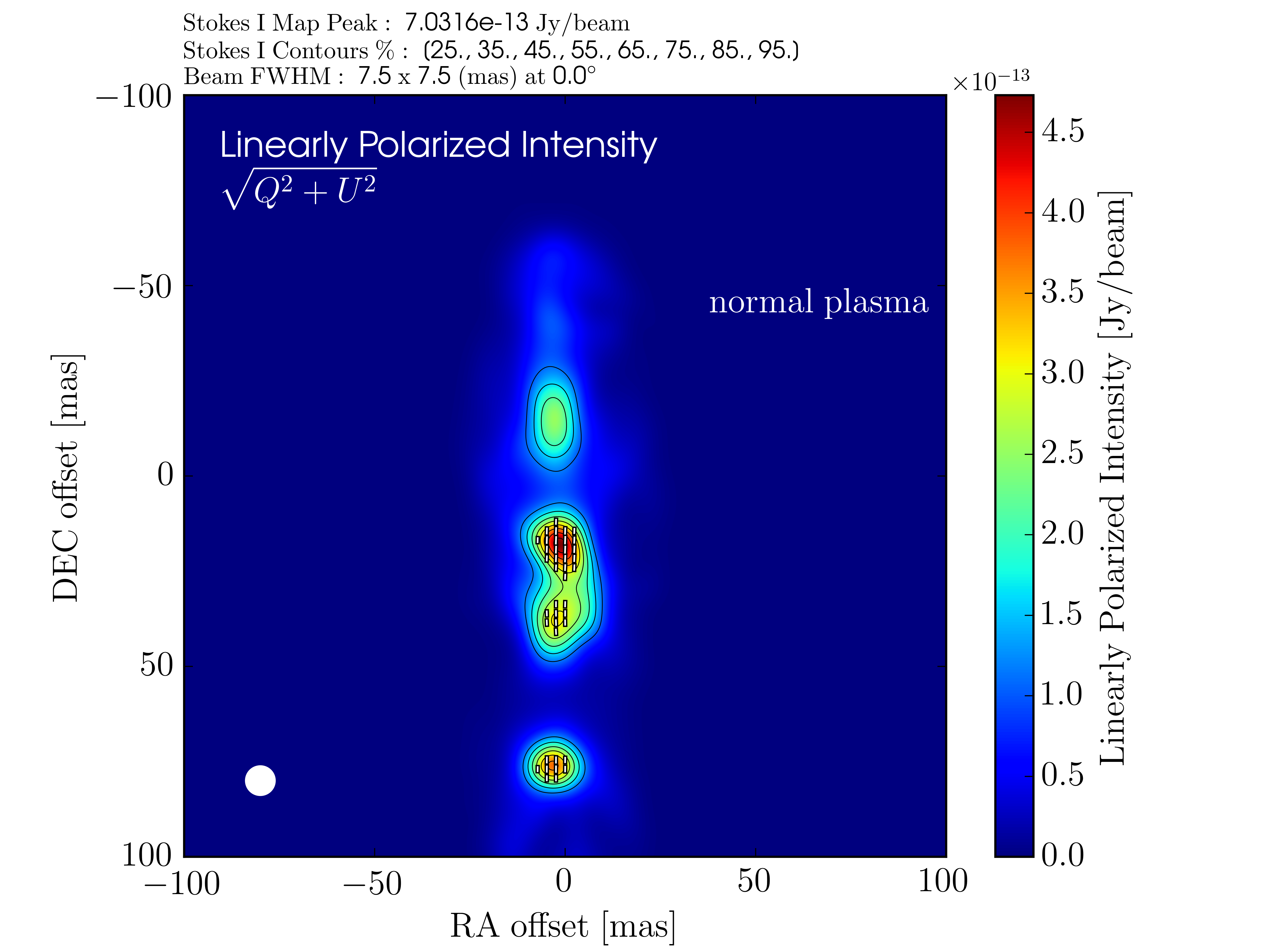}
\includegraphics[width=0.49\linewidth]{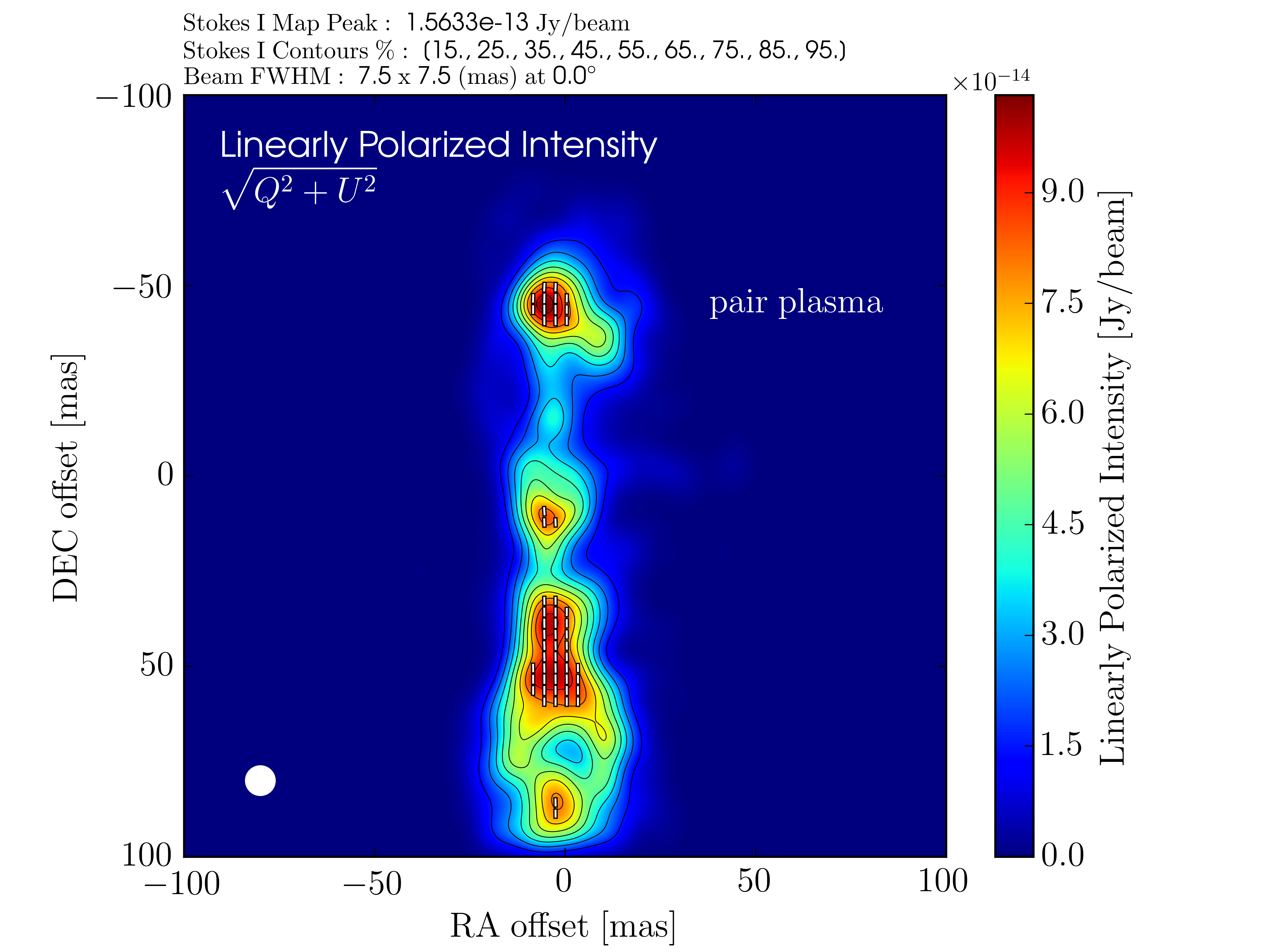}
\includegraphics[width=0.49\linewidth]{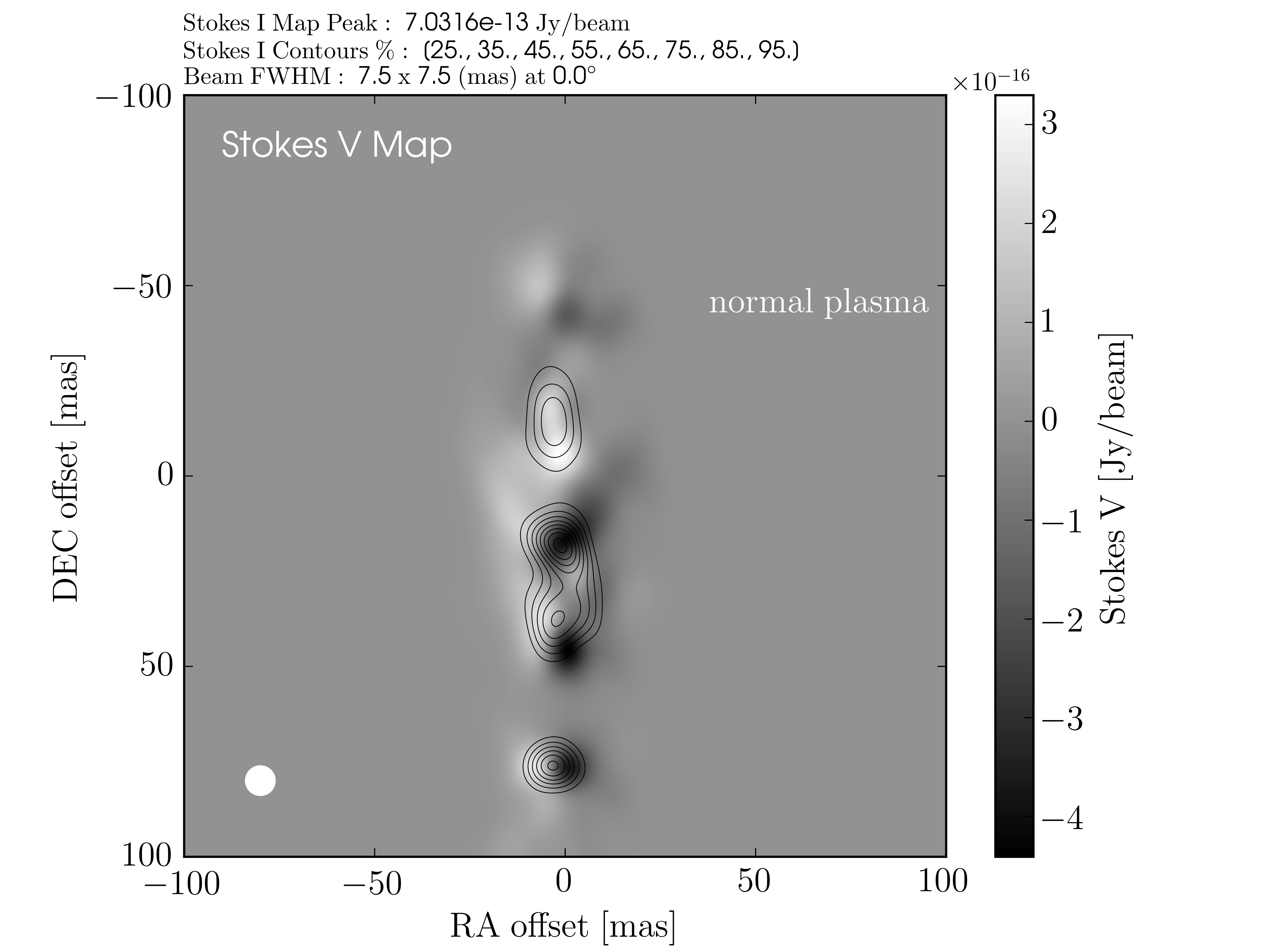}
\includegraphics[width=0.49\linewidth]{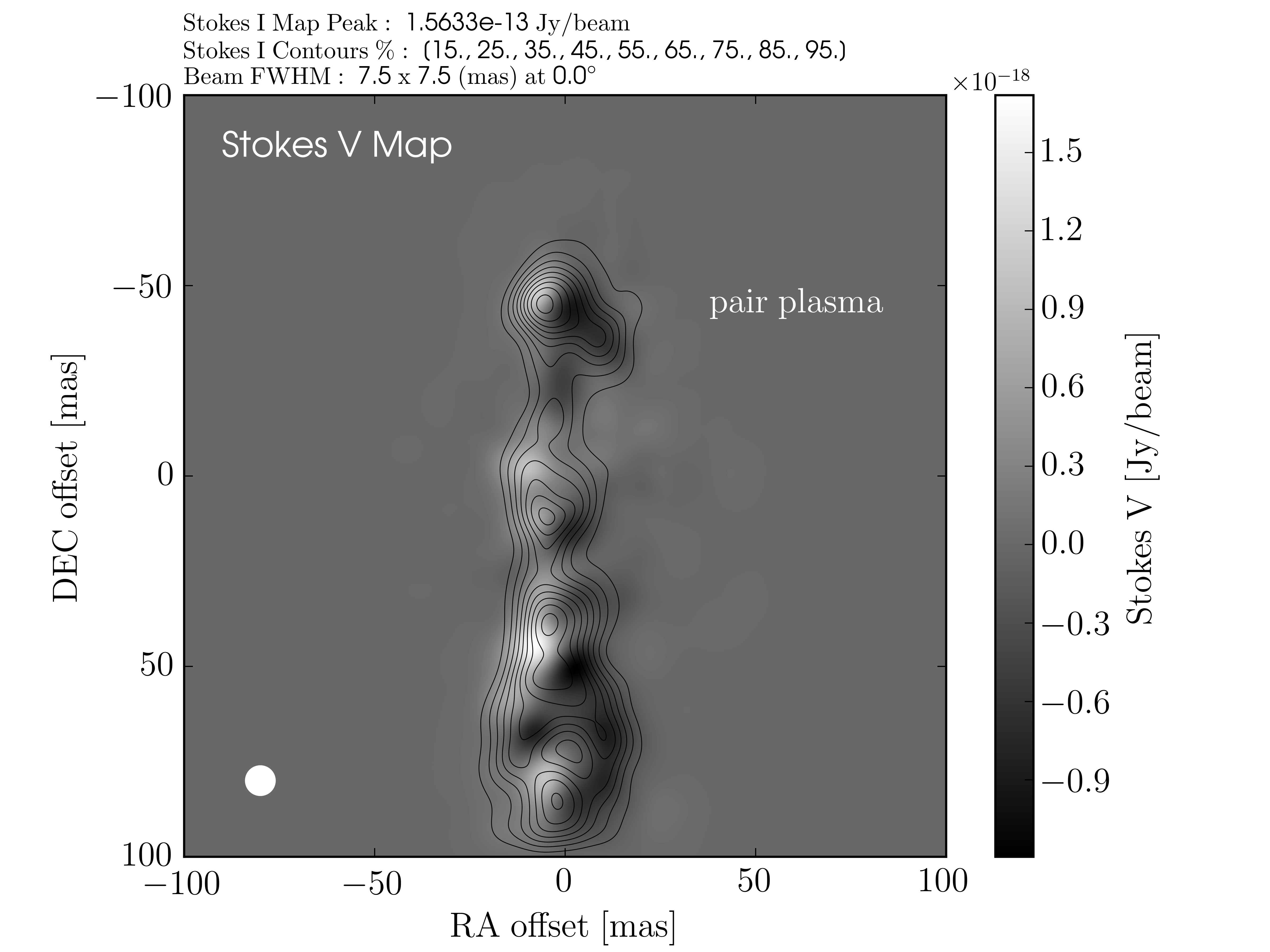}
\caption{\label{fig6}\textbf{Upper left panel:} Slow-light interpolated Stokes $I$ image (at an observing frequency of $\nu_{\rm{obs}} = 1 ~ \rm{GHz}$) of the normal plasma ($e^{-}$-$p^{+}$) jet. \textbf{Upper right panel:} Corresponding slow-light interpolated Stokes $I$ image of the pair plasma ($e^{-}$-$e^{+}$) jet. Hybrid computational grids (constructed using the scheme illustrated in Figure \ref{fig2}) were used in each ray-tracing calculation. \textbf{Middle left panel:} Rendering of the LP intensity of the normal plasma jet. White line segments indicate the electric vector position angles (EVPAs) as projected onto the plane of the sky. The effects of relativistic aberration (see \citealt{lyutikov05}) on the orientation of these EVPAs have been included in these calculations. \textbf{Middle right panel:} Corresponding LP intensity image of the pair plasma jet. \textbf{Lower left panel:} A Stokes V image of the normal plasma jet highlighting the different regions within the jet that produce positive and negative CP. \textbf{Lower right panel:} Corresponding Stokes V image of the pair plasma jet. All six images have been convolved with a circular Gaussian beam of FWHM $7.5 ~\rm{mas}$ (shown in the lower left of each panel).}
\end{figure*}

The results of our slow-light interpolated ray-tracing calculations are presented in Figure \ref{fig6}. The slow-light interpolation results in an \textit{averaging} of the emission along each sightline. In contrast to the fast-light images, the morphology of the resultant emission becomes \textit{blurred} and the distinct morphologies present between the normal plasma and pair plasma jet compositions (evident in the fast-light images shown in Figure \ref{fig3}) are less pronounced in the corresponding slow-light images illustrated in Figure \ref{fig6}.

In particular, the \textit{hybrid} computational grids (constructed via the algorithm illustrated in Figure \ref{fig2}), through which we ray-trace, are composed of $\sim50$ distinct/stratified jet epochs (each separated in time by 10 code time steps). This number of epochs corresponds to the number of distinct plasma cells that a plane-parallel ray encounters when each jet is imaged at right angles to the jet axis. As discussed in \S\ref{Slow-Light Interpolation}, as each ray propagates across the jet, it encounters newer upstream values of jet plasma due to the finite light crossing time across each PIC plasma cell (i.e., ten time steps - see Equation \ref{eqn15}). Further computational time (i.e., more epochs) is required in order to produce slow-light images of these simulations when each jet is viewed edge-on (i.e., with the orientation of a blazar). 

We finally point out that \textit{features/blobs} in our slow-light images, in contrast to the fast-light images, do not necessarily map to individual plasma structures within the jet flow and are instead a mixture of emission from multiple plasma components along various sightlines through the jet.

\section{Summary \& Conclusions}
\label{Summary and Conclusions}

We have carried out full Stokes polarized radiative transfer calculations (via ray-tracing) through 3D relativistic PIC simulations of a normal plasma ($e^{-}$-$p^{+}$) and of a pair plasma ($e^{-}$-$e^{+}$) jet. We generate two sets of images of each jet simulation, one in which we invoke the fast-light approximation (see Figure \ref{fig3}), and the other in which we implement slow-light interpolation (see Figure \ref{fig6}). It is clear, upon comparison of Figures \ref{fig3} and \ref{fig6}, that slow-light interpolation has a discernible effect on the emission emanating from within each jet. In particular:
\begin{itemize}
      \item[$\bullet$] The finite light-crossing times through our relativistic jet simulations results in `blending' of various plasma emission features along each sightline. 
\end{itemize} 
It is also clear that there are differences both in the morphology and in the fractional level of polarization emanating from the two jet plasma compositions. Specifically: 
\begin{itemize}
      \item[$\bullet$] The pair plasma jet exhibits a more \textit{filamentary} synchrotron emission morphology in comparison with the normal plasma jet.
      \item[$\bullet$] The normal plasma jet has a much larger value of integrated fractional circular polarization in comparison with the pair plasma jet.
\end{itemize}
These differences are a reflection of the distinct plasma dynamics occurring within each jet: the pair plasma jet is prone to larger kinetic instabilities (e.g., Kelvin Helmholtz and Weibel instabilities) in comparison with the normal plasma jet and is, as a result, less stable, resulting in a more filamentary emission structure. The lower levels of circular polarization in the pair plasma jet are in keeping with the synchrotron theory we have incorporated into our ray-tracing calculations. 

We emphasize that the applicability of our calculations to jets on parsec scales is quite limited. In particular: 
\begin{itemize}
      \item[$\bullet$] Both jet simulations, when scaled into physical units, are only hundreds of kilometers in extent and are extremely tenuous in nature.
\end{itemize} 
We present these calculations, despite this limitation in scale, for two main purposes: (i) to provide a point of comparison for future polarimetric imaging of larger-scale PIC simulations, and (ii) to emphasize that kinetic scale jet dynamics can produce distinct morphologies in the jet's synchrotron radiation. These calculations reveal how the micro physics of the jet affect the macro emission we detect in the radio. This relationship between the micro and the macro is not commonly addressed in most relativistic jet simulations. Refinement of: (i) the jet injection scheme, (ii) the grid size, and (iii) the jet particle content are planned for future analysis.    

This work clearly emphasizes the \textit{dire} need for \textit{vastly} larger computational grid sizes and particle populations when attempting to model kinetic scale effects in relativistic plasmas on astrophysical length scales. Hybrid techniques (e.g., \citealt{mignone18, vaidya18, davelaar19, parfrey19, bacchini20}) will be crucial for future jet simulations. We also highlight the care that must be taken when interpreting features in interferometric radio maps of relativistic jets, especially if the light crossing time of the jet exceeds the relevant dynamical timescales of the emitting plasma. 

\begin{acknowledgements}

We are grateful to I.~Myserlis for fruitful discussions regarding the intricacies of radiative transfer, E.~Ros for invaluable feedback regarding VLBI, and to the anonymous referee for a thorough review of this manuscript. We also acknowledge J.~L.~G\'{o}mez for the initial idea to pursue this line of research. K.-I. Nishikawa has been supported by NASA research grants NNG05GK73G and NNX13AP14G. The PIC simulations were performed on Pleiades at the NASA Advanced Supercomputing Division (NAS) as well as on Gordon and Comet at The San Diego Supercomputer Center (SDSC), and on Bridges at the Pittsburgh Supercomputing Center (PSC), which are supported by the NSF. We have made use of elements of radmc3dPy: a python package/user-interface to the RADMC-3D code developed by A. Juhasz.

\end{acknowledgements}

\appendix
\newpage
\onecolumn
\section{Comparison of Electron Larmor Radii to the PIC Grid Cell Sizes}
\label{appA}

\begin{figure*}
\centering
\includegraphics[width=0.49\linewidth]{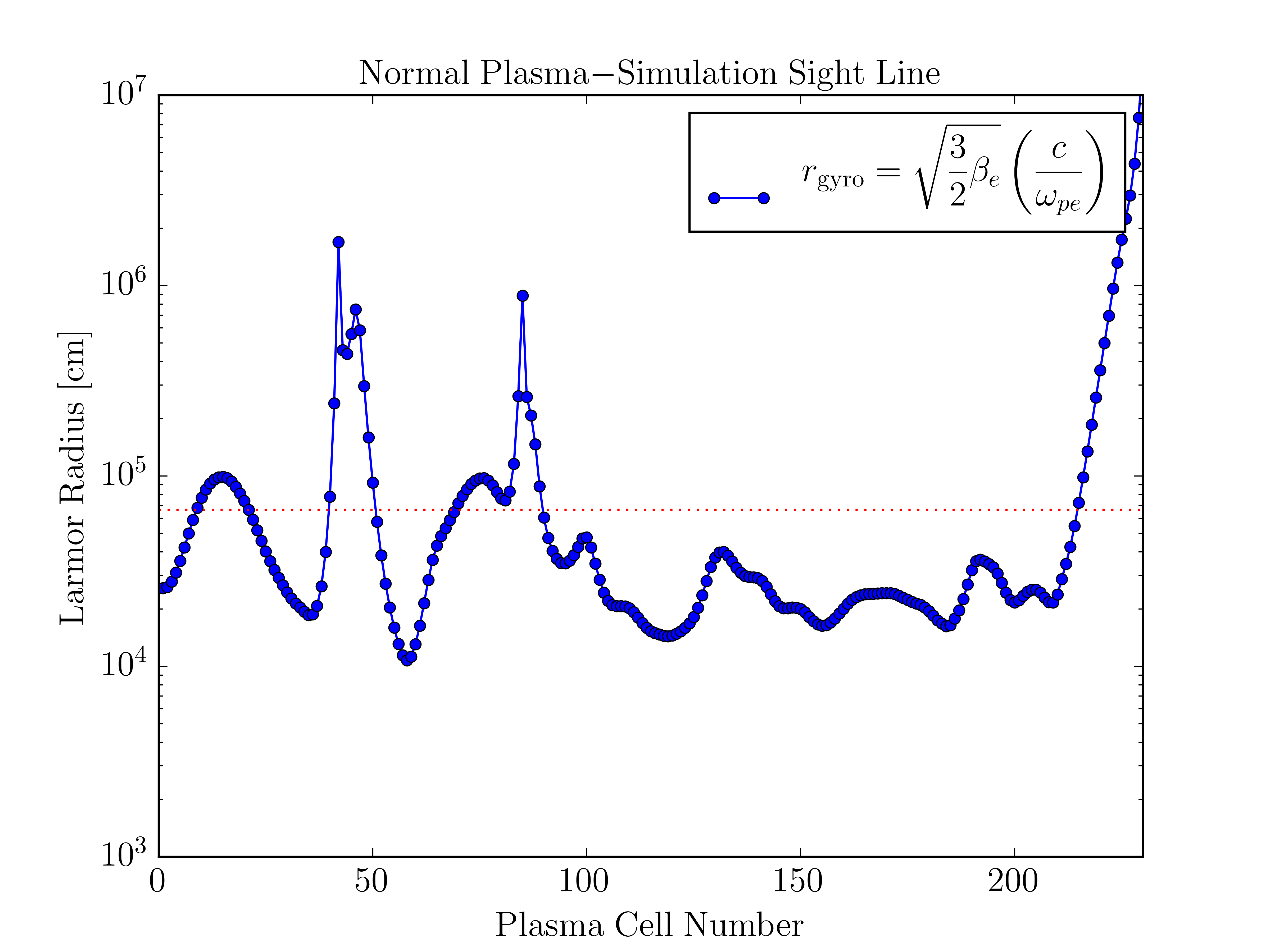}
\includegraphics[width=0.49\linewidth]{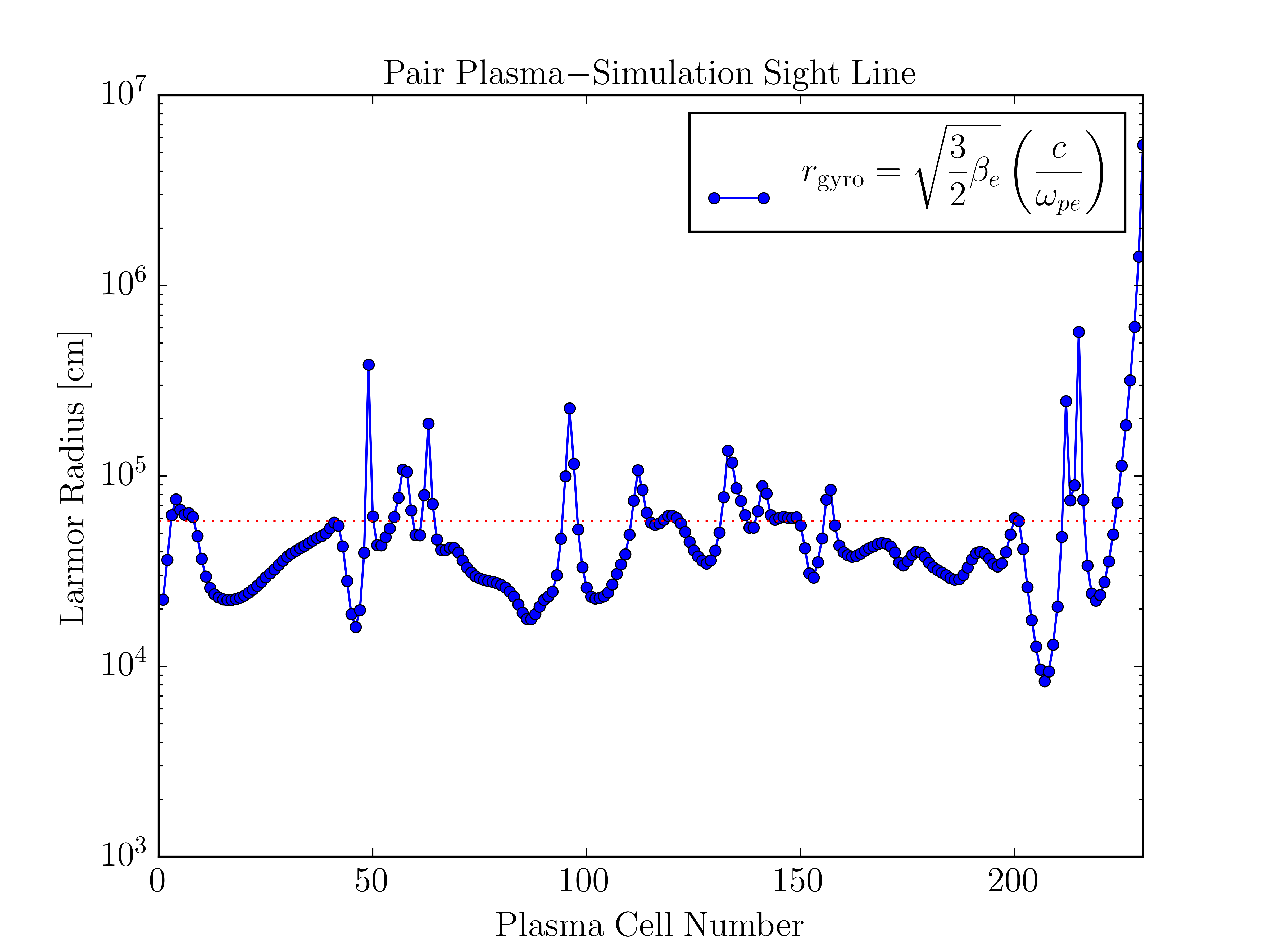}
\caption{\label{figA1}\textbf{Left panel:} Variation of the electron Larmor radii ($r_{\rm{gyro}}$) within the normal plasma ($e^{-}$-$p^{+}$) jet along the sightline indicated by the red dot in the lower left panel of Figure \ref{fig4}. The radiative transfer progresses from left to right: starting on the far side of the jet relative to the observer (cell 0), and then advancing through the jet plasma toward the near side of the jet (cell 240). The red dashed line in each panel demarcates the scaled PIC grid cell size ($\Delta l_{\rm{scale}}$). \textbf{Right panel:} Corresponding variation of the pair plasma ($e^{-}$-$e^{+}$) jet Larmor radii along the sightline indicated by the red dot in the lower right panel of Figure \ref{fig4}.}
\end{figure*}

\noindent Due to the very small length scales within our PIC simulations, care must be taken in the implementation of the physical scaling relations (\S\ref{Scaling}) to ensure that the individual electron Larmor radii do not vastly exceed the plasma grid cell sizes. This is a necessary criterion for modeling isotropic synchrotron emission in the radiative transfer (equation \ref{eqn14}) that we apply along each ray in our transfer calculations. The radiative transfer is \textit{extremely} sensitive to the lower cutoff in the electron power-law distribution ($\gamma_{\rm{min}}$). For a power-law distribution of synchrotron emitting electrons: $n(\gamma) \propto \gamma^{-s}$, the synchrotron emissivity is a function of the integral $\int_{\gamma_{\rm{min}}}^{\gamma_{\rm{max}}} n(\gamma) ~ d\gamma \propto \gamma_{\rm{max}}^{~ -(s-1)} - \gamma_{\rm{min}}^{~ -(s-1)}$. Provided that $s > 1$ ($s \simeq 2.3$ in our calculations) and $\gamma_{\rm{max}} \gg \gamma_{\rm{min}}$, the contribution from the high-energy end of the power-law is minimal. Values of $\gamma_{\rm{min}}$ along sightlines through the $e^{-}$-$p^{+}$ and $e^{-}$-$e^{+}$ jet plasmas are illustrated in the lower panels of Figure \ref{figB1}. In particular, at the base of the normal plasma jet: $\gamma_{\rm{min}} \simeq 15$ and $B \simeq 10^{0} ~ \rm{Gauss}$. This implies that the low-energy electrons in this region will emit synchrotron radiation at a critical frequency ($\nu_{c}$) of:
\begin{equation}\label{eqnA1}
\nu_{\rm{c}} \simeq \gamma_{e}^{3} ~ \nu_{\rm{gyro}} = \gamma_{e}^{3} ~ \left( \frac{ e B }{ 2 \pi ~ \gamma_{e} m_{e} c } \right) \simeq 6.3 \times 10^{8} ~ \rm{Hz} ~,	 
\end{equation}
which corresponds to Doppler beamed GHz emission in the observer's frame when a bulk Lorentz factor $\Gamma = 15$ jet is oriented edge-on to our line of sight (i.e., with the orientation of a blazar). The corresponding Larmor radii of these low-energy electrons is:
\begin{equation}\label{eqnA2}
r_{\rm{gyro}} = \frac{ \gamma_{e} m_{e} c^{2} }{ e B } \simeq 2.6 \times 10^{4} ~ \rm{cm} ~.	 
\end{equation}
The above Larmor radius is (\textit{by design}) smaller than the scaled PIC plasma cell sizes ($\Delta l_{\rm{scale}} \simeq 6.5 \times 10^{4} ~ \rm{cm}$, Equation \ref{eqn7}). Our choice of physical scaling (i.e., the fiducial values of $n_{e} \simeq 10^{1} ~ \rm{cm}^{-3}$; Equation \ref{eqn6} and $\beta_{e} \simeq 10^{-3}$; Equation \ref{eqn9}) is motivated by: (i) the ability of our PIC simulations to produce numerically tractable levels of synchrotron radiation, and (ii) to ensure that $r_{\rm{gyro}}(\gamma_{\rm{min}}) \leq \Delta l_{\rm{scale}}$. Equation \ref{eqnA2}, however, parameterizes the gyration of electrons about a single ordered magnetic field line. The magnetic fields within our PIC simulations, in contrast, are highly turbulent in nature. A more applicable expression for the electron Larmor radii within our PIC simulations can be recast in terms of the plasma beta ($\beta_{e}$) and the electron plasma frequency ($\omega_{pe}$): 
\begin{equation}\label{eqnA3}
r_{\rm{gyro}} = \sqrt{ \frac{3}{2} \beta_{e} } \left( \frac{ c }{ \omega_{pe} } \right) 
\end{equation}
(see, e.g., \citealt{sironi15a}). Equation \ref{eqnA3} can be derived using the \textit{ratios} of the electron (e) to proton (p) cyclotron frequencies: $\omega_{ce}/\omega_{cp} = \gamma_{e}^{-1} ( m_{p}/m_{e} )$, the plasma frequencies: $\omega_{pe}/\omega_{pp} = \sqrt{ \gamma_{e}^{-1} ( n_{e}/n_{p} )( m_{p}/m_{e} ) }$, and the ratio of the Alfv\'{e}n speed to the speed of light: $v_{A}/c = \omega_{cp}/\omega_{pp}$, assuming relativistic electrons and nonrelativistic protons. Combining these ratios allows one to re-formulate the electron Larmor radius ($r_{\rm{gyro}} \equiv v_{e}/\omega_{ce}$) into the form of Equation \ref{eqnA3} in the relativistic limit where: $kT_{e} \simeq m_{e} \gamma_{e} v_{e}^2 / 3$. In Figure \ref{figA1} we plot individual electron Larmor radii computed using Equation \ref{eqnA3} (for the normal plasma jet - left panel, and the pair plasma jet - right panel) along rays highlighted by the red dots in the lower panels of Figure \ref{fig4}. The red dashed horizontal line in each panel highlights the scaled PIC grid cell size ($\Delta l_{\rm{scale}}$). These ray profiles verify that the Larmor radii of our simulated PIC electrons are largely \textit{contained} within our numerical cells. The regions where the electron Larmor radii exceed the plasma cell size occur where the magnetic field drops precipitously (see the top panels of Figure \ref{figB1}). These zones contribute minimally to the integrated synchrotron emission along each ray since the synchrotron emissivity is $\propto B^{2}$.

\newpage
\onecolumn
\section{Electron Phase-Space \& Energy Distributions}
\label{appG}

\begin{figure*}
\centering
\includegraphics[trim={    1.0cm     0.0cm      2.0cm       0.0cm      }, clip,width=0.33\linewidth]{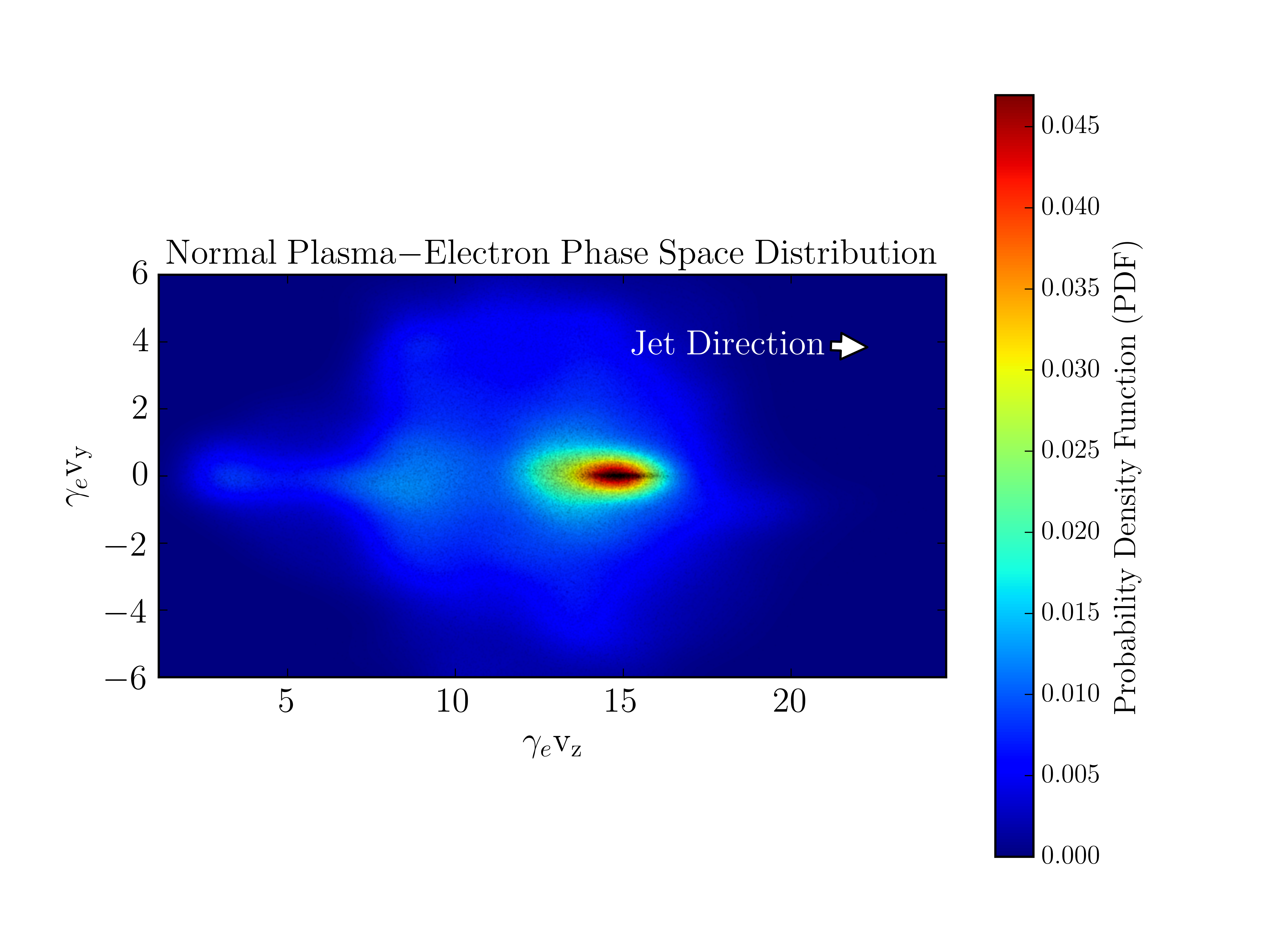}
\includegraphics[trim={    1.3cm     0.0cm      1.9cm       0.0cm      }, clip,width=0.33\linewidth]{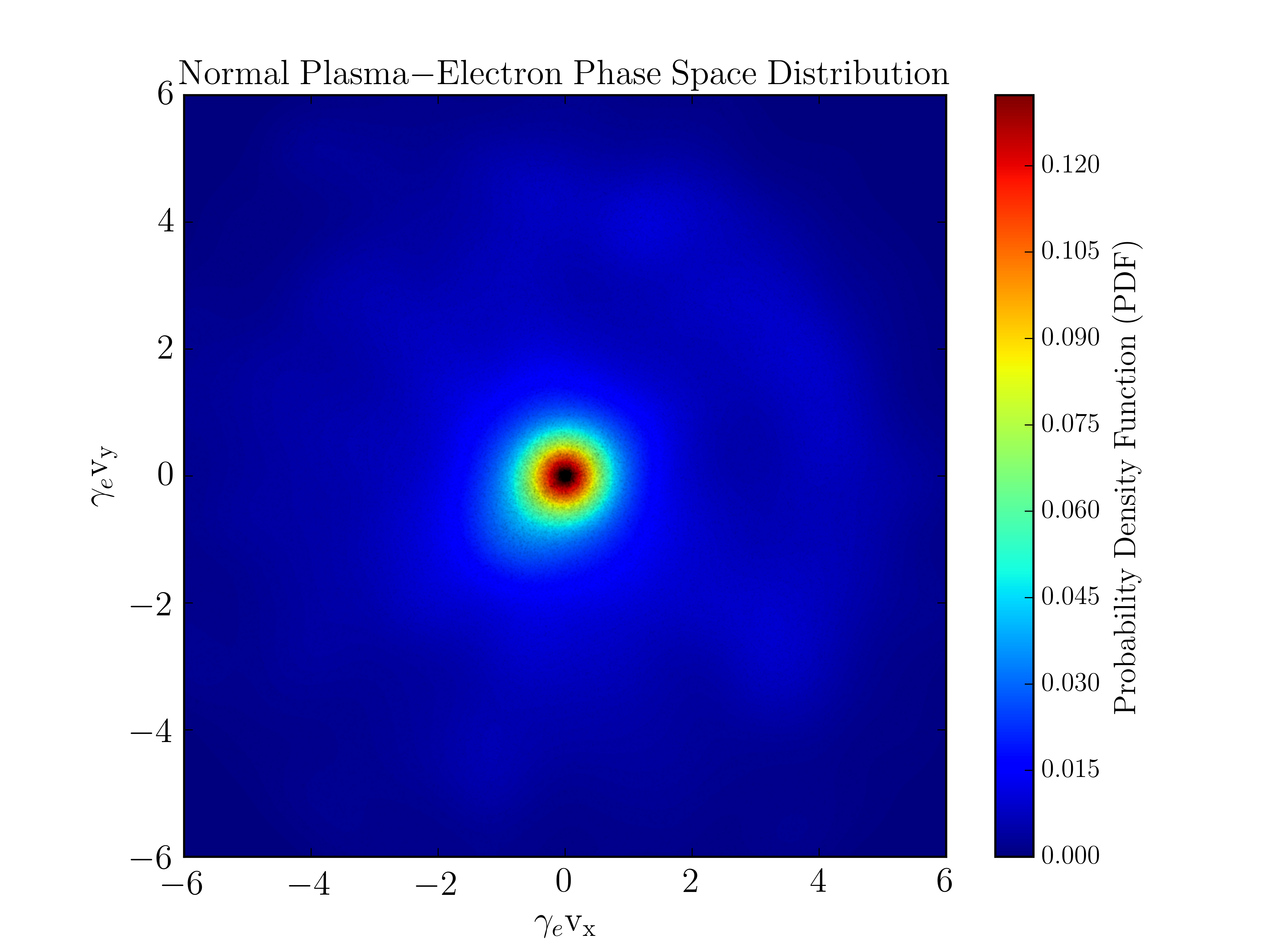}
\includegraphics[trim={    0.8cm     0.0cm      2.0cm       0.0cm      }, clip,width=0.33\linewidth]{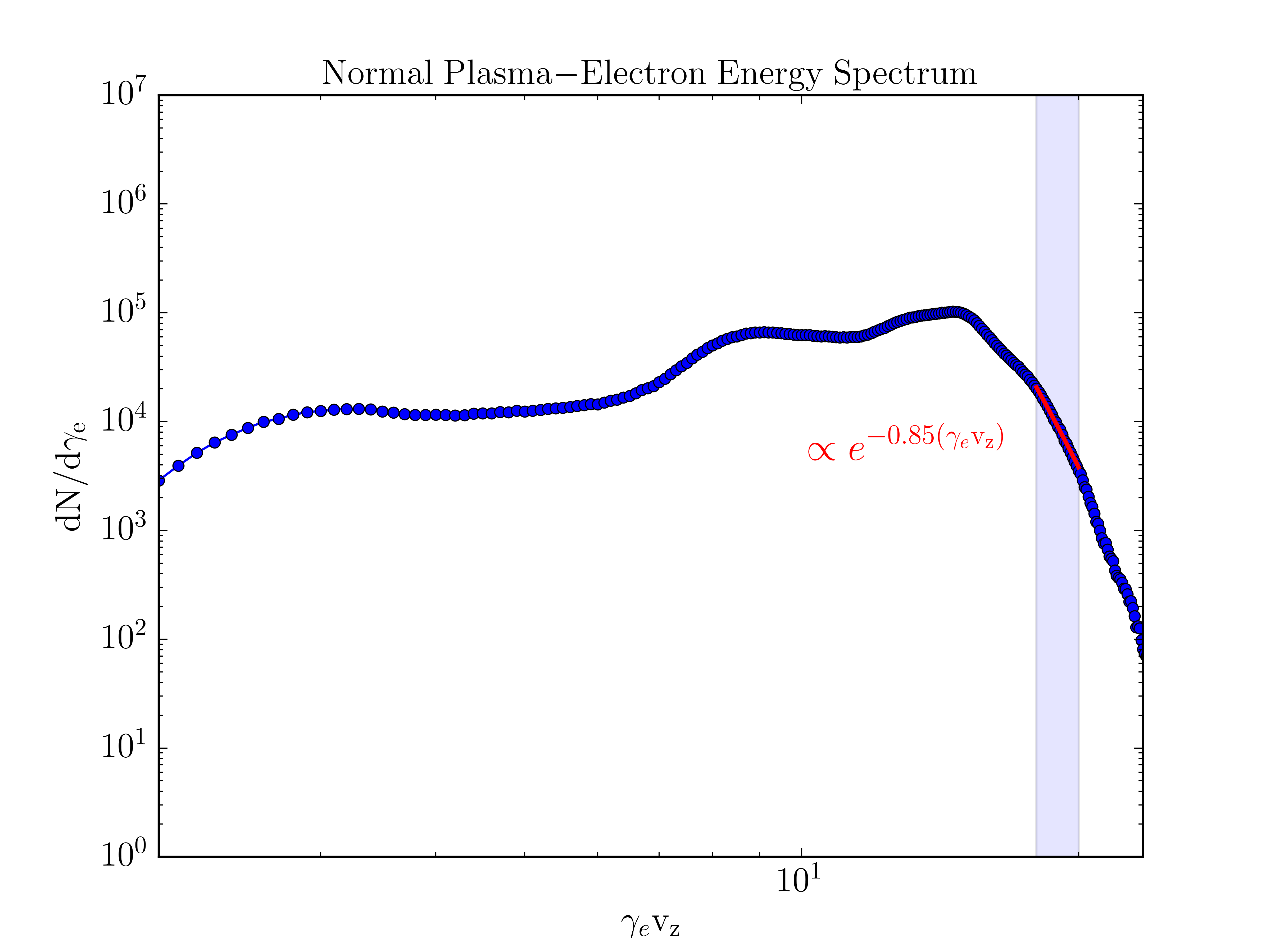}
\caption{\label{figG1}\textbf{Left panel:} An electron phase-space distribution plot along the jet axis - individual electrons are plotted as points. The underlying color scheme highlights the corresponding probability density function (PDF). \textbf{Middle panel:} An electron phase-space distribution plot perpendicular to the jet axis. \textbf{Right panel:} Electron energy spectrum for the particles contained within our PIC simulations. The high-energy tail exhibits an exponential cut-off in energy (highlighted in red).}
\end{figure*}

The synchrotron emission/absorption coefficients (contained in Equation \ref{eqn14}) assume that the underlying electron distribution within the jet is isotropic. While we have attempted to scale our PIC simulations in a manner that ensures the individual electron Larmor radii are largely contained within our grid cells (see Appendix \ref{appA}), this scaling does not guarantee that the particle distribution is indeed isotropic in nature. To explore this point further, we have generated 2D phase-space distribution plots of the electrons contained within the jet regions we generate synchrotron emission from (see the left and middle panels of Figure \ref{figG1}). Clearly, anisotropy is present within the underlying particle distribution used in our emission calculations (especially along the jet axis - Figure \ref{figG1} left panel). Despite this anisotropy, we compute the synchrotron emission using the fully isotropic emission coefficients of Equation \ref{eqn14} in order to gain a `first approximation' of the synchrotron emission emanating from our PIC simulations.

\bigskip
\vspace{-0.1cm}

\noindent The synchrotron emission is calculated in a post-process fashion. We first obtain (within each cell) the local 3D fluid velocity and magnetic field components which are averaged over 27 neighboring cells with weighting to help ensure good statistical properties within each cell. These values are then used to compute the cell's `local' magnetic field strength and `local' electron Lorentz factor. In addition, the number of particles are counted and scaled (via Equation \ref{eqn12}) to obtain a `local' electron number density. 

\bigskip
\vspace{-0.1cm}

\noindent As a further simplification in our emission calculations, we have arbitrarily fixed a constant power-law index ($s=2.3$) within each plasma cell. Future refinement of our algorithm will be needed in order to connect the spectral indices used in our synchrotron emission calculations more closely to the `local' particle energy spectrum within each cell. To explore this point further, we have generated a particle energy spectrum of the electrons contained within the jet regions we compute synchrotron emission from (see the right panel of Figure \ref{figG1}). Clearly, the particle spectrum (especially the high-energy tail) is much steeper than the power-law index of $s=2.3$ used in our emission calculations. This remains a limitation/discrepancy between our PIC models and our synchrotron emission calculations.

\newpage
\onecolumn
\section{Faraday Rotation \& Conversion Depths}
\label{appB}

The Faraday rotation ($\tau_{F}$) and conversion ($\tau_{C}$) depths within an individual plasma cell may be written as: $\tau_{F} = \zeta_{V}^{*} ~ \tau$ and $\tau_{C} = \zeta_{Q}^{*} ~ \tau$ (see \citealt{wardle03, homan09, osullivan13}). The term $\tau$ denotes the synchrotron opacity of the plasma. The terms $\zeta_{Q}^{*}$ and $\zeta_{V}^{*}$ denote normalized plasma absorption coefficients: $\zeta_{(Q,V)}^{*} \equiv \kappa^{*}_{(Q,V)}/\kappa_{I}$ (i.e., the $\kappa$'s in the matrix shown in \S\ref{Polarized Radiative Transfer}) and may be written as: 

\begin{align}
              \zeta_{Q}^{*} &=-\zeta_{\alpha}^{*Q} ~ ( \nu/\nu_{\rm{min}} )^{\alpha - \frac{1}{2}} ~ \left\{ \left[  1 - \left( \frac{\nu_{\rm{min}}}{\nu} \right)^{ \alpha - \frac{1}{2} } \right] \left( \alpha - \frac{1}{2} \right)^{-1} \right\} ~ \rm{for} ~ \alpha > 0.5\label{eqnB1}  \\
              \zeta_{V}^{*} &=\; \; \: \zeta_{\alpha}^{*V} ~ ( \nu/\nu_{\rm{min}} )^{\alpha + \frac{1}{2}} ~ \frac{ \rm{ln} ~ \gamma_{\rm{min}} }{ \gamma_{\rm min} } ~ \frac{ ( ~ n_{e^{-}} - n_{e^{+}} ~ ) }{ ( ~ n_{e^{-}} + n_{e^{+}} ~ )} ~ \rm{cot}( \vartheta )  \left[  1 + \frac{ \alpha + 2 }{ 2 \alpha + 3 } \right]
\label{eqnB2} ,
\end{align}

\noindent where $\zeta_{\alpha}^{*Q}$ and $\zeta_{\alpha}^{*V}$ are physical constants (of order unity) that are tabulated in \cite{jones77a} and depend on the value of the spectral index $\alpha$.  The quantity $\nu_{\rm{min}} \equiv \gamma_{\rm{min}}^{2} \nu_{B \perp}$, where $\nu_{B \perp}$ is given by:

\begin{equation}\label{eqnB3} 
\nu_{B \perp} = \frac{ e B ~ \rm{sin} \vartheta }{ 2 \pi ~ m_{e} c } ~ .  
\end{equation}

\begin{figure*}
\centering
\includegraphics[width=0.49\linewidth]{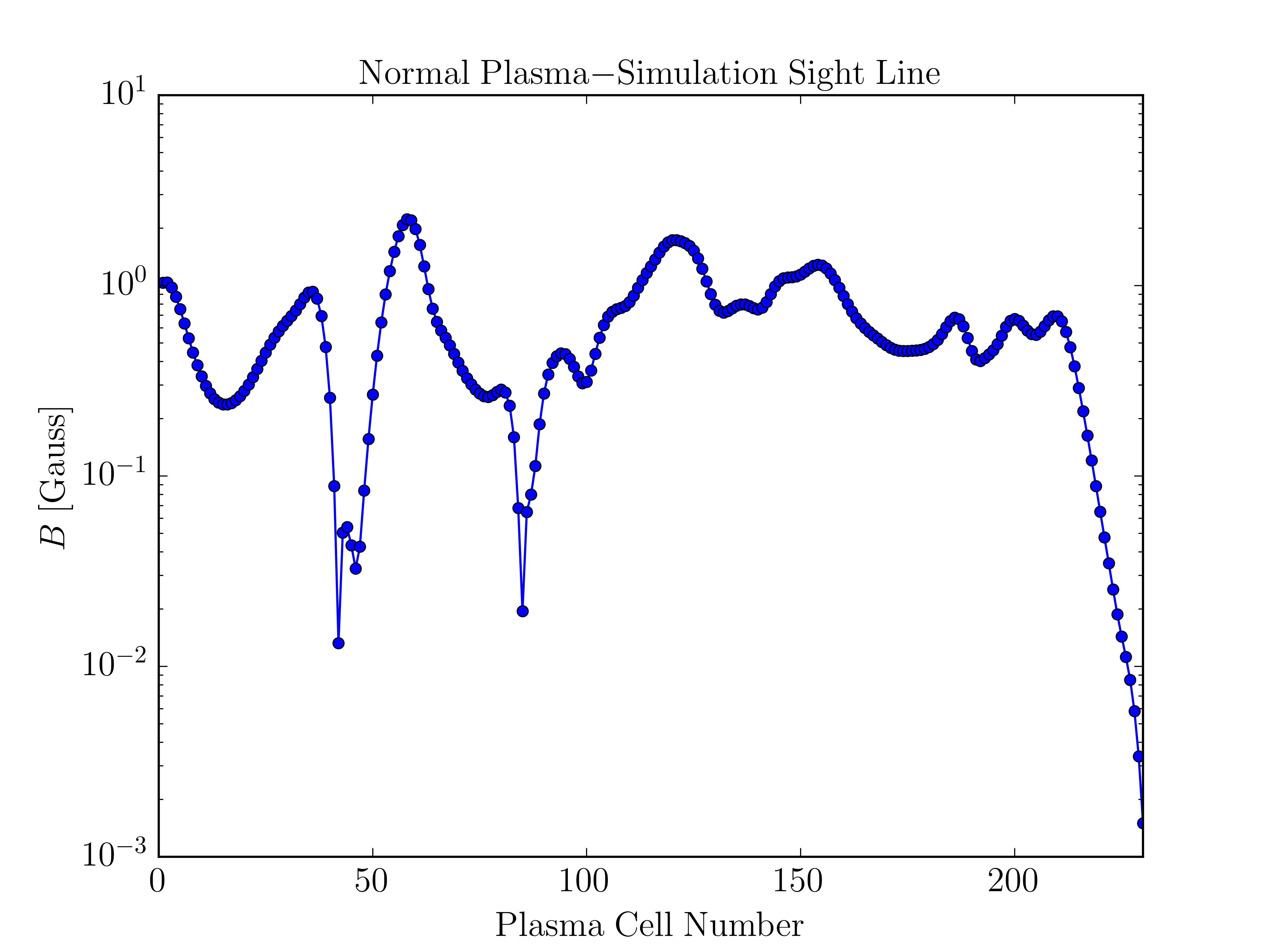}
\includegraphics[width=0.49\linewidth]{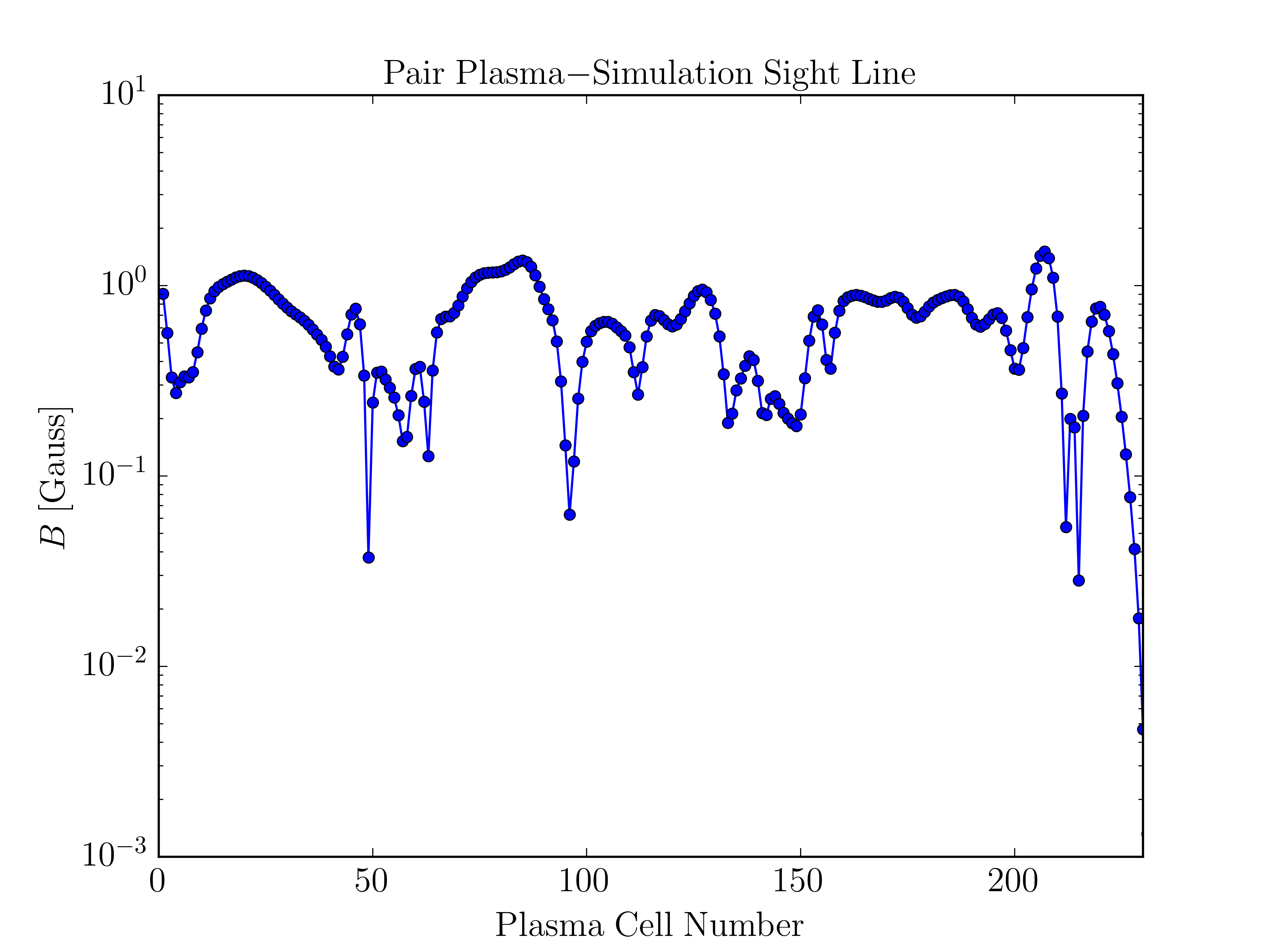}
\includegraphics[width=0.49\linewidth]{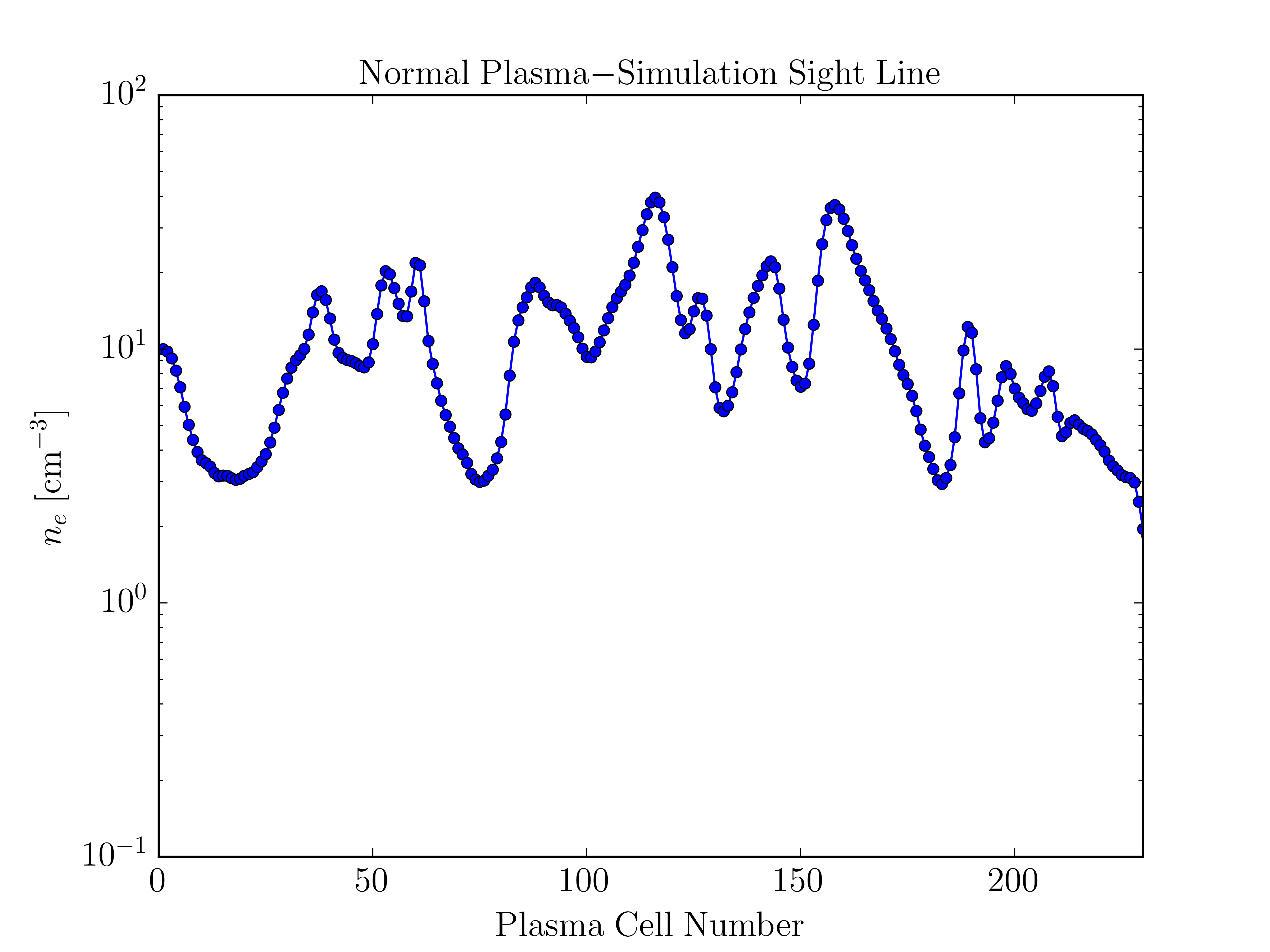}
\includegraphics[width=0.49\linewidth]{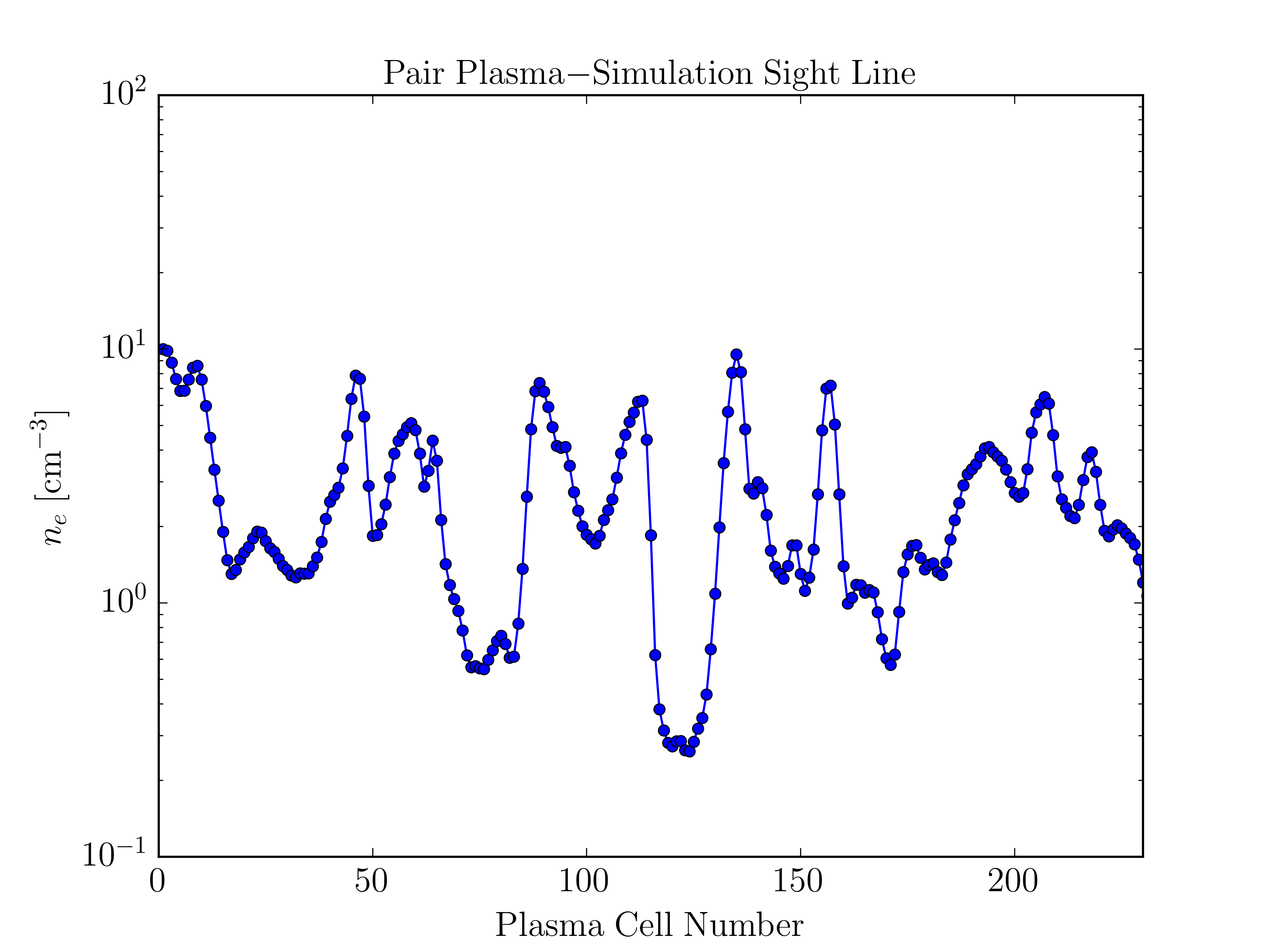}
\includegraphics[width=0.49\linewidth]{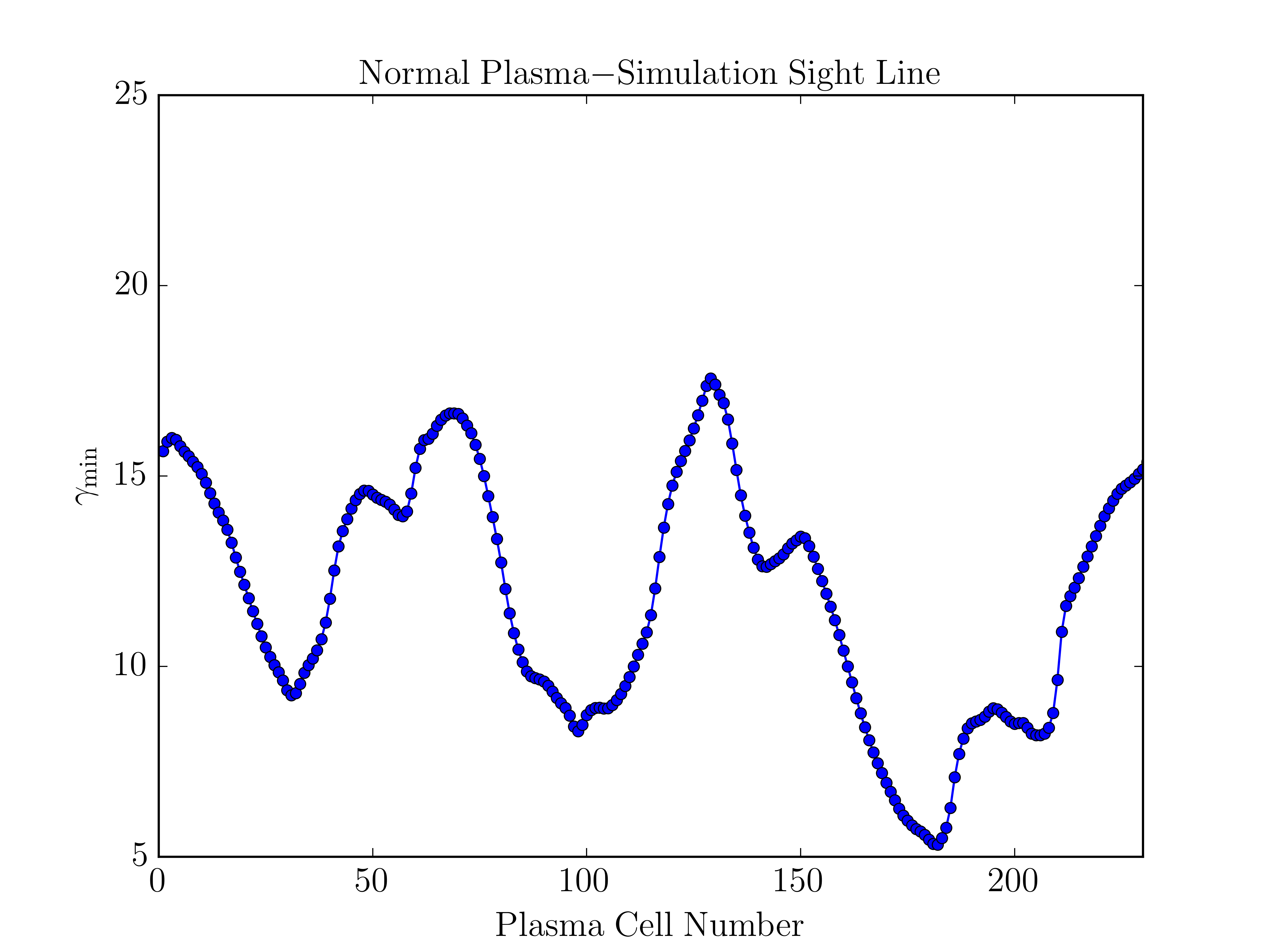}
\includegraphics[width=0.49\linewidth]{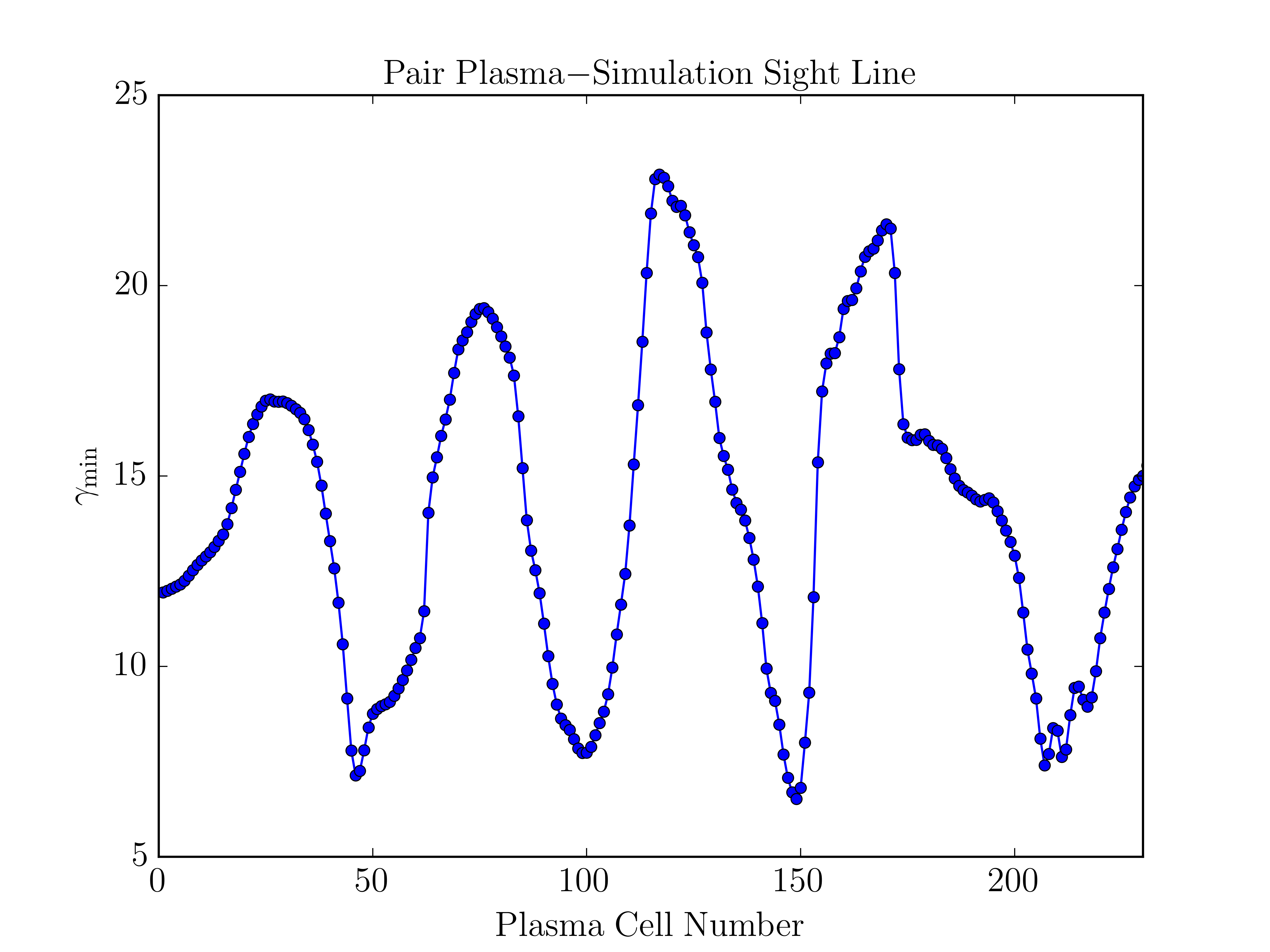}
\caption{\label{figB1}\textbf{Upper left panel:} Variation of the scaled magnetic field strength ($B$) within the normal plasma ($e^{-}$-$p^{+}$) jet along the sightline indicated by the red dot in the lower left panel of Figure \ref{fig4}. The radiative transfer progresses from left to right: starting on the far side of the jet relative to the observer (cell 0), and then advancing through the jet plasma toward the near side of the jet (cell 240). \textbf{Upper right panel:} Corresponding variation of the scaled magnetic field strength along the pair plasma ($e^{-}$-$e^{+}$) jet sightline indicated by the red dot in the lower right panel of Figure \ref{fig4}. \textbf{Middle left panel:} Variation of the scaled electron number density ($n_{e}$) along the normal plasma jet sightline. \textbf{Middle right panel:} Corresponding variation of the scaled electron number density along the pair plasma jet sightline. \textbf{Lower left panel:} Variation of the minimum electron Lorentz factor ($\gamma_{\rm{min}}$) along the normal plasma jet sightline. \textbf{Lower right panel:} Corresponding variation of the minimum electron Lorentz factor along the pair plasma jet sightline. These panels highlight the physical scaling (i.e., \S\ref{Scaling}) that we have applied to our dimensionless PIC grid values.}
\end{figure*}

\noindent The angle $\vartheta$ denotes the angle that each sightline within our ray-tracing calculations makes with respect to the local magnetic field vector in the co-moving frame of each plasma cell. Under the assumption of charge neutrality ($n_{e^{-}} \simeq n_{e^{+}} + n_{p^{+}}$), the term $( ~ n_{e^{-}} - n_{e^{+}} ~)/( ~ n_{e^{-}} + n_{e^{+}} ~ )$ in Equation \ref{eqnB2} may be re-written as $( ~ 1+ 2/r_{+} ~ )^{-1}$, where $r_{+} \equiv n_{p^{+}}/n_{e^{+}}$ and parameterizes the plasma composition of each cell (see \S\ref{Fast-light Images}). In all of the computations presented in this paper, we have assumed a constant optically thin spectral index of $\alpha = 0.65$. In future, we plan on refining our algorithm to vary $\alpha$ cell-to-cell and compute it directly from the local electron power-law index ($s$): $\alpha = (s-1)/2$. The synchrotron opacity ($\tau$) of an individual plasma cell may be written as: 

\begin{align}\label{eqnB4}
\tau &= \int \kappa ~ \rm{d}l  \nonumber \\
        &= \kappa_{\alpha} ~ ( r_{e}c ) ~ \nu_{B \perp}^{ ~ ~ ~ -1} ~ [ ~ 4 \pi g(\vartheta) ~ ] ~ [ ~ n_{e} ~ ] ~ ( \nu_{B \perp}/\nu )^{\alpha + 5/2} ~ l ~ ,
\end{align}

\noindent where $l$ is the path length of the ray through the plasma cell (computed using RADMC-3D) and where $n_{e}$ is the electron number density within the plasma cell.  The length scale of the PIC plasma cells in the simulations presented in this paper is $\simeq 6.5 \times 10^{4} ~ \rm{cm}$ (see Equation \ref{eqn7}). The parameter $\kappa_{ \alpha }$ is a physical constant (of order unity) that is tabulated in \cite{jones77a} and depends on the value of the spectral index $\alpha$. The term $g(\vartheta)$, represents the electron pitch angle distribution. It is evident, upon inspection of equations: \ref{eqnB1}, \ref{eqnB2}, \ref{eqnB3}, and \ref{eqnB4}, that $\tau_{F}$, $\tau_{C}$, and $\tau$ are together themselves functions of each plasma cell's: $B$, $n_{e}$, and $\gamma_{\rm{min}}$. The ray profiles (similar to those presented in Figure \ref{fig5}) of these three variables (for the sightlines indicated by the red dots in the lower panels of Figure \ref{fig4}) are illustrated in Figure \ref{figB1}. Upon inspection of these panels one can see that our choice of physical scaling (i.e., \S\ref{Scaling}) results in magnetic field strengths that range from $B \simeq 10^{-2} - 10^{0} ~ \rm{Gauss}$ and electron number densities that range from $n_{e} \simeq 10^{0} - 10^{2} ~ \rm{cm}^{-3}$. These values are in rough agreement with the plasma conditions that have been inferred in blazar jets from theoretical modeling of shock acceleration and turbulence (see, e.g., \citealt{marscher14}).

\bigskip
\vspace{-0.1cm}

\noindent We present here an alternate physical scaling method (commonly used in magnetohydrodynamic simulations) which generates a similar magnetic field scale factor for comparison to our PIC approach (\S\ref{Scaling}). In an RMHD simulation one typically defines the following three scale factors:

\begin{align}
&\rm{UNIT}\_\rm{DENSITY} ~ ~ ~ ~ ~ ~ ~ ~ ~ ~ \rho_{o} ~ ~ ~ ~ ~ ~ ~ ~ ~ [ ~ \rm{g} ~ \rm{cm}^{-3} ~ ] \label{eqnB5}  \\
&\rm{UNIT}\_\rm{LENGTH} ~ ~ ~ ~ ~ ~ ~ ~ ~ ~ ~ l_{o} ~ ~ ~ ~ ~ ~ ~ ~ ~ ~ [ ~ \rm{cm} ~ ] \label{eqnB6} \\
&\rm{UNIT}\_\rm{VELOCITY} ~ ~ ~ ~ ~ ~ ~ v_{o} ~ ~ ~ ~ ~ ~ ~ ~ ~ [ ~ \rm{cm} ~ \rm{s}^{-1} ~ ] \label{eqnB7} ~ ,
\end{align}

\noindent (see the PLUTO\footnote{\href{https://ascl.net/1010.045}{https://ascl.net/1010.045}} code manual for further discussion). All other physical scaling quantities can be derived from these three unit values. In particular, the magnetic field strength scale factor (in cgs) may be written as:

\begin{equation}\label{eqnB8} 
~ ~ ~ ~ ~ ~ ~ ~ ~ ~ ~ ~ ~ ~ ~ ~ ~ ~ ~ B_{o} = v_{o} \sqrt{ 4 \pi \rho_{o} } ~ ~ ~ ~ ~ ~ ~ ~ ~ [ ~ \rm{Gauss} ~ ] ~.  
\end{equation}

\noindent As discussed in \S\ref{Scaling}, we set $v_{o} = 3 \times 10^{10} ~ \rm{cm} ~ \rm{s}^{-1}$. Again, under the assumption of charge neutrality ($n_{e^{-}} \simeq n_{e^{+}} + n_{p^{+}}$), and invoking the normal plasma case, in which $n_{p^{+}} \gg n_{e^{+}}$, one may make the assumption that: $n_{p^{+}} \simeq n_{e^{-}}$. While the electrons dominate the nonthermal synchrotron emission, the protons (in contrast) dominate the thermal fluid dynamics. It follows that: $\rho_{o} = n_{o} m_{p}$, where $n_{o} \simeq 10^{1} ~ \rm{cm}^{-3}$ (i.e., setting the proton number density equal to the electron value from \S\ref{Scaling}). This results in $\rho_{o} \simeq 1.7 \times 10^{-23} ~ \rm{g} ~ \rm{cm}^{-3}$, having used $m_{p} = 1.6726231 \times 10^{-23} \rm{g} ~ \rm{cm}^{-3}$. Inserting $v_{o}$ and $\rho_{o}$ into Equation \ref{eqnB8} yields a magnetic field strength scaling factor of $B_{o} \simeq 0.4 ~ \rm{Gauss}$ which is in agreement with the scaling value we obtain in \S\ref{Scaling} (Equation \ref{eqn10}) after invoking a fiducial jet plasma beta of $\beta_{e} \simeq 10^{-3}$.

\newpage
\onecolumn
\section{Radiative Transfer Angles}
\label{appF}

\begin{figure*}
\centering
\scalebox{0.67}{\includegraphics[trim={0cm 0cm 0cm 0cm}, clip, width=1.0\columnwidth,clip]{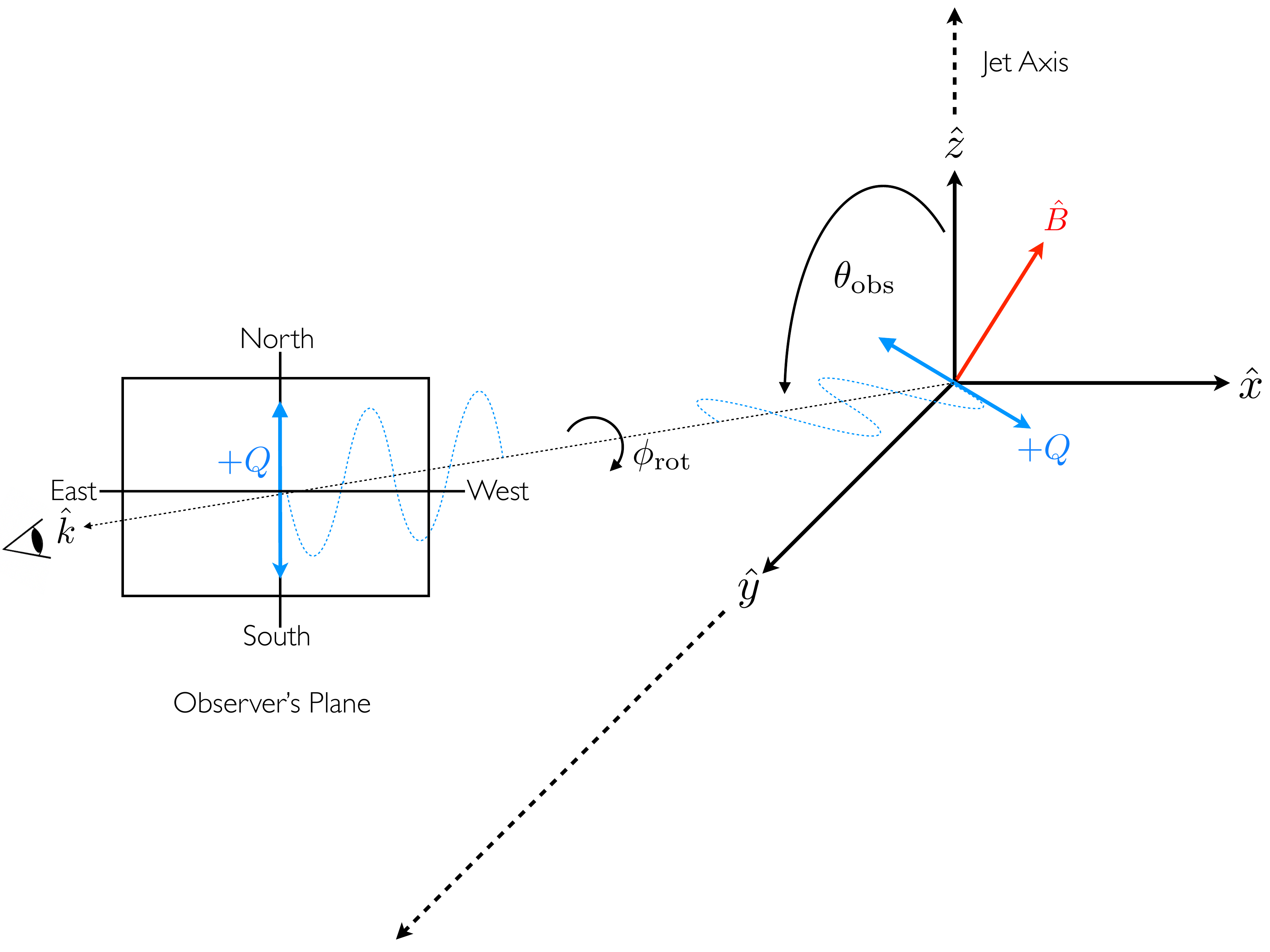}}
\caption{\label{figF1} A schematic representation of the angles involved in our radiative transfer calculations. The observer's sightline (wave vector $\hat{k}$) makes an angle $\theta_{\rm{obs}}$ with respect to the jet/$\hat{z}$ axis. The magnetic field unit vector, $\hat{B} = ( ~ B_{x}, ~ B_{y}, ~ B_{z} ~ )$, of an individual plasma cell is depicted in red. The radiative transfer is carried out in a cell-specific `Stokes U = 0' linear polarization basis (depicted in blue). The cell-specific linear polarization basis is then rotated by an angle $\phi_{\rm{rot}}$ onto a generalized observer's plane (shown to the left).}
\end{figure*}

The synchrotron emission/absorption coefficients (contained in Equation \ref{eqn14}) are functions of the angle ($\vartheta$) each ray makes with respect to the local magnetic field vector in the co-moving frame of each plasma cell. When computing the Stokes parameters, we execute the following numerical steps (cell-by-cell) to ensure a proper treatment of the angles involved in the radiative transfer: 

\begin{enumerate}[label=(\roman*)]

\item We first compute the angle our line-of-sight makes with respect to each cell's local magnetic field vector accounting for relativistic aberration due to the bulk flow of the jet. In particular, the following Lorentz transformation maps our sightline unit vector $\hat{k}$ in the observer's frame to the corresponding unit vector $\hat{k}^{\prime}$ in the co-moving frame of the jet plasma:

\begin{ceqn}
\begin{equation}\label{eqnF1} 
\hat{k}^{\prime} = \frac{ \hat{k} + \Gamma \vv{ \beta } ~ \left[ ~ \frac{ \Gamma }{ \Gamma + 1 } \beta ~ \rm{cos}( \theta_{\rm{obs}} ) - 1 ~  \right] }{ \Gamma ~ [ ~ 1 - \beta ~ \rm{cos}( \theta_{\rm{obs}} ) ~ ] }
\end{equation}
\end{ceqn}

\noindent (see \citealt{lyutikov05}), where $\Gamma$ is the bulk Lorentz factor of the jet flow and $\beta \equiv v_{\rm{ jet }}/c = \sqrt{ 1 - \frac{ 1 }{ \Gamma^{2} } }$. We make the simplifying assumption that there exists a common `jet frame'; namely, that all cells within our simulations propagate in a laminar fashion along the jet-axis and share a common bulk Lorentz factor ($\Gamma = 15$ in this case). This simplification makes the radiative transfer more tractable and permits us to compute the synchrotron emission along plane parallel rays (depicted in Figure \ref{fig2}). Future refinement of our algorithm will be required to account properly for cell-to-cell variations in the relativistic motions of the plasma within the jet.

\bigskip
\vspace{-0.1cm}

\noindent As illustrated in Figure \ref{figF1}, we orient the jet along the $\hat{z}$-axis and therefore $\vv{ \beta } = \big\{~ 0, ~ 0, ~ \beta ~ \big\}$.  From Figure \ref{figF1} it also follows that: $\hat{k} = \big\{~  0, ~ \rm{sin}( \theta_{\rm{obs}} ), ~ \rm{cos}( \theta_{\rm{obs}} ) ~ \big\}$, where $\theta_{\rm{obs}}$ is the angle the observer's line-of-sight makes with respect to the jet axis.  Recalling the definition of the relativistic Doppler boosting factor: $\delta \equiv \frac{ 1 }{ \Gamma(~ 1 - \beta \rm{cos}( \theta_{\rm{obs}} ) ~) }$, Equation \ref{eqnF1} can be rewritten in component form as:

\begin{ceqn}
\begin{align}\label{eqnF2}
\hat{k}^{\prime} &= \delta ~ \left\{~ 0, ~ \rm{sin}( \theta_{\rm{obs}} ), ~ \rm{cos}( \theta_{\rm{obs}} ) +  \Gamma \beta ~ \left[ ~ \frac{ \Gamma }{ \Gamma + 1 } \beta \rm{cos}( \theta_{\rm{obs}} ) - 1 ~  \right] ~ \right\}\nonumber \\
			&= \delta ~ \Bigg\{~ 0, ~ \rm{sin}( \theta_{\rm{obs}} ), ~ \Gamma \big[ ~ \rm{cos}( \theta_{\rm{obs}} ) - \beta ~ \big]  ~ \Bigg\}
\end{align}
\end{ceqn} 

\noindent after combining terms and simplifying. The local magnetic field unit vector is given by $\hat{B} = ( ~ B_{x}, ~ B_{y}, ~ B_{z} ~ )$.  It then follows that:

\begin{ceqn}
\begin{equation}\label{eqnF3}
\rm{cos}( \vartheta ) = \hat{k}^{\prime} \cdot \hat{B} \rightarrow \vartheta = \rm{cos}^{-1} \Biggl\{ \delta ~ \rm{sin}( \theta_{\rm{obs}} ) ~ B_{y} + \delta ~ \Gamma [  ~ \rm{cos}(\theta_{\rm{obs}}) - \beta ~ ] ~ B_{z} \Biggl\} ~ .
\end{equation}
\end{ceqn}

\item With $\vartheta$ determined (cell-by-cell) the radiative transfer is then computed across each cell. These calculations are done in a cell specific `Stokes U = 0' linear polarization basis. The analytic solution used in our computations is presented in \cite{jones77a} (Appendix B) and summarized in \cite{macdonald18} (Appendix A).

\item Once the radiative transfer across a cell is complete, we then rotate the cell specific linear polarization basis onto a generalized observer's plane (illustrated in Figure \ref{figF1}). In particular, the angle of this rotation ($\phi_{\rm{rot}}$) is:

\begin{equation}\label{eqnF4}
\phi_{\rm{rot}}  = \rm{cos}^{-1}  \left\{  \frac{ B_{x} }{ \sqrt{ B_{x}^{2} + \Big[ ~ B_{y} -  \rm{sin} ( \theta_{\rm{obs}} ) \big\llbracket ~ B_{y} ~ \rm{sin}(\theta_{\rm{obs}}) - B_{z} ~ \rm{cos}(\theta_{\rm{obs}}) ~ \big\rrbracket  ~ \Big]^{2} + \Big[  ~ B_{z} -  \rm{cos}(\theta_{\rm{obs}}) \big\llbracket ~ B_{z} ~ \rm{cos}(\theta_{\rm{obs}}) - B_{y} ~ \rm{sin}(\theta_{\rm{obs}}) ~ \big\rrbracket  ~ \Big]^{2} } } \right\} ~ ,
\end{equation}

\noindent and is equal to the angle the projected magnetic field $\big( ~ \hat{B}_{\rm{proj}} = \hat{B} - ( ~ \hat{B} \cdot  \hat{k} ~ ) \hat{k} ~ \big)$ makes with respect to the x-axis (i.e., west) on the observer's plane. This rotation is then applied to the Stokes parameters in the following manner:

\begin{ceqn}
\begin{equation}\label{eqnF5}\normalsize{
\left( \begin{array}{c}
I_{\nu^{\rm{rot}}} \\[10pt]
Q_{\nu^{\rm{rot}}} \\[10pt]
U_{\nu^{\rm{rot}}} \\[10pt]
V_{\nu^{\rm{rot}}} \end{array} \right) = \left( \begin{array}{cccc}
1 & 0 & 0 & 0 \\[10pt]
0 & \rm{cos}(2\phi_{\rm{rot}}) & \rm{sin}(2\phi_{\rm{rot}}) & 0 \\[10pt]
0 & -\rm{sin}(2\phi_{\rm{rot}}) & \rm{cos}(2\phi_{\rm{rot}}) & 0 \\[10pt]
0 &  0 & 0 & 1 \end{array} \right)  
\left( \begin{array}{c}
I_{\nu} \\[10pt]
Q_{\nu} \\[10pt]
U_{\nu} \\[10pt]
V_{\nu} \end{array} \right) }~.
\end{equation}
\end{ceqn}

\item These `generalized' Stokes parameters are then recorded and passed on to the next cell via our slow-light interpolation scheme at which point we rotate back into the jet plasma frame and repeat steps (i-iii) for the next plasma cell along each sightline. 

\end{enumerate}

\newpage
\onecolumn
\section{Radiative Transfer Limits}
\label{appC}

\begin{figure*}
\centering
\includegraphics[width=0.49\linewidth]{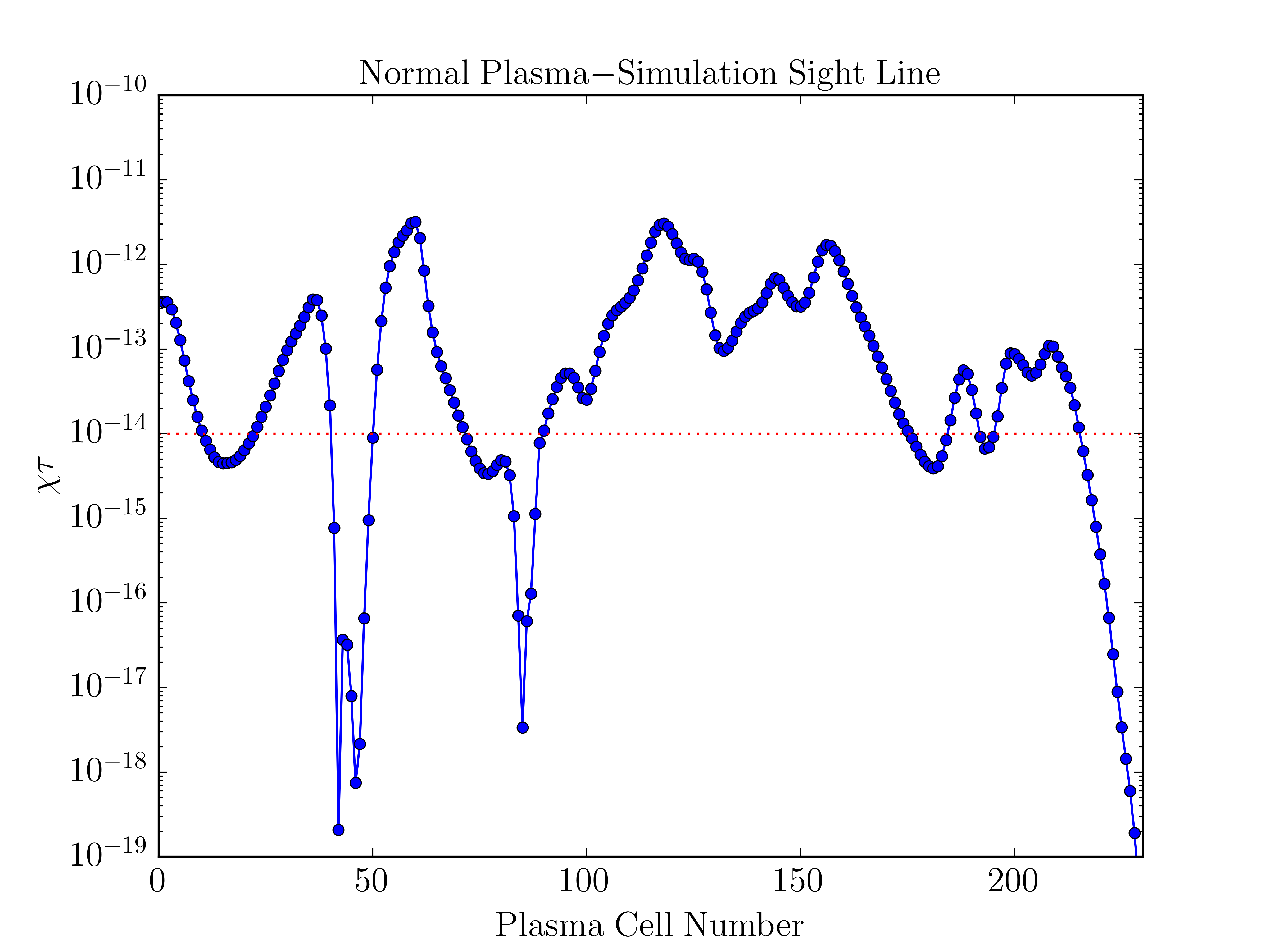}
\includegraphics[width=0.49\linewidth]{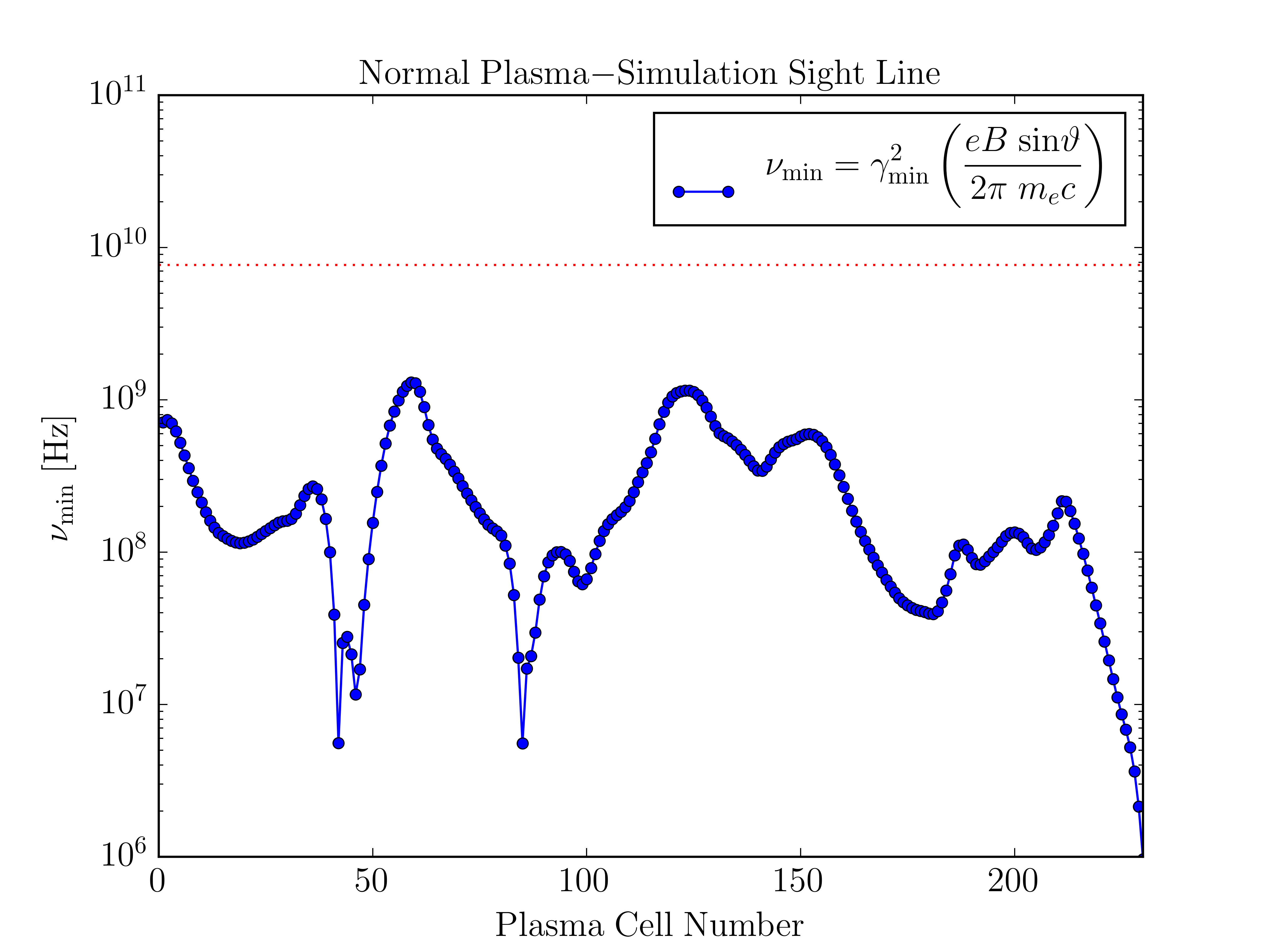}
\caption{\label{figC1}\textbf{Left panel:} Variation of the term $\chi \tau$ (synchrotron opacity) within the normal plasma ($e^{-}$-$p^{+}$) jet along the sightline indicated by the red dot in the lower left panel of Figure \ref{fig4}. The red dashed line indicates a numerical limit below which we enforce the criterion $I_{\nu},Q_{\nu},U_{\nu},V_{\nu} \rightarrow I_{\nu}^{~ 0},Q_{\nu}^{~ 0},U_{\nu}^{~ 0},V_{\nu}^{~ 0}$ directly in our PIC ray-tracing calculations. \textbf{Right panel:} Variation of the minimum frequency ($\nu_{\rm{min}}$) within the normal plasma ($e^{-}$-$p^{+}$) jet along the sightline indicated by the red dot in the lower left panel of Figure \ref{fig4}. The red dashed line indicates the emitted frequency $\nu_{\rm{em}}$ in the plasma frame corresponding to $\nu_{\rm{obs}} = 230 ~ \rm{GHz}$. In both panels, the radiative transfer progresses from left to right: starting on the far side of the jet relative to the observer (cell 0), and then advancing through the plasma toward the near side of the jet (cell 240).}
\end{figure*}

\noindent Care must be taken when implementing the analytic solution to the polarized radiative transfer equation presented in Equation \ref{eqn14}. As discussed in \cite{janett19}, there are many numerical challenges one faces when attempting to implement polarized radiative transfer through plasma simulations. The analytic solution to the matrix in Equation \ref{eqn14} (which is presented in Appendix B of \citealt{jones77a}) is laced with hyperbolic sine and cosine functions whose arguments consist of the expression $\chi \tau$. Here, $\chi$ is a term (of order unity) related to the Faraday rotation and conversion depths of the plasma (also presented in Appendix B of \citealt{jones77a}) and $\tau$ is the synchrotron opacity of the plasma (see Appendix \ref{appB} of this manuscript). Given the extremely small cell sizes within our PIC simulations, the path length $l$ (see Equation \ref{eqnB4}) of each ray through our plasma cells is many orders of magnitude smaller than typical path lengths through RMHD grids. This results in extremely small synchrotron opacities along the sightlines through our PIC jets (see the left panel of Figure \ref{figC1}). Fortran's hyperbolic \textit{cosh} and \textit{sinh} functions do not go exactly to 1.0 and 0.0 respectively, when there arguments (i.e., $\chi \tau$) fall below $1.0 \times 10^{-14}$ (i.e., the red dashed line in the left panel of Figure \ref{figC1}). Performing radiative transfer beyond this numerical limit can introduce erroneous values in the Stokes parameters. As discussed in \cite{macdonald18} (see there Appendix A), as $\tau \rightarrow 0$ within a given plasma cell the Stokes parameters should analytically equal the values from the previous plasma cell along each ray's path (i.e., $I_{\nu},Q_{\nu},U_{\nu},V_{\nu} \rightarrow I_{\nu}^{~ 0},Q_{\nu}^{~ 0},U_{\nu}^{~ 0},V_{\nu}^{~ 0}$). We enforce this criterion directly in our PIC ray-tracing calculations when a ray encounters a plasma cell in which $\chi \tau \le 1.0 \times 10^{-14}$. Also, when viewing our PIC simulations at right angles (i.e., $\theta_{\rm{obs}} = 90^{\circ}$) sightlines through the jet becomes increasingly optically thin because each ray intersects less jet plasma. Since $\tau \propto \nu^{-(\alpha + 5/2)}$, where $\alpha = 0.65$, observing at lower frequencies helps to increase the synchrotron opacity and mitigate this numerical limitation in the radiative transfer. This is why, when imaging our plasma jets at right angles, we tune $\nu_{\rm{obs}} = 1 ~ \rm{GHz}$. 

\bigskip
\vspace{-0.1cm}

\noindent As stated in \cite{jones77a}, the frequency dependent synchrotron emission/absorption coefficients contained in Equation \ref{eqn14} are only valid provided that the frequency of emission $\nu_{\rm{em}}$ (in the co-moving frame of the plasma) exceeds the minimum frequency $\nu_{\rm{min}}$ of the plasma (defined in Appendix \ref{appB}). We check, cell-by-cell, that this criterion is met in all of our ray-tracing calculations (see the right panel of Figure \ref{figC1}). In particular, $\nu_{\rm{obs}} = \delta ~ \nu_{\rm{em}}$, where $\delta$ is the Doppler factor and is given by $\delta \equiv ( ~ \Gamma( 1 - \beta ~ \rm{cos} ~ \theta_{\rm{obs}} ) ~ )^{-1}$. Here $\Gamma$ is the bulk Lorentz factor of the jet, $\beta \equiv v/c = (  1 - \Gamma^{-2} )^{1/2}$, and $\theta_{\rm{obs}}$ is the angle between the jet axis and our line-of-sight. The minimum frequency within each plasma cell is a function of: (i) the cell's minimum electron Lorentz factor, (ii) the local magnetic field strength, and (iii) the angle our line-of-sight makes with respect to each plasma cell's local magnetic field vector, $\nu_{\rm{min}}( \gamma_{\rm{min}}, B, \vartheta )$. 
Within our PIC simulations, $\nu_{\rm{min}}$ typically ranges in value from $\sim10^{7}-10^{9} ~ \rm{Hz}$. In order to ensure the validity of the radiative transfer (i.e.,$ \nu_{\rm{min}} < \nu_{\rm{em}}$) when imaging our plasma jets at $\theta_{\rm{obs}} = 0^{\circ}$, we tune $\nu_{\rm{obs}} = 230 ~ \rm{GHz}$ (which corresponds to an emitted frequency of $\nu_{\rm{em}} \simeq 7.7 \times 10^{9} ~ \rm{Hz}$ in the rest frame of the plasma and is highlighted by the red dashed lined in the right panel of Figure \ref{figC1}).  

\bigskip
\vspace{-0.1cm}

\noindent We finally highlight an argument presented in \cite{bjornsson19} \& \cite{bjornsson20}, that questions the applicability of the analytic expressions we use to model the polarization of relativistic jets (namely, by treating our PIC cells as piecewise homogeneous regions of magnetized plasma). It is argued that instead of performing the radiative transfer on the Stokes parameters (i.e., Equation \ref{eqn14}) one should instead implement the method of \textit{characteristic waves} and carry out the polarized radiative transfer on the electric and magnetic fields within the plasma (computing the Stokes parameters only as a final step). While both methods should, in principle, be interchangeable, the contributing terms to CP are more easily identifiable in the characteristic wave method. In future, we plan on implementing the characteristic wave method of polarized radiative transfer in our ray-tracing algorithm for comparison to the calculations presented here. We also plan on implementing an integration scheme (similar to the methods presented in \citealt{ruszkowski02, dexter16, moscibrodzka18}) to solve Equation \ref{eqn14} numerically, thus providing a further comparison to the analytic expressions of \cite{jones77a} used in this manuscript.

\newpage
\onecolumn
\section{Particle-in-cell Spectropolarimetry}
\label{appD}

\begin{figure*}
\centering
\includegraphics[width=0.49\linewidth]{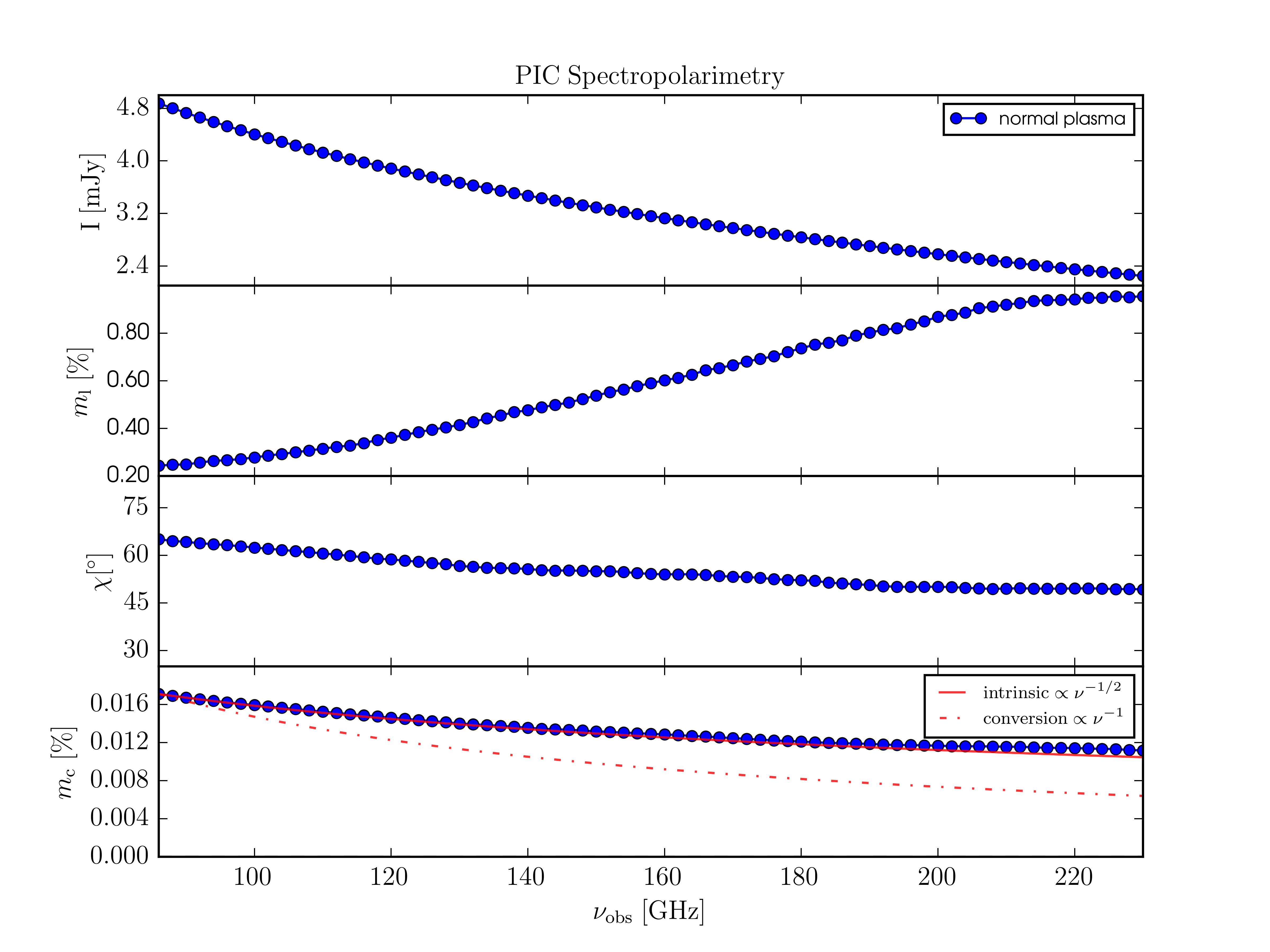}
\includegraphics[width=0.49\linewidth]{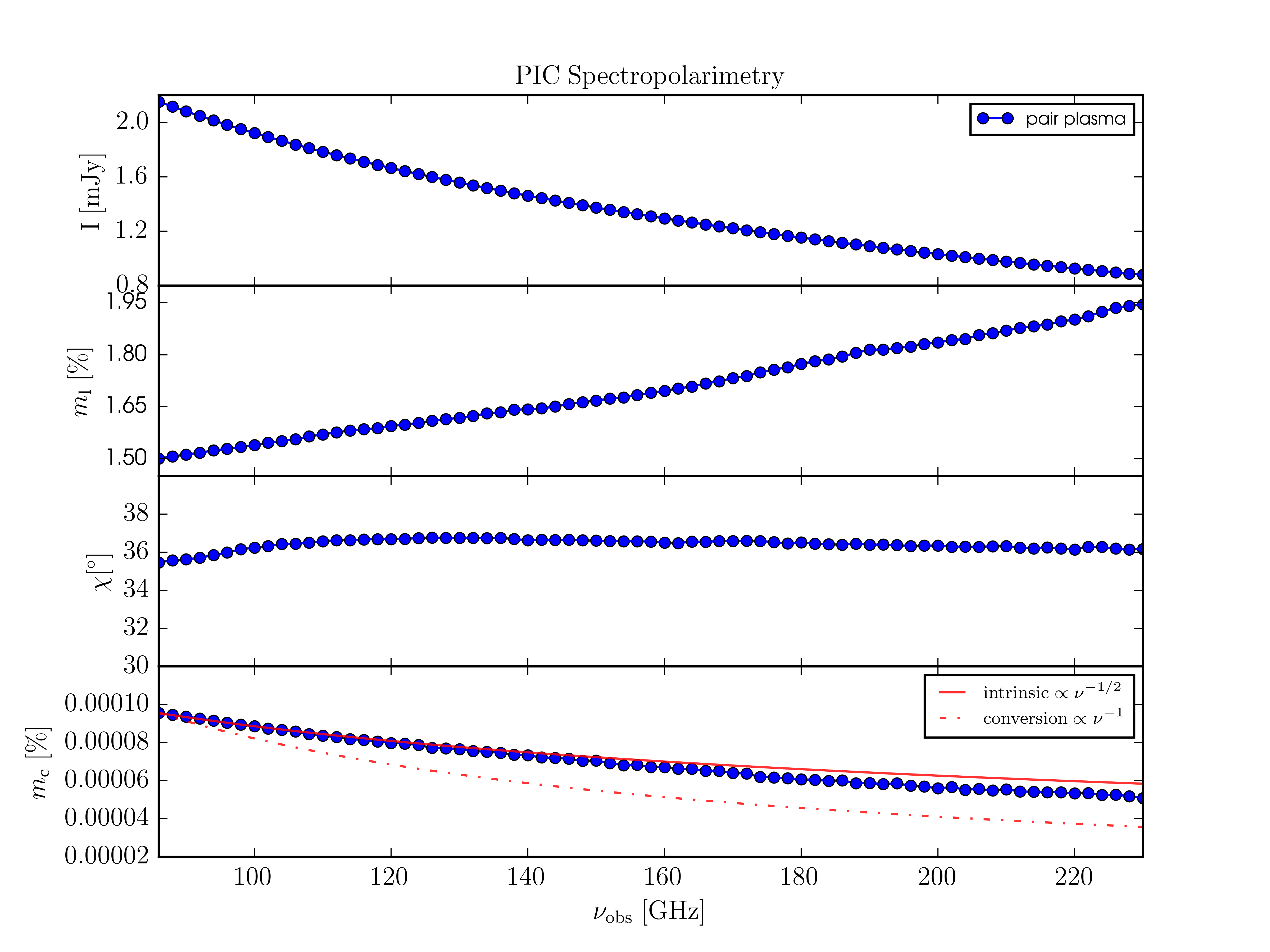}
\caption{\label{figD1}\textbf{Left panel:} Spectropolarimetry of the normal plasma jet: (top) integrated total intensity - $I ~ $, (upper middle) integrated fractional linear polarization - $m_{\rm l} ~ $, (lower middle) integrated EVPA - $\chi ~ $, and (bottom) integrated fractional circular polarization - $m_{\rm c} ~ $, over the frequency range $\nu_{\rm{obs}} = 86 - 230 ~ \rm{GHz}$. \textbf{Right panel:} Corresponding spectropolarimetry of the pair plasma jet. The solid red and dashed red lines in the lower panels highlight theoretical predictions for the frequency dependence of $m_{\rm c}(\nu)$ in the limits of intrinsic emission and emission dominated by Faraday conversion in the high-rotation limit, respectively.}
\end{figure*}

\noindent As a further comparison between the radiative properties of the normal plasma jet and the pair plasma jet we generate spectra of each jet's: (i) integrated total intensity - $I ~ $, (ii) integrated fractional linear polarization - $m_{\rm l} ~ $, (iii) integrated EVPA - $\chi ~ $, and (iv) integrated fractional circular polarization - $m_{\rm c} ~ $, over the frequency range $\nu_{\rm{obs}} = 86 - 230 ~ \rm{GHz}$ (see Figure \ref{figD1}), when each jet is oriented edge-on to the observer as shown in Figure \ref{fig4} (i.e., like blazars). In particular, we follow the formalism of \cite{kim19} and compute \textit{integrated} Stokes values for each simulation in the following manner: 

\begin{align}
I_{\rm int} &= \sum I_{i,j} ~ \times A_{\rm pixel}/A_{\rm beam} ~ [\rm{Jy}] \label{eqnD1}  \\
Q_{\rm int} &= \sum Q_{i,j} \times A_{\rm pixel}/A_{\rm beam} ~ [\rm{Jy}]   \label{eqnD2} \\
U_{\rm int} &= \sum U_{i,j} \times A_{\rm pixel}/A_{\rm beam} ~ [\rm{Jy}]  \label{eqnD3} \\
V_{\rm int} &= \sum V_{i,j} \times A_{\rm pixel}/A_{\rm beam}  ~ [\rm{Jy}]  \label{eqnD4} ~,
\end{align}

\noindent where $\sum I_{i,j}$, $\sum Q_{i,j}$, $\sum U_{i,j}$, and $\sum V_{i,j}$ are summations of the map pixel values contained within the outermost Stokes I contour of each ray-traced image (conservatively set at $20\%$ of each map's peak value). The term $A_{\rm pixel} ~ (= 0.4 \times 0.4 ~ \rm{square} ~ \rm{mas})$ denotes the angular extent of each pixel in our ray-traced images and is computed from the ratio of the RADMC-3D pixel size ($\sim 3 \times 10^{4}  \rm{cm}$) to the source distance ($\sim 1.5 \times 10^{13} \rm{cm}$; 1 AU). The term $A_{\rm beam} ~ (= \pi ~ \psi_{\rm{maj}} \times \psi_{\rm{min}}/4 ~ \rm{ln} ~ 2)$ denotes the angular extent of the convolving beam, where $\psi_{\rm{maj}}$ and $\psi_{\rm{min}}$ are the FWHM of the major and minor beam axes, respectively. For this analysis we have used a \textit{fixed} circular beam of $\psi_{\rm{maj}} = \psi_{\rm{min}} = 3.5 ~ \rm{mas}$ (illustrated in the lower left of each panel in Figure \ref{fig4}). From these integrated Stokes values we compute the following polarimetric quantities: 

\begin{align}
I &= I_{\rm{int}} \times 1000 ~ [\rm{mJy}]  \label{eqnD5} \\
m_{\rm{l}} &= \sqrt{ Q_{\rm{int}}^{2} + U_{\rm{int}}^{2} }/ I_{\rm{int}} ~ \times 100 ~ [\%]  \label{eqnD6} \\
\chi &=  1/2 ~ \rm{arctan} ( ~ U_{\rm{int}}/Q_{\rm{int}} ~ ) \times 180/\pi ~ [^\circ] \label{eqnD7} \\
m_{\rm{c}} &= -V_{\rm{int}}/I_{\rm{int}} \times 100 ~ [\%]  \label{eqnD8} ~.
\end{align}
The above quantities are then recorded for each ray-tracing calculation at individual frequencies ranging from $\nu_{\rm{obs}} = 86 - 230 ~ \rm{GHz}$ (see Figure \ref{figD1}).

\bigskip
\vspace{-0.1cm}

\noindent The normal plasma jet and the pair plasma jet are both optically thin in this frequency range. Both spectra exhibit a general trend of decreasing intensity (top panels) with increasing fractional linear polarization (upper middle panels) accompanied by decreasing fractional circular polarization (bottom panels) in agreement with synchrotron theory. An anticorrelation between linear and circular polarization across frequency has been detected in the quasar PKS B2126-158 with high-precision CP measurements made using the Australian Compact Telescope Array (ATCA) (\citealt{osullivan13}). The solid red and dashed red lines in the lower panels highlight theoretical predictions for the frequency dependence of $m_{\rm c}(\nu)$ in the limits of intrinsic emission and emission dominated by Faraday conversion in the high-rotation limit, respectively (see \citealt{pacholczyk70, jones77a}). Clearly, the emission produced in our PIC models is intrinsic in origin due to the very small Faraday depths through the jet plasma along each sightline (see Figure \ref{fig5}). 

\newpage
\onecolumn
\section{Interferometric Array Sensitivity}
\label{appE}

\begin{figure*}
\centering
\includegraphics[width=0.33\linewidth]{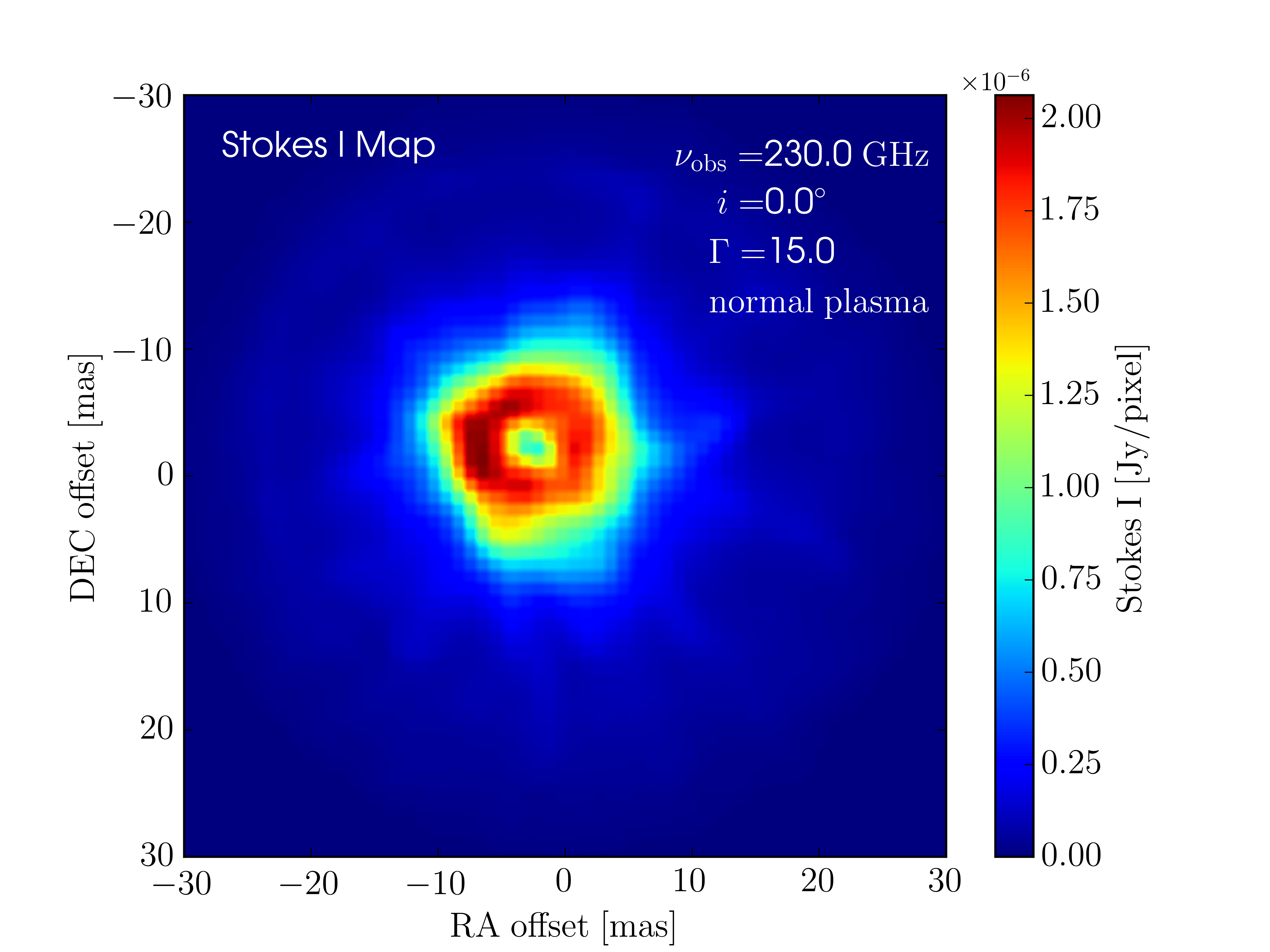}
\includegraphics[width=0.33\linewidth]{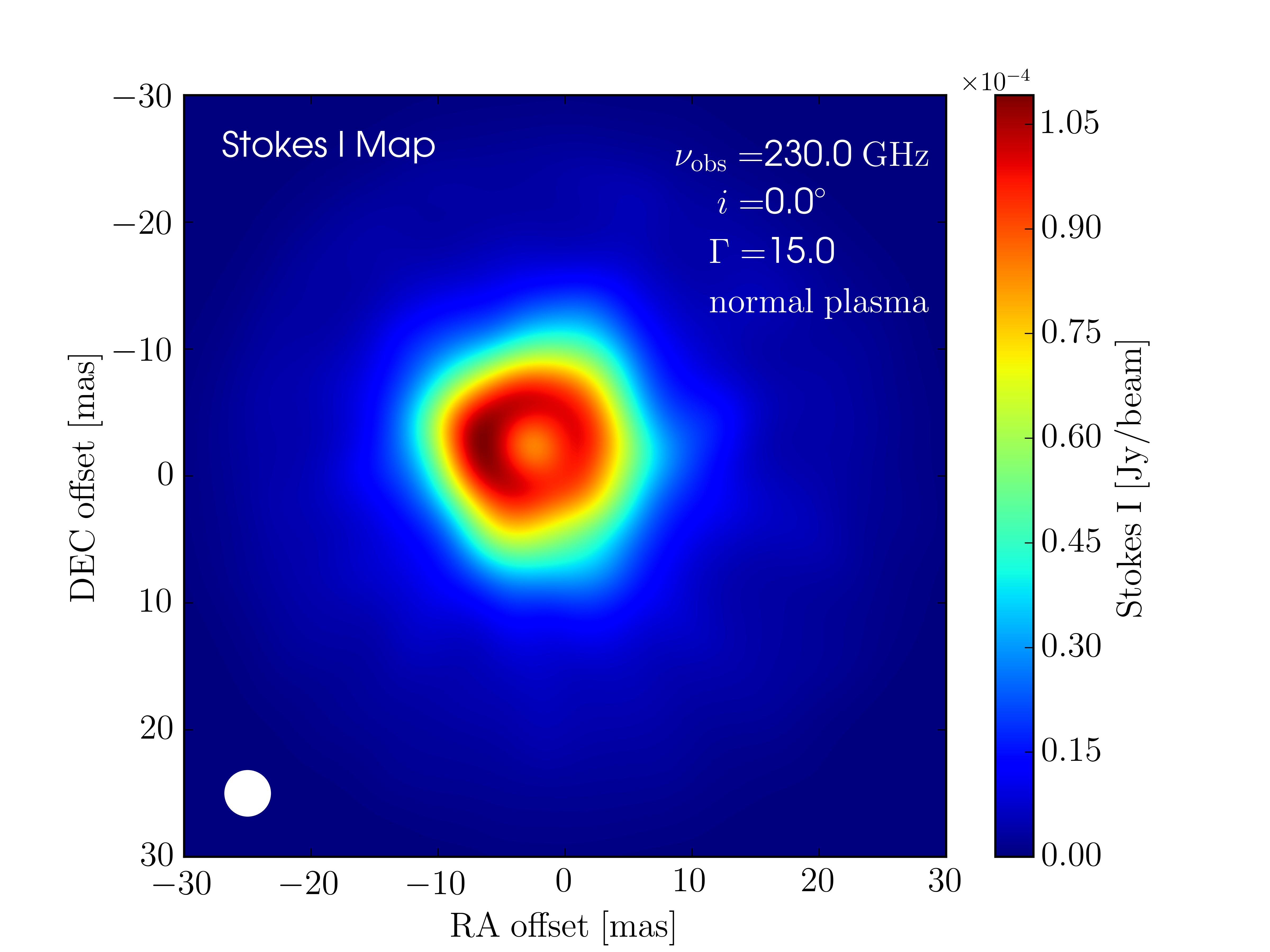}
\includegraphics[width=0.33\linewidth]{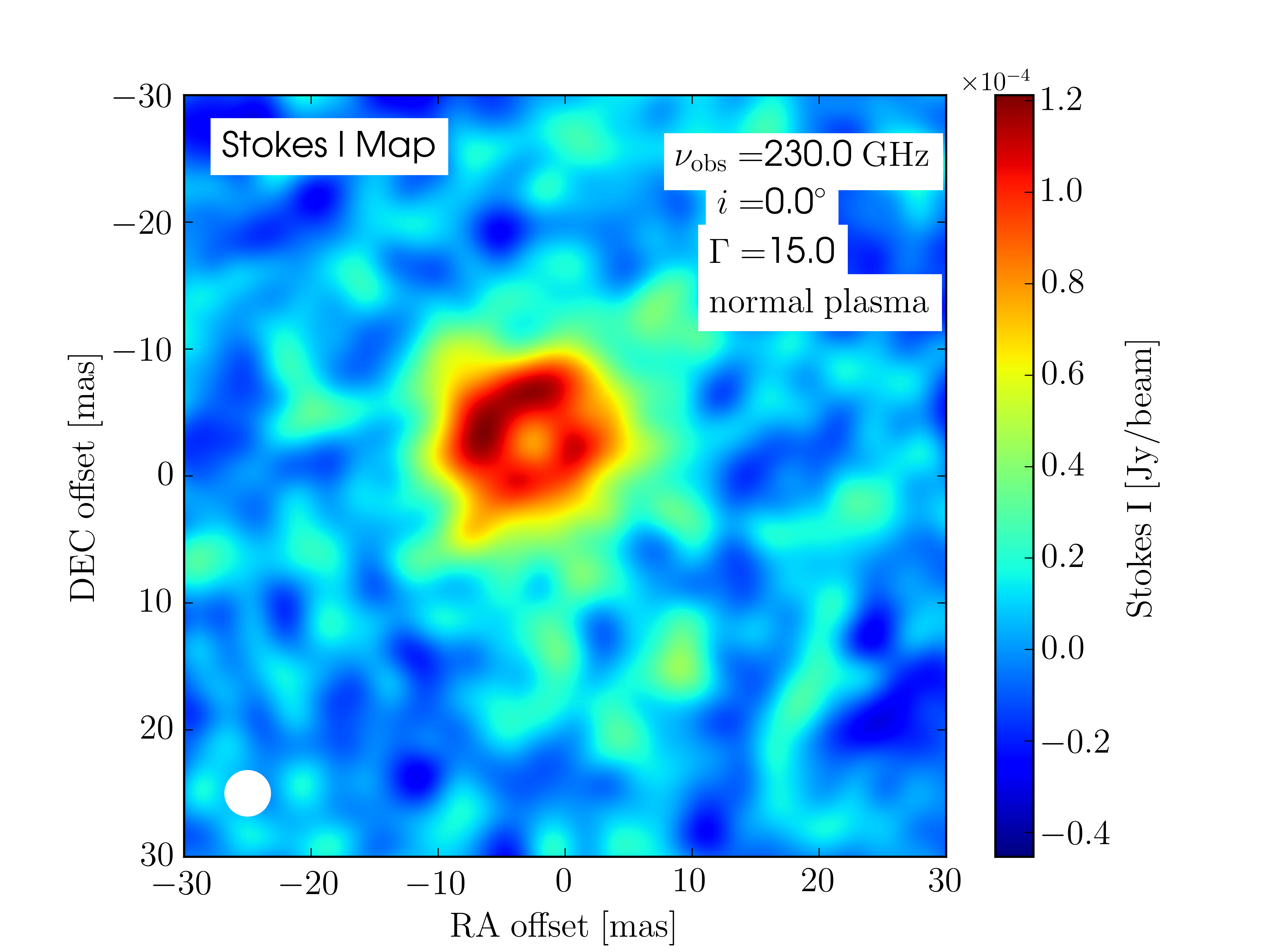}
\caption{\label{figE1}\textbf{Left panel:} Resolved fast-light Stokes $I$ image (at an observing frequency of $\nu_{\rm{obs}} = 230 ~ \rm{GHz}$) of the normal plasma ($e^{-}$-$p^{+}$) jet simulation. The orientation of the jet axis to our line-of-sight is $\theta_{\rm{obs}} = 0^{\circ}$. \textbf{Middle panel:} Fast-light Stokes $I$ image of the normal plasma jet, but in which the image has been convolved with a circular Gaussian beam of FWHM $3.5 ~\rm{mas}$ (shown in the lower left of the panel). \textbf{Right panel:} Fast-light Stokes $I$ image of the normal plasma jet, but in which an artificial Gaussian noise floor (set at $\sim 10^{-6} ~\rm{Jy}$/beam) has been included to illustrate the \textit{extreme} interferometric array sensitivity needed to detect our PIC jets even at such close proximity.}
\end{figure*}

In addition to being able to mimic the resolution of an interferometric array (via beam convolution), our ray-tracing algorithm can also mimic the sensitivity limit of an interferometric array through the introduction of a Gaussian noise floor into the synthetic images we produce. We illustrate this effect in Figure \ref{figE1}, in which we contrast images of the same simulation epoch; one in which we have no resolution/sensitivity limit (Left panel), one in which we apply a resolution limit by convolving our resolved image with a circular Gaussian beam of FWHM $3.5 ~\rm{mas}$ (Middle panel), and the other in which we additionally introduce a \textit{infinitesimally small} Gaussian noise floor of $\sim10^{-6} ~\rm{Jy}$/beam (Right panel). Clearly, the Doppler de-beamed jets presented in Figures \ref{fig3} and \ref{fig6} would not be detectable by existing ground-based interferometric arrays. We also point out that all of the images presented in this paper assume complete uv-coverage and, as such, represent highly idealized radio images. Future refinement of our ray-tracing algorithm is required in order to include the effects of finite uv-coverage in our synthetic maps.

\end{document}